\documentclass[aps,prx, twocolumn,notitlepage, showkeys,showpacs,superscriptaddress]{revtex4-1}
\usepackage{epsf}
\usepackage{bm}
\usepackage{times}
\usepackage{amsfonts}
\usepackage{amssymb}
\usepackage{amsmath,mathtools}
\usepackage{array}
\usepackage{enumerate,dsfont,here}
\usepackage{dcolumn,multirow}
\usepackage{tikz-cd}
\usepackage{bbding}
\usepackage[colorlinks=true,citecolor=blue,linkcolor=blue]{hyperref} 
\usepackage[normalem]{ulem}

\newcommand{\bk}{\bm{k}}

\newcommand{\br}{\bm{r}}
\newcommand{\bp}{\bm{p}}
\newcommand{\bb}{\bm{b}}
\newcommand{\ba}{\bm{a}}

\newcommand{\Xf}{X_f}

\newcommand{\mZ}{\mathbb{Z}}
\newcommand{\calT}{\mathcal{T}}
\newcommand{\calC}{\mathcal{C}}
\newcommand{\calG}{\mathcal{G}}
\newcommand{\calM}{\mathcal{M}}

\newcommand{\calA}{\mathcal{A}}

\newcommand{\calP}{\mathcal{P}}

\newcommand{\XBdG}{X_{\text{BdG}}}

\newcommand{\BS}{\{\mathrm{BS}\}}
\newcommand{\AI}{\{\mathrm{AI}\}}
\newcommand{\tiG}{\tilde{\mathcal{G}}}

\begin{document}

\title{$\mathbb{Z}_2$-enriched symmetry indicators for topological superconductors in the 1651 magnetic space groups}
\author{Seishiro Ono}
\affiliation{Department of Applied Physics, University of Tokyo, Tokyo 113-8656, Japan}
\author{Hoi Chun Po}
\affiliation{Department of Physics, Massachusetts Institute of Technology, Cambridge, Massachusetts 02139, USA}
\author{Ken Shiozaki}
\affiliation{Yukawa Institute for Theoretical Physics, Kyoto University, Kyoto 606-8502, Japan}

\begin{abstract}
While the symmetry-based diagnosis of topological insulators and semimetals has enabled large-scale discovery of topological materials candidates, the extension of these approaches to the diagnosis of topological superconductors remains a major open question.
One important new ingredient in the analysis of topological superconductivity is the presence of $\mathbb Z_2$-valued Pfaffian invariants associated with certain high-symmetry momenta. Such topological invariants lie beyond the conventional scope of symmetry representation theory for band structures, and as such they are nontrivial to incorporate into the systematic calculations of the symmetry indicators of band topology. Here, we overcome this challenge and report the full computation of the $\mathbb Z_2$-enriched symmetry indicators for superconductors in all symmetry settings. Our results indicate that incorporating the $\mathbb Z_2$ band labels enhance the diagnostic power of the scheme in roughly $60\%$ of the symmetry settings.
Our framework can also be readily integrated with first-principles calculations to elucidate on the possible properties of unconventional superconductivity in a given compound. As a demonstration, we analyze explicitly the interplay between pairing symmetry and topological superconductivity in the recently discovered superconductors CaPtAs and CaSb$_2$.
\end{abstract}

\maketitle

While spatial symmetries have always been a critical component in the analysis of unconventional pairing, their role in protecting distinct classes of topological materials has only been understood in the past decade. The rich set of phases protected by spatial symmetries includes, for instance, the mirror Chern insulators~\cite{Tanaka:2012aa,Hsieh:2012aa}, higher-order topological phases~\cite{Schindlereaat0346,PhysRevLett.119.246402,PhysRevLett.119.246401,Benalcazar61,PhysRevB.96.245115, PhysRevB.98.205129, PhysRevX.9.011012,Fang2017, 2005.13507}, and topological semimetals~\cite{RevModPhys.90.015001}. 
Recent theoretical frameworks have also enabled the classification of all topological crystalline insulators in the symmetry settings most relevant for materials~\cite{PhysRevX.8.011040, PhysRevX.8.031070, Shiozaki2018,1810.00801, Song2018, PhysRevB.99.075105, PhysRevB.99.085127, PhysRevB.99.115116,1907.09354}. However, there is generally a disconnect between the more abstract classification results and the concrete prediction of nontrivial topology in materials, and more direct methods for diagnosing band topology in first principle calculations are indispensable in topological materials discovery.

As is exemplified by the Fu-Kane parity criterion~\cite{PhysRevB.76.045302} and the partial detection of the Chern number by rotation eigenvalues \cite{PhysRevB.83.245132,PhysRevB.86.115112,PhysRevB.85.165120}, the representations of crystalline symmetries furnished by the Bloch states could indicate the presence of nontrivial band topology. 
This paradigm was developed further in Refs.~\cite{Po2017, TQC}, which provided general recipes for utilizing symmetry representations to isolate topological band structures from their trivial, atomic counterparts. These theories have subsequently led to comprehensive surveys of topological materials among crystal structure databases~\cite{Tang2019_NP, Tangeaau8725, Zhang2019,Vergniory2019,Tang2019}, and an enormous number of candidates of topological (crystalline) insulators and semimetals candidates have been discovered.

The discovery of topological insulators and semimetals has also shed a new light on the properties of unconventional superconductors~\cite{RevModPhys.80.1083,RevModPhys.87.137,PhysRevLett.113.046401,SatoFujimoto,PhysRevB.95.224514,PhysRevLett.122.227001}.
In particular, the intense research effort in understanding topological superconductors has also enabled the identification and analysis of stable nodal structures in the pairing amplitude~\cite{PhysRevB.90.024516,AZ-node,PhysRevB.97.180504,PhysRevB.97.134512,PhysRevB.99.134513}. Correspondingly,
the theory of symmetry indicators~\cite{Po2017} has also been generalized to superconductors~\cite{Ono-Watanabe2018,Ono-Yanase-Watanabe2019,Ahn2019,Skurativska2019,SI_Shiozaki,Ono-Po-Watanabe2020,SI_Luka}.
A new ingredient in the theory is the possible existence of $\mZ_2$-valued Pfaffian invariants at high-symmetry momenta~\cite{SI_Shiozaki, SI_Luka}. For example, Ref.~\onlinecite{SI_Luka} found that such $\mZ_2$-enriched symmetry indicators for $C_4$ symmetric systems with $^1E$ or $^2E$ pairings can detect the Chern number modulo eight, not four~\cite{PhysRevB.86.115112}. Although such $\mZ_2$-valued symmetry data could be related to the number of filled bands in the normal-state band structure under the weak-pairing assumption~\cite{PhysRevB.81.134508,PhysRevB.81.220504,PhysRevLett.105.097001,Ono-Po-Watanabe2020,SI_Luka,SI_Shiozaki}, they correspond to a new set of band labels which cannot be readily analyzed using the original framework of symmetry-indicators designed for non-superconducting band structure. Consequently, such $\mZ_2$ band labels were not incorporated into the systematic calculation in Ref.~\onlinecite{Ono-Po-Watanabe2020}.

In this work, we establish an efficient and general scheme for computing the $\mZ_2$-enriched symmetry indicators for topological superconductors. This is achieved by defining an auxiliary, but mathematically equivalent, problem in which the $\mZ_2$ labels could be treated as integer-valued. The full set of indicator groups we obtained is tabulated in Supplementary Materials (SM). 
We highlight that the $\mZ_2$-enrichment improved the diagnostic power of $589$ out of the $980$ symmetry settings relevant for spin-orbit-coupled superconductors with time-reversal symmetry (class DIII).
Furthermore, our formulation is readily applicable to realistic materials using density functional theory (DFT), and we illustrate the diagnosis algorithm using two recently discovered superconductors CaPtAs~\cite{PhysRevLett.124.207001} and CaSb$_2$~\cite{PhysRevMaterials.4.041801} as concrete examples.

\textit{Set up}.---In this work, we focus on superconductors which can be described by the Bardeen-Cooper-Schrieffer (BCS) theory. The Bogoliubov–de Gennes (BdG) Hamiltonian takes the form 
\begin{align}
	\label{eq:BdG}
	H_{\bk}^{\text{BdG}} \equiv \begin{pmatrix}
		H_{\bk}& \Delta_{\bk} \\
		\Delta_{\bk}^{\dagger} & -H_{-\bm{k}}^{*}
	\end{pmatrix},
\end{align}
where $H_{\bk}$ and $\Delta_{\bk}$ denote the Hamiltonian of the normal phase and the superconducting gap function, respectively. 


We recall the symmetries of $H_{\bk}^{\text{BdG}}$. Suppose that the normal phase has a magnetic space group (MSG) {$\calM$. An element $g\in \calM$ maps a point $\br$ in the real space to $g\br = p_g\br+\bm{t}_g$, where $p_g$ is an element of $\text{O}(3)$, and $\bm{t}_g$ is not always a lattice vector. There are four types of MSGs. 230 of them have only unitary symmetries, which are referred to as type I MSG, i.e., they are just space groups (SGs). Other MSGs are generally denoted by $\calM =\tiG +\calA$, where $\tiG$ is a SG and $\calA$ is the anti-unitary part of $\calM$. Furthermore, $\calA$ can be written as $\tiG g_0 \calT$, where $g_0$ is a unitary symmetry operation, and $\calT$ is the time-reversal (TR) operation. When $g_0$ is an element of $\tiG$, the MSG is called type II MSG, which is a direct product of $\tiG$ and $\{e, \calT\}$ with $e$ denoting the identity operation. On the other hand, $g_0$ of a type III MSG is not an element of $\tiG$ but $g_{0}^{2}$ is an element of $\tiG$. A type IV MSG has a half-translation as $g_0$.
The normal-state Hamiltonian $H_{\bk}$ and unitary matrices $U_{\bk}(g)$ and $U_{\bk}(a)$ for $\forall g\in\tiG$ and $\forall a\in\calA$ satisfy  $U_{\bk}(g)H_{\bk}=H_{g\bm{k}}U_{\bk}(g)$ and $U_{\bk}(a)H_{\bk}^{*}=H_{a\bm{k}}U_{\bk}(a)$. 
If the superconducting phase is also symmetric under $g$ and $a$, $\Delta_{\bk}$ should satisfy
\begin{align}
	U_{\bk}(g)\Delta_{\bk}U_{-\bk}^{T}(g) = \chi_g \Delta_{g\bk},\\
	U_{\bk}(a)\Delta_{\bk}^{*}U_{-\bk}^{T}(a) = \chi_a \Delta_{a\bk},
\end{align}
where $\chi_g,\chi_a \in \text{U}(1)$ characterize the symmetry property of superconducting gaps. Then, representations of $g$ in the superconducting phase is denoted by
 \begin{align}
 	U_{\bk}^{\text{BdG}}(g) \equiv \begin{pmatrix}
 	U_{\bk}(g) &  \\
 	& \chi_g U_{-\bm{k}}^{*}(g)
 	\end{pmatrix},\\
 	U_{\bk}^{\text{BdG}}(a) \equiv \begin{pmatrix}
 		U_{\bk}(g) &  \\
 		& \chi_a U_{-\bm{k}}^{*}(a)
 		\end{pmatrix},
 \end{align}
which satisfy $U_{\bk}^{\text{BdG}}(g)H_{\bk}^{\text{BdG}}=H_{g\bm{k}}^{\text{BdG}}U_{\bk}^{\text{BdG}}(g)$, $U_{\bk}^{\text{BdG}}(a)\left(H_{\bk}^{\text{BdG}}\right)^{*}=H_{a\bm{k}}^{\text{BdG}}U_{\bk}^{\text{BdG}}(a)$. 

Let us consider a point $\bk$ in the Brillouin zone (BZ). Then, we define a subgroup $\tiG_{\bk}$ of $\tiG$ by $ \{h \in  \tiG| h\bk = \bk +^\exists\bm{G}\}$, where $\bm{G}$ is a reciprocal lattice vector, and this subgroup is called ``little group.'' As discussed in Ref.~\onlinecite{SI_Luka}, we can always choose a basis such that
\begin{align}
	U_{\bk}^{\text{BdG}}(h) &=\text{diag}\left[U_{\bk}^{\alpha_1}(h)\otimes\mathds{1}_{N_1}, \cdots, U_{\bk}^{\alpha_n}(h)\otimes\mathds{1}_{N_n}\right],\\
	H_{\bk}^{\text{BdG}}&=\text{diag}\left[\mathds{1}_{D^{\alpha_1}}\otimes H^{\alpha_1}_{\bk}, \cdots, \mathds{1}_{D^{\alpha_n}}\otimes H^{\alpha_n}_{\bk}\right],
\end{align}
where $U_{\bk}^{\alpha}(h)$ is an irreducible representations of $\tiG_{\bk}$~\cite{SM}. Here, $D^{\alpha}$ and $N_{\alpha}$ are dimensions of $U_{\bk}^{\alpha}(h)$ and $H^{\alpha}_{\bk}$, respectively.

In addition to magnetic space group symmetries, the BdG Hamiltonian always has the particle-hole symmetry (PHS) $\calC$. When $U^{\text{BdG}}(\calC)$ denotes a unitary part of $\calC$, $U^{\text{BdG}}(\calC)$ satisfies $U^{\text{BdG}}(\calC)[H_{\bk}^{\text{BdG}}]^*=-H_{-\bm{k}}^{\text{BdG}}U^{\text{BdG}}(\calC)$ and $U^{\text{BdG}}(\calC)(U^{\text{BdG}}(\calC))^{*} = \eta_{\calC} \mathds{1}$, where $\eta_{\calC}=\pm 1$.
The full symmetry group $G$ is denoted by
\begin{align}
	G &= \calM + \calM \calC\\
	&= \begin{cases}
	 \tiG + \tiG \calC \quad \text{(type I MSGs)}\\
	 \tiG + \tiG g_0 \calT+ \tiG \calC  + \tiG g_0 \Gamma \ \ \text{(Other MSGs)}
	\end{cases}
\end{align}
where $\Gamma$ is the chiral symmetry coming from the product of $\calT$ and $\calC$. 

In general, representations can be projective. In other words, for $a, a' \in \tiG A$ ($A$ is an anti-unitary symmetry) and $g, g' \in \calG_0$ ($\calG_0$ is a unitary symmetry group), the representations satisfy
\begin{align}
\label{eq:factor-1}
U_{g'\bm{k}}^{\text{BdG}}(g)U_{\bk}^{\text{BdG}}(g') = z_{g,g'}U_{\bk}^{\text{BdG}}(gg'), \\
U_{a\bm{k}}^{\text{BdG}}(g)U_{\bk}^{\text{BdG}}(a) = z_{g,a}U_{\bk}^{\text{BdG}}(ga), \\
U_{g\bm{k}}^{\text{BdG}}(a)[U_{\bk}^{\text{BdG}}(g)]^* = z_{a,g}U_{\bk}^{\text{BdG}}(ag), \\
\label{eq:factor-2}
U_{a'\bm{k}}^{\text{BdG}}(a)[U_{\bk}^{\text{BdG}}(a')]^* = z_{a,a'}U_{\bk}^{\text{BdG}}(aa') ,
\end{align}
where $z_{g, g}, z_{g,a}, z_{a,g}, z_{a,a'}\in \text{U}(1)$ are projective factors. 

\textit{Emergent Altland-Zirnbauer symmetry classes}.---
Here, we discuss the effects of additional symmetry elements in $\tilde{\calT} = \tiG g_0 \calT$, $\tilde{\calC} =\tiG\calC$, and $\tilde{\Gamma} =\tiG g_0 \Gamma$}. For the part $\tilde{X}\ (X=\calT, \calC, \Gamma)$ of $G$, we define a subset 
$X_{\bk}$ of $\tilde{X}$ by $X_{\bk} = \{x \in  \tilde{X}| x\bk = \bk +^\exists\bm{G}\}$. As seen in the following discussions, one can determine effective internal symmetry classes by using the information of $X_{\bk}$.



Additional symmetries sometimes relate $H^{\alpha}_{\bk}$ to another sector $H^{\beta}_{\bk}$. To determine the action of these symmetries on $H^{\alpha}_{\bk}$,  we define the emergent Altland-Zirnbauer (EAZ) symmetry class for each sector by the following criteria~\cite{Bradley,Shiozaki2018}:
\begin{align}
&W^{\alpha}_{\bk}(\calC) =\frac{1}{\vert \calC_{\bk}/T \vert}\sum_{c \in \calC_{\bk}/T }\omega_{\bk}^{\calC}(c, c)\tilde{\chi}_{\bk}^{\alpha}(c^2) \in \{0, \pm 1\},
\end{align}
where $\omega_{\bk}^{\calC}(c, c)=e^{-i \bk \cdot (p_c \bm{t}_{c} + \bm{t}_{c})} z_{c,c}$ and $\tilde{\chi}_{\bk}^{\alpha}(g) = \mathrm{tr}[U_{\bk}^{\alpha}(g)e^{i\bk\cdot\bm{t}_g}]$. For MSGs except for type I, 
\begin{align}
W^{\alpha}_{\bk}(\calT) &=\frac{1}{\vert \calT_{\bk}/T \vert}\sum_{a \in \calT_{\bk}/T }\omega_{\bk}^{\calT}(a, a)\tilde{\chi}_{\bk}^{\alpha}(a^2) \in \{0, \pm 1\},\nonumber\\
W^{\alpha}_{\bk}(\Gamma) &= \frac{1}{\vert \tiG_{\bk}/T \vert} \sum_{g \in \tiG_{\bk}/T } \frac{\omega_{\bk}(\gamma, \gamma^{-1} g \gamma)}{\omega_{\bk}(g, \gamma)}[\tilde{\chi}^{\alpha}_{\bk}(\gamma^{-1} g \gamma)]^{*}\tilde{\chi}^{\alpha}_{\bk}(g),\nonumber\\
&\in\{0,1\},
\end{align}
where $\omega^{\calT}_{\bk}(a, a)=e^{-i \bk \cdot (p_a \bm{t}_{a} + \bm{t}_{a})} z_{a,a'}$, $\omega_{\bk}(g, g')=e^{-i \bk \cdot (p_g \bm{t}_{g'} - \bm{t}_{g'})} z_{g,g'}$, and $\gamma = g_0\Gamma$.
The EAZ symmetry class for $H_{\bk}^{\alpha}$ is determined by $(W^{\alpha}_{\bk}(\calT) , W^{\alpha}_{\bk}(\calC), W^{\alpha}_{\bk}(\Gamma))$. As discussed in Ref.~\onlinecite{SI_Luka}, we characterize band structures by  the Pfaffian invariants $p_{\bk}^{\alpha}\in\{0,1\}$ and irreducible representations $N_{\bk}^{\alpha}\in\mZ$ depending on the EAZ symmetry classes. Table~\ref{tab:EAZ} shows the correspondence between EAZ classes and topological indices. 

\begin{table}
	\begin{center}
		\caption{\label{tab:EAZ}The classification of EAZ symmetry classes. }
		\begin{tabular}{c|c|c|c}
			\hline
			EAZ & $(W^{\alpha}_{\bk}(\calT) , W^{\alpha}_{\bk}(\calC), W^{\alpha}_{\bk}(\Gamma)$ & classification & index  \\
			\hline\hline
			A & $(0,0,0)$ & $\mZ$ & $N_{\bk}^{\alpha}$ \\
			AIII & $(0,0,1)$ & $0$ & None \\
			AI & $(1,0,0)$ & $\mZ$ & $N_{\bk}^{\alpha}$ \\
			BDI & $(1,1,1)$ & $\mZ_2$ & $p_{\bk}^{\alpha}$ \\
			D & $(0,1,0)$ & $\mZ_2$ & $p_{\bk}^{\alpha}$ \\
			DIII & $(-1,1,1)$ & $0$ & None \\
			AII & $(-1,0,0)$ & $\mZ$ & $\tilde{N}_{\bk}^{\alpha}=N_{\bk}^{\alpha}/2$ \\
			CII & $(-1,-1,1)$ & $0$ & None \\
			C & $(0,-1,0)$ & $0$ & None \\
			CI & $(1,-1,1)$ & $0$ & None \\
			\hline
		\end{tabular}
	\end{center}
\end{table}

\textit{Linear algebra with $\mathbb Z_2$ entries}.---
Once we identify the EAZ symmetry classes for $H_{\bk}^{\alpha}$ at all high-symmetry points, we can map a band structure to a set of topological indices 
\begin{align}
\label{eq:vecn}
	\bm{n}=(n_1, n_2, \cdots, n_D).
\end{align}
For simplicity, we order the entries such that the first $D_P$ elements are $\mZ_2$ numbers and the rest are integers.

\textit{$\BS$}.--- In general, the elements of $\bm{n}$ should satisfy various kinds of constraints, called compatibility relations~\cite{Shiozaki2018,Po2017, TQC, SI_Luka, PhysRevX.7.041069}. Unlike the case of insulators, for the present case of superconductors with the Pfaffian invariants some compatibility relations are defined modulo two~\cite{SI_Luka}. We call them ``$\mZ_2$-compatibility relations.'' We suppose that there are $d$ compatibility relations, and $d_p$ out of $d$ are $\mZ_2$-compatibility relations. 
We can represent these relations by a $d \times D$ integer matrix $C$, for which the first $d_p$ rows are $\mZ_2$-compatibility relations.

The solution space to the compatibility relations can be identified with the kernel of $C$, which could be defined more concretely as  $\BS \equiv \ker C \cap \left(\mZ_{2}^{D_P} \times \mZ^{D-D_P} \right)$ \cite{Po2017}. A major difficulty in the systematic computation of $\BS$ here stems from the $\mZ_2$ nature of the Pfaffian invariants, and in the following we describe how the problem could be overcome.

The key idea is to notice that the computation can be performed by solving an auxiliary problem in which all the $\mZ_2$-values and relations are promoted to integer-valued ones.
First, we reinterpret all rows of $C$ as integer-valued compatibility relations. In the following discussion, ``prime'' indicates that we 
are considering such an auxiliary problem. For example, when we forget about $\mZ_2$-ness in the compatibility relations, we represent the matrix $C$ by $C'$ . To capture the $\mZ_2$-ness of original compatibility relations, we introduce $\tilde{\calP} = (\tilde{\bp}_1, \tilde{\bp}_2, \cdots, \tilde{\bp}_{d_p})$, where $\tilde{\bp}_i$ is a $d$-dimensional vector whose $j$-th component is $2\delta_{ij}$.

To see how $\tilde{\calP}$ works, let us suppose that $\bm{n}$ satisfies the compatibility relations when the $\mZ_2$-ness is taken into account. When we instead treat $\bm{n}$ as integer vectors, the first $d_p$ elements of $C'\bm{n}'$ can be non-zero and even, i.e., 
\begin{align}
\label{eq:comp-1}
C'\bm{n}' &= \sum_{i=1}^{d_p} m_i \tilde{\bp}_i \quad (m_i \in \mZ), 
\end{align} 
Therefore, we can rewrite this equation as 
\begin{equation}
\left(\begin{array}{cc} C' & \tilde{\calP}\end{array}\right)\tilde{\bm{n}}= \bm{0},
\end{equation}
where $\tilde{n}=(\begin{array}{c}n'_1, \cdots, n'_D,-m_1, \cdots,-m_{d_p}\end{array})$.

The computation of $\BS$ is now reduced to that of the null space of $\left(\begin{array}{cc} C' & \tilde{\calP}\end{array}\right)$. Suppose we get $l$ vectors, which are $(D+d_p)$-dimensional. 
By construction, only the first $D$ entries of these vectors are meaningful, and so we 
restrict ourselves to them. 
Furthermore, the first $D_P$ entries are initially $\mZ_2$-valued, and so we can replace them with their values modulo two~\cite{SM}.

Finally, we construct $\BS \simeq \mZ_{2}^{m} \times \mZ^n$. From the above vectors, we can find a set of independent vectors $\{\bb_i\}_{i=1}^{d_{\text{BS}}}$, and some of these vectors are generators of $\mZ_2$ part of $\BS$. Therefore, coefficients of such generators take $\mZ_2$-values. Then, we get
\begin{align}
\label{eq:BS}
\{\mathrm{BS}\} &= \left \{ \sum_{i=1}^{d_{\rm BS}}  r_i \bb_i\ :\ r_i \in \mZ_2\ \text{or } \mZ  \right\}.
\end{align}

\textit{Computation of symmetry indicators}.---Topological superconductors can be exposed by first isolating the subset of atomic-limit solutions to the compatibility relations, which we denote by $\AI$.
The systematic construction of the collections of atomic-limit Hamiltonian $\AI$ including the Pfaffian invariants can be performed in the same way as Refs.~\onlinecite{Po2017,Ono-Po-Watanabe2020,SI_Luka}. 
We now illustrate how to compute the $\mZ_2$-enriched symmetry indicators, mathematically defined through the quotient group $\XBdG = \BS/ \AI$.
The difficulty associated with the $\mZ_2$ nature of the Pfaffian invariants can again be overcome through the construction of an auxiliary, integer-valued problem.

Similar to before, to keep track of the $\mZ_2$-ness we define vectors 
$\bp_i$ whose $j$-th component takes $2\delta_{ij}$. Here, each $\bp_i$ is $d$-dimensional, and $i = 1,\dots, D_P$.
The auxiliary problem is defined through
\begin{align}
	\label{eq:BSf}
	\BS_f &= \left \{ \sum_{i=1}^{d_{\rm BS}}  r'_i \bb_i' + \sum_{j=1}^{D_P} m_j \bm{p}_j~:~ r'_i, m_j \in \mathbb Z \right\}.
\end{align}
Let $P$ be the group formed by the sums of $\bb_i$ with integer coefficients.
By construction, $P \subseteq \BS_f$ and $\BS=\BS_f/P$. Similarly, we define
\begin{align}
\label{eq:AIf}
\AI_f &= \left \{ \sum_{i=1}^{d_{\rm AI}}  s'_i \ba_i' + \sum_{j=1}^{D_P} l_j \bm{p}_j~:~ s'_i, l_j \in \mathbb Z \right\}.
\end{align}
Again, we find $P \subseteq \AI_f$ and $\AI=\AI_f/P$.


Since $P \subseteq \AI_f\subseteq \BS_f$~\cite{SM}, we can apply the third isomorphism theorem~\cite{Isomorphism} to $\XBdG$, and we get
\begin{align}
	\label{eq:X_f}
	\XBdG =\frac{\{ {\rm BS}\}}{\{ {\rm AI}\}} = \frac{\{ {\rm BS}\}_f/P}{\{ {\rm AI}\}_f/P} = \frac{\{ {\rm BS}\}_f}{\{ {\rm AI}\}_f} = X_f.
\end{align}
Therefore, what we have to do is to compute a quotient group between free abelian groups $\mZ^{N}$, which can be solved using the Smith normal form~\cite{Po2017,SI_Adrian}. 


Using the above scheme, we compute $X_f$ for MSGs $\calM$, one-dimensional representations $\chi_g$, signs of the square of PHS $\calC$ ($\eta_{\calC}=\pm 1$), and spinful/spinless systems. The classification tables are included in SM. In SM, we discuss what $\XBdG$ indicates for several examples. As shown in SM, symmetry indicators with Pfaffian invariants detect not only fully gapped topological (crystalline) superconductors, but also various nodal superconductors such as the Bogoliubov Fermi surfaces~\cite{PhysRevLett.118.127001}.

\textit{Application to realistic materials}.--- 
Having developed an algorithm for the systematic computation of the $\mZ_2$-enriched symmetry indicators for any symmetry setting, we could now apply it to investigate the topological properties of realistic superconducting materials.
We will first discuss the general scheme for materials diagnosis, and then illustrate it with two materials as examples: CaPtAs~\cite{PhysRevLett.124.207001} and CaSb$_2$~\cite{PhysRevMaterials.4.041801}.
Here, we perform density functional theory (DFT) calculations using QUANTUM-ESPRESSO~\cite{qe1,qe2}.
 We also use \textit{qeirreps}~\cite{qeirreps} to compute irreducible representations. 


\textit{General scheme}.---
The first step to diagnose topological superconducting phases in DFT calculations is to get the vector $\bm{n}$ defined in Eq.~\eqref{eq:vecn}.
Invoking the weak-pairing assumption, the entries in $\bm{n}$ can be inferred from the counts of irreducible representations at the various high-symmetry momenta~\cite{PhysRevB.81.134508,PhysRevB.81.220504,Ono-Po-Watanabe2020,SI_Luka}. 
More concretely, let $n_{\bk}^{\alpha}\vert_{\text{occ}}$ be the number of irreducible representations $U_{\bk}^{\alpha}$ among the occupied bands. Then we can define a Pfaffian-like $\mathbb Z_2$ label $p_{\bk}^{\alpha}=n_{\bk}^{\alpha}\vert_{\text{occ}} \mod 2$~\cite{SI_Luka}, as well as an integer-valued label $N_{\bk}^{\alpha}=n_{\bk}^{\alpha}\vert_{\text{occ}} - n_{-\bk}^{\bar{\alpha}}\vert_{\text{occ}}$~\cite{Ono-Po-Watanabe2020}.
Next, we check if the vector satisfies all compatibility relations. When $\bm{n}$ satisfies compatibility relations,  $\bm{n}$  can be expanded by the basis of $\BS$, i.e., $\bm{n} = \sum_{i=1}^{d_{\mathrm{BS}}}r_i \bb_i$ with $r_i \in \mZ_2 \text{ or }\mZ$. If the vector cannot be expanded, the target material should be a nodal superconductor. 

The final step is to judge whether the set of topological invariants at high-symmetry points is equivalent to a trivial one. To see this, we again expand the vector by the basis of $\AI$. When $\bm{n}$ cannot be expanded 
by the $\AI$ basis using integer coefficients, the target is a topologically nontrivial superconductor.
Note that, due to the $\mZ_2$-enrichment, the dimension of $\BS$ is different from that of $\AI$ in general. 
This is a new feature in the present scheme, and hence, unlike in  Refs.~\cite{Tang2019_NP, Tangeaau8725, Tang2019}, simply expanding $\bm{n}$ by the basis of $\AI$ does not allow one to extract all the information about the system within the symmetry indicator scheme.


\begin{figure}
	\begin{center}
		\includegraphics[width=0.99\columnwidth]{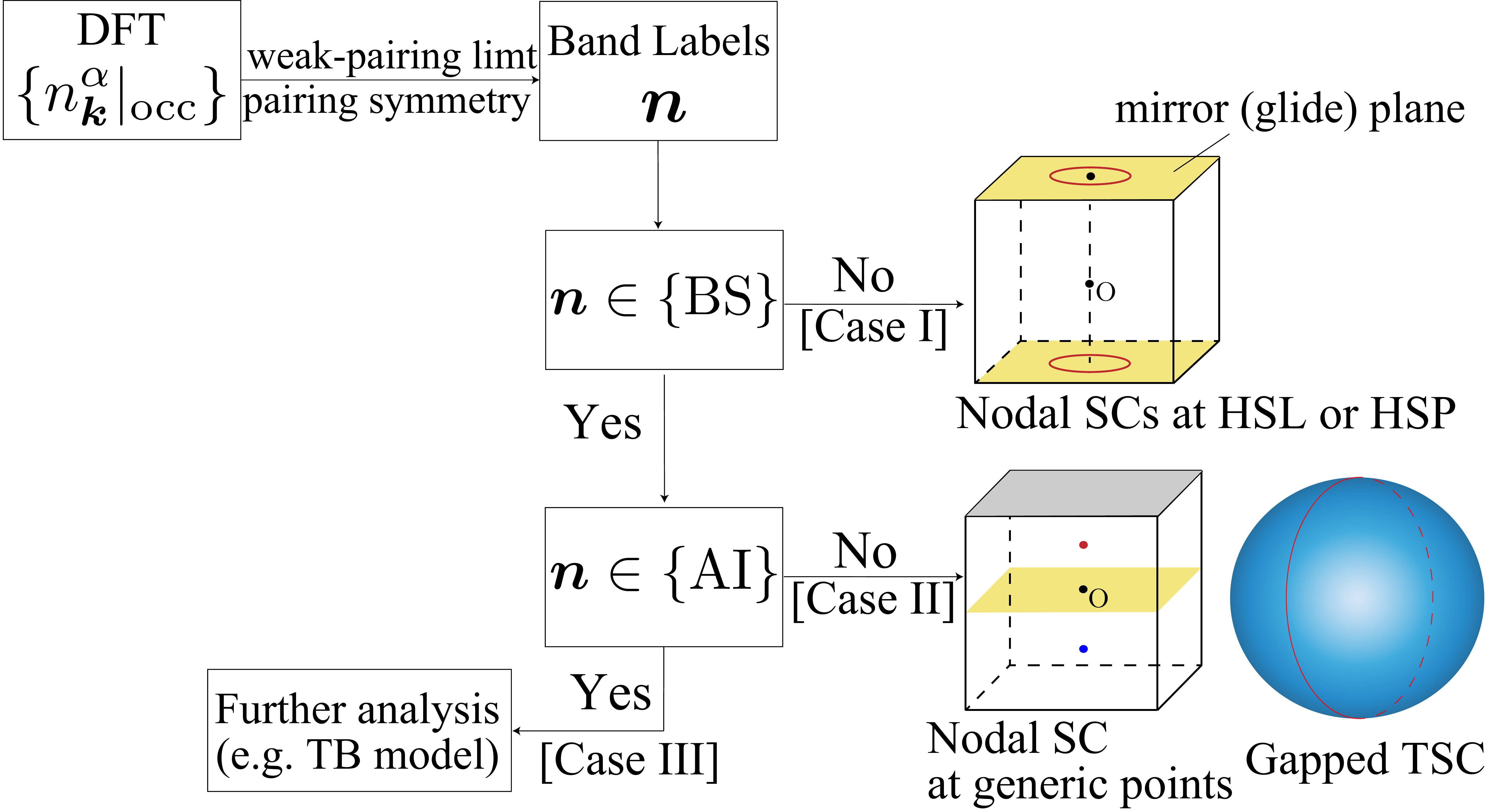}
		\caption{\label{fig:algorithm}Illustration of material investigations. First, we get irreducible representations in the normal phase using DFT. Second, we construct the vector $\bm{n}$ which is a set of band labels with assuming the weak-pairing assumption and pairing symmetry $\chi_g$. Third, we check if $\bm{n} \in \BS$. When $\bm{n} \notin \BS$, the superconductor should have nodal structures at high-symmetry lines (HSLs) or planes (HSPs). Finally, we ask if $\bm{n}\in \AI$. If the answer is no, the superconductor is topologically nontrivial. Note that this is a necessary condition, i.e., we cannot say the superconductor is trivial when $\bm{n}\in \AI$.
		}
	\end{center}
\end{figure}

We now apply the aforementioned method to the superconducting phases of CaPtAs and CaSb$_2$. The results are summarized in Table~\ref{tab:material}. In the following, let us highlight  the case of $B_{g}$ pairing for CaSb$_2$, which illustrates the importance of the $\mZ_2$ enrichment.

\textit{CaSb$_2$ with $B_{g}$ order parameter}.--- This material is a nodal-line metal even in the presence of the spin-orbit coupling~\cite{CaSb2_DFT}. Its space group is $P2_{1}/m$ (SG 11), which contains the inversion symmetry $I$, the two-fold screw symmetry $S_{2y}$, and the glide symmetry $G_y$. 
Recently, superconductivity has been reported~\cite{PhysRevMaterials.4.041801}. We find that this material with $B_{g}$ order parameter $(\chi_I=+1, \chi_{S_{2y}}=-1)$ is a symmetry-enforced nodal-line supercondutor. This is because compatibility relations along the $\Gamma$-Z, $\Gamma$-B, Y-C, and Y-A lines are violated. In the following discussions, we elaborate on the expected nodal structures of this pairing. 

%

\begin{figure}[t]
	\begin{center}
		\includegraphics[width=0.8\columnwidth]{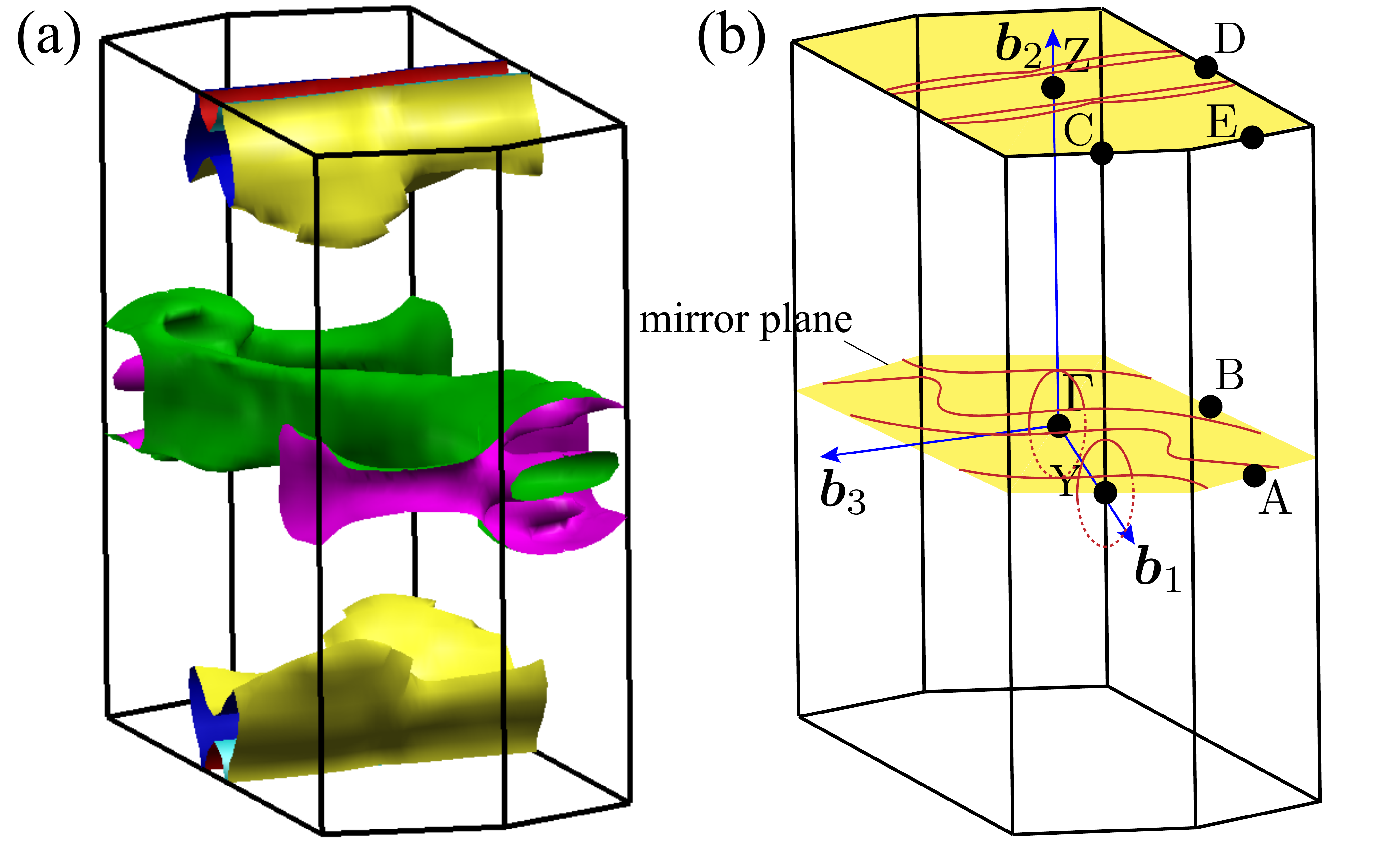}
		\caption{\label{fig:FS}Fermi surfaces of CaSb$_2$ (a) and a expected nodal structure (b). The red lines in (b) represent a expected nodal structure. Here, we assume that the \textit{strict} weak-pairing limit holds, and so the nodal structures in the mirror plane correspond to the Fermi surfaces.
		}
	\end{center}
\end{figure}

The main features are nodal lines in the mirror plane. In the $k_y=0$ plane, we can define the Pfaffian invariants for each mirror eigenvalue $p_{\bk}^{\pm i}$ at high-symmetry points $\Gamma$, Y, A, and B. When the system is fully gapped, $p_{\bk}^{\pm i}$ should be the same in this plane. However, one can find that the Pfaffian invariants at $\Gamma$ and Y are different from those at A and B~\cite{SM}, which indicates the violations of $\mZ_2$-compatibility relations at $\Gamma$-B, $\Gamma$-A, Y-B, and Y-A. These violations imply that there exist nodal lines separating regions with $p_{\bk}^{\pm i} =1$ and $p_{\bk}^{\pm i} =0$. 
On the other hand, since the EAZ class for all of Z, C, E, and D is  DIII, no band labels are assigned to these points. However, the EAZ classes at generic points in the $k_y = \pi$ plane are in class AII, and so we define integer-valued band labels $\tilde{N}_{\bk}^{\pm i}$ in Table~\ref{tab:EAZ}. If the system satisfies all compatibility relations, $\tilde{N}_{\bk}^{\pm i} = 0$. By computing irreducible representations at suitable points in this plane, we have identified regions where $\tilde{N}_{\bk}^{\pm i} \neq 0$~\cite{SM}. In fact, from the Fermi surfaces in Fig.~\ref{fig:FS}, one can see that there are two such regions bounded again by nodal lines. Yet, from the analysis on the high-symmetry momenta we see that the regions with $\tilde{N}_{\bk}^{\pm i} \neq 0$ could be shrunk out of existence, i.e., these nodal lines can be pair-annihilated by continuous deformations~\cite{SM}. Such annihilation of the nodal lines descending from the normal-state Fermi surface could be realized when the pairing energy scale exceeds that of the spin-orbit coupling.

Similar to the above case, a violated $\mZ_2$-compatibility relation at a two-fold rotation symmetric line is related to a nodal line pinned to the line, as discussed in Ref.~\cite{PhysRevB.99.134513}. Therefore, the violations of compatibility relations at $\Gamma$-Z and Y-C lines indicate the existence of such line nodes. In Fig.~\ref{fig:FS}, we show the Fermi surfaces and a expected nodal structure in the \textit{strict} weak-pairing limit.

\begin{table}
	\begin{center}
		\caption{\label{tab:material}Summary of material diagnosis. The first and second column lists materials and their pairing symmetries. The third and fourth column represents the categories in Fig.~\ref{fig:algorithm} diagnosed by the method in Ref.~\cite{Ono-Po-Watanabe2020} and this work, respectively. The fifth column indicates which types of topologies should appear. Detailed information is included in SM. 
		}
		\begin{tabular}{ccccc}
			\hline
			Materials & $\Delta_{\bk}$ & Ref.~\cite{Ono-Po-Watanabe2020} & This work & Topology \\
			\hline\hline
			CaPtAs~\cite{PhysRevLett.124.207001} & $^{1}E$ or $^{2}E$ & Case I & Case I~\cite{Note3} & Weyl SC~\cite{Note1}\\
			\hline
			\multirow{4}{*}{CaSb$_2$~\cite{PhysRevMaterials.4.041801}} & $A_u$ & Case I & Case I & Nodal-line SC~\cite{Note2} \\
			& $B_u$ & Case II & Case II & HOTSC \\
			& $A_g$ & Case III & Case III & $-$ \\
			& $B_g$ & Case III & Case I & Nodal-line SC\\
			\hline
		\end{tabular}
	\end{center}
\end{table}

\textit{Conclusion and Outlook}.---
In this work, we establish a systematic method to compute symmetry indicators taking into account the $\mZ_2$-valued Pfaffian invariants unique to superconductors~\cite{SI_Shiozaki,SI_Luka}. This is achieved by solving an auxiliary problem in which all the $\mZ_2$-invariants and compatibility relations are lifted to integer-valued ones. 
We have also computed exhaustively the $\mZ_2$-enriched symmetry indicators for symmetry classes relevant for (possibly magnetic) superconductors, and found that the diagnostic power of the symmetry indicator scheme is improved in about 60\% of the symmetry classes. We also illustrate how the symmetry indicator analysis can be integrated with DFT calculations to diagnose the topological and nodal features of unconventional superconductors using CaPtAs~\cite{PhysRevLett.124.207001} and CaSb$_2$~\cite{PhysRevMaterials.4.041801} as concrete examples.

Our results open up various possibilities for future studies. As discussed in Refs.~\cite{QuantitativeMappings,PhysRevX.8.031070}, symmetry indicators can be directly related to surface, hinge, and corner states and our results could guide the experimental effort in discovering materials candidates for higher-order topological superconductors. Furthermore, in the weak-pairing limit, band labels are reduced to irreducible representations in the normal phases. Hence, one can establish various connections between Fermi surfaces topologies and the descendent superconducting topological phases. 
Lastly, the demonstrated integration of the $\mZ_2$-enriched symmetry indicators with DFT-based analysis opens the way for performing comprehensive surveys for topological superconductor materials candidates, similar to those carried out for insulators and semimetals in Refs.~\cite{Tang2019_NP, Tangeaau8725, Zhang2019,Vergniory2019,Tang2019}.


\begin{acknowledgments}
	SO would like to thank 
	Akishi Matsugatani, Shuntaro Sumita, Nomoto Takuya, and Haruki Watanabe for valuable discussions. 
	HCP thanks Dominic Else for helpful discussions.
	A part of DFT calculations in this work has been done using the facilities of the Supercomputer Center, the Institute for Solid State Physics, the University of Tokyo.
	The work of SO is supported by Materials Education program for the future leaders in Research, Industry, and Technology (MERIT), The ANRI Fellowship, and  KAKENHI Grant No. JP20J21692 from the JSPS. 
	The work of HCP is supported by a Pappalardo Fellowship at MIT and a Croucher Foundation Fellowship.
	The work of KS is supported by PRESTO, JST (Grant No. JPMJPR18L4).
\end{acknowledgments}

\bibliography{ref}
\clearpage
\onecolumngrid

\begin{center}
	\textbf{\large Supplemental Materials for ``$\mZ_2$-enriched symmetry indicators for topological superconductors in the 1651 magnetic space groups''}
\end{center}

\section{Detailed computations}
\label{app:detailed}
In this section, we prove some statements in the main text. 
\subsection{Proof of canonical form}
\label{app:canonical}
In the main text,  we choose the basis such that
\begin{align}
\label{eq:U_cano}
U_{\bk}^{\text{BdG}}(h)&=\text{diag}\left[U_{\bk}^{\alpha_1}(h)\otimes\mathds{1}_{N_1}, \cdots, U_{\bk}^{\alpha_n}(h)\otimes\mathds{1}_{N_n}\right],\\
\label{eq:H_cano}
H_{\bk}^{\text{BdG}}&=\text{diag}\left[\mathds{1}_{D^{\alpha_1}}\otimes H^{\alpha_1}_{\bk}, \cdots, \mathds{1}_{D^{\alpha_n}}\otimes H^{\alpha_n}_{\bk}\right].
\end{align}
Here, we prove that we can do it.

First, $H_{\bk}^{\text{BdG}}$ and $U_{\bk}^{\text{BdG}}(h) $ can be (block-)diagonalized by
\begin{align}
\label{eq:BdGHam_2}
H_{\bk}^{\text{BdG}} &= \Psi_{\bk}
\begin{pmatrix}
\epsilon_{\bk1}\mathds{1}_{\alpha_1} & 0 & \cdots & 0 \\
0 &  \epsilon_{\bk2}\mathds{1}_{\alpha_2}  & \cdots & 0 \\
\vdots & \vdots & \ddots & \cdots \\
0 & 0 & \cdots & \epsilon_{ \bk M_{\bk}}\mathds{1}_{\alpha_{M_{\bk}}}
\end{pmatrix}\Psi_{\bk}^{\dagger},\\
\label{eq:Rep_TB}
U_{\bk}^{\text{BdG}}(h) &= \Psi_{\bk}
\begin{pmatrix}
U_{\bk}^{\alpha_1}(h) & 0 & \cdots & 0 \\
0 &  U_{\bk}^{\alpha_2}(h) & \cdots & 0 \\
\vdots & \vdots & \ddots & \cdots \\
0 & 0 & \cdots & U_{\bk}^{\alpha_{M_{\bk}}}(h) 
\end{pmatrix}\Psi_{\bk}^{\dagger},
\end{align}
where $\epsilon_{\bk n}$ is the $n$-th eigenenergy of $H_{\bk}^{\text{BdG}}$, and $\Psi_{\bk}$ is a unitary matrix composed of all eigenvectors of $H_{\bk}^{\text{BdG}}$.  
Then, we introduce the projection operator in Eq.~(4.38) of Ref.~\onlinecite{Springer}
\begin{align}
P^{\alpha}_{\bk} &=\frac{D^\alpha}{\vert H \vert}\sum_{h\in  H}\left[\chi_{\bk}^{\alpha}(h)\right]^{*}U_{\bk}^{\text{BdG}}(h).
\end{align}
For simplicity, we omit ``BdG'' and $\bk$ from here. From the Eq.\eqref{eq:Rep_TB}, we can rewrite the projection operator by
\begin{align}
P^{\alpha} &=\frac{D^\alpha}{\vert H \vert}\sum_{h\in H}\left[\chi^{\alpha}(h)\right]^{*}\begin{pmatrix}
U_{\bk}^{\alpha_1}(h) & 0 & \cdots & 0 \\
0 &  U_{\bk}^{\alpha_2}(h) & \cdots & 0 \\
\vdots & \vdots & \ddots & \cdots \\
0 & 0 & \cdots & U_{\bk}^{\alpha_{M_{\bk}}}(h) 
\end{pmatrix}.
\end{align}
Furthermore, $\frac{D^\alpha}{\vert H \vert}\sum_{h\in H }\left[\chi^{\alpha}(h)\right]^{*}U^{\beta}(h)=\mathds{1}_{D^{\alpha}}\delta^{\alpha\beta}$. From using this relation and exchanging some rows and columns, we can choose the basis such that
\begin{align}
P^{\alpha_n} &= \begin{pmatrix}
0& 0 & \cdots & 0 \\
0 &  \ddots & \cdots & 0 \\
0 & \cdots &  \mathds{1}_{D^{\alpha_n}} & 0\\
\vdots & \vdots & \ddots & \cdots \\
0 & 0 & 0 & 0
\end{pmatrix}\\
&= \mathds{1}_{D^{\alpha_n}}\otimes \tilde{\mathds{1}}_n,
\end{align}
where $\tilde{\mathds{1}}_n$ represents block-diagonal matrix whose $n$-th block is $N_n$-dimensional identity matrix. 

Finally, we can get the forms in Eqs.~\eqref{eq:U_cano}  and~\eqref{eq:H_cano} by 
\begin{align}
\left(\sum_{\alpha}P^{\alpha}\right)\left(V^{\dagger}\Psi^\dagger H^{\text{BdG}} \Psi V\right)\left(\sum_{\alpha}P^{\alpha}\right),\\
\left(\sum_{\alpha}P^{\alpha}\right)\left(V^{\dagger}\Psi^\dagger U^{\text{BdG}} \Psi V\right)\left(\sum_{\alpha}P^{\alpha}\right),
\end{align}
where $V$ denotes the product of elementary matrices. 

\subsection{Mathematical aspect on the computation of $\BS$}
Let us consider the following commutative diagram:
\begin{equation}\label{eq:Snake}
\begin{tikzcd}[row sep=small]
0 \arrow[r]  &  P \arrow[r,"i"]  \arrow[d,"C'\vert_{P}"]& \mZ^D \arrow[d,"C'"] \arrow[r,"\pi_1"] &  \mZ^D/P \arrow[d,"C"]\\
0 \arrow[r]  &  \tilde{P} \arrow[r,"\tilde{i}"] &  \mZ^d  \arrow[r,"\pi_2"] &  \mZ^d/\tilde{P}
\end{tikzcd},
\end{equation}
where 
\begin{align}
P &= \mathrm{span}_\mZ\left\{\bm{p}_{i} \ :\ i =1,\dots, D_P \right\} \quad ([\bm{p}_{i} ]_j = 2\delta_{ij}),\\
\tilde{P} &= \mathrm{span}_\mZ\left\{\tilde{\bm{p}}_{i} \ :\ i =1,\dots, d_p \right\} \quad ([\tilde{\bm{p}}_{i} ]_j = 2\delta_{ij}).
\end{align}
Moreover, from the above definitions, $\ker \pi_{1} = P$ and $\ker \pi_{2} = \tilde{P}$.

Ref.~\onlinecite{SI_Shiozaki} shows how to compute the kenerl of general homomorphism. Applying the discussion to our setting, we can compute the $\ker C$ by 
\begin{align}
\ker C &= \tilde{\pi}\left(\ker(C' \oplus \tilde{i}) \right)/i(P),\\
&= \pi_{1}\left(\tilde{\pi}\left(\ker(C' \oplus \tilde{i}) \right)\right),
\end{align}
where $\tilde{\pi}: \mZ^D  \oplus \tilde{P} \rightarrow \mZ^D $ is a projection. 

In the main text, we perform three steps to compute $\ker C$: (i) computing $\ker \left(\begin{array}{cc} C' & \tilde{\calP}\end{array}\right)$, (ii) picking the first $D$ integers from the basis of $\ker \left(\begin{array}{cc} C' & \tilde{\calP}\end{array}\right)$, and (iii) replacing the first $D_P$ integers with their values modulo two. The step (i), (ii), and (iii) correspond to  $\ker(C' \oplus \tilde{i})$, $\tilde{\pi}$, and $\pi_1$.

\subsection{Proof of $\AI_f \subseteq \BS_f$}
\label{app:proof}
Here we prove $\AI_f \subseteq \BS_f$. By definition, $\ba_i \in \AI$ must be expanded by the basis of $\BS$, i.e.,
\begin{align}
\ba_i = \sum_{i=1}^{d_{\text{BS}}}r_i \bb_i. 
\end{align}
When we forget about $\mZ_2$-ness in these vectors, 
\begin{align}
\ba'_i = \sum_{i=1}^{d_{\text{BS}}}r'_i \bb'_i + \sum_{j=1}^{D_P} m_j \bm{p}_j. 
\end{align}
Therefore, $\sum_{i=1}^{d_{\rm AI}}  s'_i \ba_i' + \sum_{j=1}^{D_P} l_j \bm{p}_j \in \AI_f$ can be written by
\begin{align}
\sum_{i=1}^{d_{\rm AI}}  s'_i \ba_i' + \sum_{j=1}^{D_P} l_j \bm{p}_j &= \left(\sum_{i=1}^{d_{\rm AI}} s'_i \right)\left(\sum_{i=1}^{d_{\text{BS}}}r'_i \bb'_i + \sum_{j=1}^{D_P} m_j \bm{p}_j\right)+\sum_{j=1}^{D_P} l_j \bm{p}_j \nonumber \\
&= \sum_{i=1}^{d_{\text{BS}}}\mathcal{N}r'_i \bb'_i + \sum_{j=1}^{D_P} (\mathcal{N}m_j  + l_j )\bm{p}_j,
\end{align}
where $\mathcal{N} \equiv \sum_{i=1}^{d_{\rm AI}} s'_i\in\mZ$. Therefore, any vector in $\AI_f$ is an element of $\BS_f$, i.e., $\AI_f \subseteq \BS_f$. 

\clearpage

\section{Computations of $X_f$ and its physical interpretation}
\label{app:example}
In this section, we present examples of the computation of $\BS$ and $X_f$. In addition, we discuss physical meaning of $\XBdG$. Through these examples, we reveal that symmetry indicators with Pfaffian invariants can indicate intriguing nodal structures. 

\subsection{2D $C_2$ symmetric systems in class DIII with $\chi_{C_2}=-1$}
\label{app:2DC2}
Let us consider two-dimensional systems orthogonal to the two-fold rotation axis with $\chi_{C_2}=-1$. There are four high symmetry points: $\Gamma=(0,0), X=(\pi,0), Y=(0,\pi)$, and $M=(\pi,\pi)$. In this symmetry setting, the BdG Hamiltonian and $U_{\bk}^{\text{BdG}}(C_2) $ can be block-diagonalized by
\begin{align}
H_{\bk}^{\text{BdG}} &= \text{diag}\left(H_{\bk}^{+},  H_{\bk}^{-}\right) \quad \ (\bk \in \text{TRIMs}),\\
U_{\bk}^{\text{BdG}}(C_2) &= \text{diag}\left(U_{\bk}^{+}(C_2) \mathds{1}_{N_+},  U_{\bk}^{-}(C_2) \mathds{1}_{N_-}\right),
\end{align}
where $U_{\bk}^{\alpha}(C_2) = i\alpha\ (\alpha=\pm 1)$ and $\mathds{1}_{N_\alpha}$ denotes the dimension of $H_{\bk}^{\alpha}$.
EAZ classes for $H_{\bk}^{+}$ and $ H_{\bk}^{-}$ are class D. Thus, the Pfaffian invariants characterize band structures, where $p_{\bk}^{\alpha}$ denotes the Pfaffian invariant for $H_{\bk}^{\alpha}$ as
\begin{align}
\label{BL-E1}
\bm{n} &= (p_{\Gamma}^{+}, p_{\Gamma}^{-},p_{X}^{+},p_{X}^{-},p_{Y}^{+},p_{Y}^{-}, p_{M}^{+},p_{M}^{-}). \nonumber
\end{align}

Due to TRS, $p_{\bk}^{\alpha}=p_{\bk}^{-\alpha}$ should be satisfied.
In addition, since the Pfaffian invariant is defined globally since $(C_{2}\calC)^2 = +1$, these indices should satisfy the following compatibility relations
\begin{align}
p_{\Gamma}^{+}+p_{\Gamma}^{-}&=p_{\bk}^{+}+p_{\bk}^{-}\mod 2\ \ (\bk \in X,Y,M),\\ 
p_{\bk}^{+}&=p_{\bk}^{-}\mod 2\ \ (\bk \in \Gamma, X,Y,M).
\end{align}
Then, $\calC'$ and $\tilde{\calP}$ are
\begin{align}
\calC' &= \begin{pmatrix}
1 & 1 & -1 & -1 & 0 & 0 & 0 & 0  \\
1 & 1 & 0 & 0 & -1 & -1 & 0 & 0 \\
1 & 1 & 0 & 0 & 0 & 0 & -1 & -1 \\
1 & 1 & 0 & 0 & 0 & 0 & 0 & 0 \\
0 & 0 & 1 & 1 & 0 & 0 & 0 & 0 \\
0 & 0 & 0 & 0 & 1 & 1 & 0 & 0 \\
0 & 0 & 0 & 0 & 0 & 0 & 1 & 1 \\
\end{pmatrix};\\
\tilde{\calP} &=\text{diag}(2,2,2,2,2,2,2).
\end{align}
From $\ker \left(\calC' \ \ \tilde{\calP}\right) $, we get $\{\text{BS}\} = (\mZ_{2})^4$, and 
\begin{align}
\BS &={\rm span}
\left\{
\begin{array}{cccccccc}
\bb_1=(1 & 1 & 0 & 0 & 0 & 0 & 0 & 0 )^\text{T} \\
\bb_2=(0 & 0 & 1 & 1 & 0 & 0 & 0 & 0 )^\text{T}  \\
\bb_3=(0 & 0 & 0 & 0 & 1 & 1 & 0 & 0 )^\text{T}  \\
\bb_4=(0 & 0 & 0 & 0 & 0 & 0 & 1 & 1)^\text{T}  \\
\end{array}
\right \} \nonumber,\\
&= \left \{ \sum_{i=1}^{4}  r_i \bb_i\ :\ r_i \in \mZ_2  \right\}\simeq (\mZ_2)^4.
\end{align}
We also have $\AI=\mZ_2$, and $\ba_1 = (1,1,1,1,1,1,1,1)$. 
Taking the quotient group, we get the symmetry indicator group $\XBdG=(\mZ_2)^3$. Two $\mZ_2$ factors of $\XBdG$ correspond to stacking Kitaev chains along $x$- or $y$-axes. The remaining $\mZ_2$ part can be understood by
\begin{align}
z_{2,xy} &=\frac{1}{2}\sum_{\bk \in \text{TRIMs}}\sum_{\alpha=\pm 1} p_{\bk}^{\alpha} \mod 2.
\end{align}

\begin{figure}[t]
	\begin{center}
		\includegraphics[width=0.99\columnwidth]{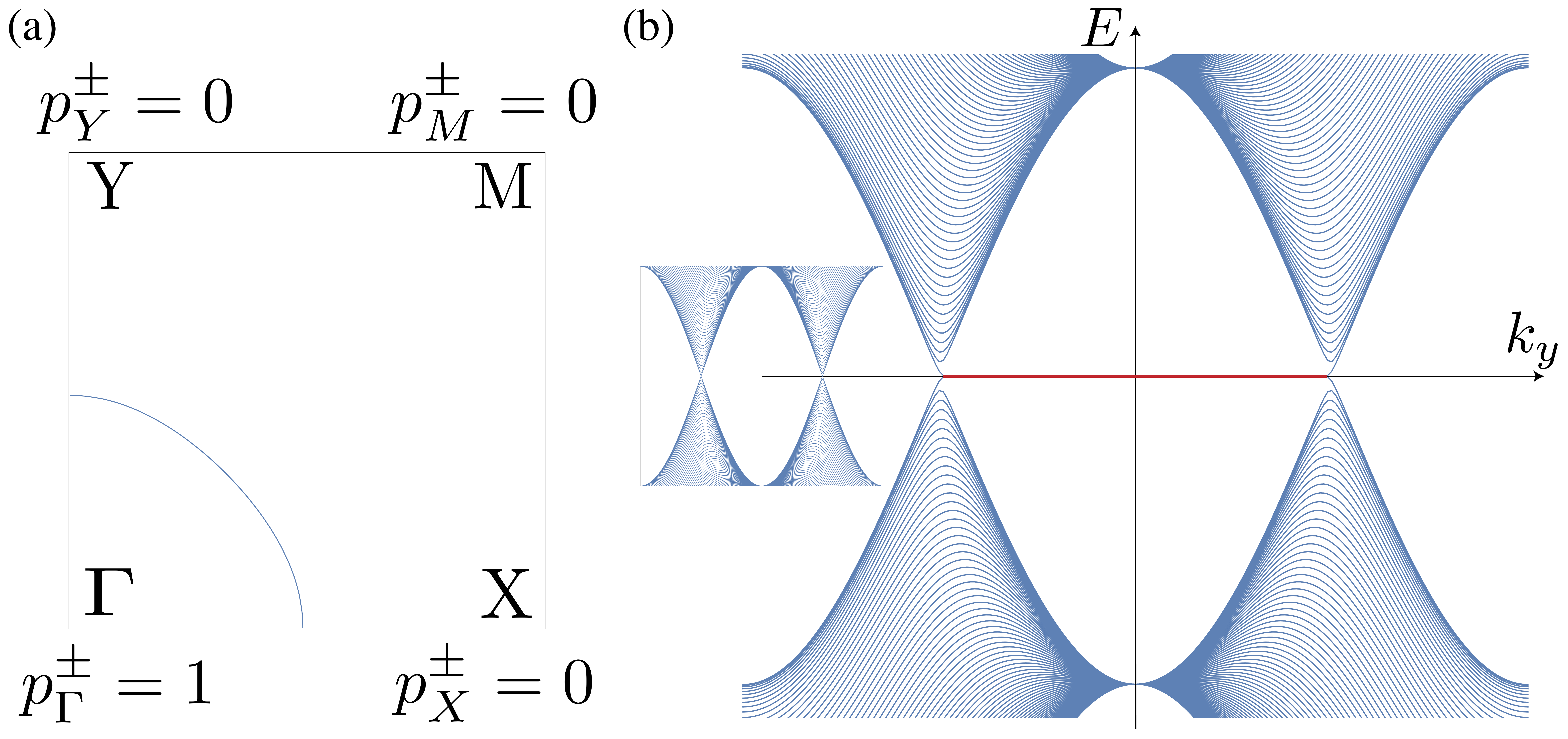}
		\caption{\label{fig:NodalC2}Numerical results of $H_{\bk}^{\text{BdG}}$ defined in Eq.~\eqref{eq:model_nodalC2}. (a) Fermi surface in the normal phase and Pfaffian invariants at high-symmetry points. (b) Quasiparticle spectrum with the open boundary condition imposed on $x$ direction. The inset is the bulk quasiparticle spectrum.}
	\end{center}
\end{figure}

From the following discussion, we reveal that $z_{2,xy}=1$ indicates point nodes at generic points. 
To see this, let us discuss the following model:
\begin{align}
\label{eq:model_nodalC2}
H_{\bk}^{\text{BdG}} &= (1-\cos k_x - \cos k_y )\sigma_0\tau_z + \sin k_x \sigma_x \tau_x,
\end{align}
which has $\bm{n} = (1,1,0,0,0,0,0,0)$ and $z_{2,xy}=1$. As seen in Fig.~\ref{fig:NodalC2} (b), this model has gapless points, and flat dispersion (red line in Fig.~\ref{fig:NodalC2} (b)) appears in the edge spectrum. 
One cannot add any additional mass term that opens the gap with keeping all symmetries. In addition, the homotopy group $\pi_1(R_1)=\mZ_2$ implies that point nodes at generic points are stable. 


\begin{table}[H]
	\begin{center}
		\caption{\label{tab:DIII-C2}Classification of node structures at generic points (GPs) for class DIII with $C_2$.}
		\begin{tabular}{c|c|c|c|c|c}
			\hline
			Symmetry & EAZ & classifying space & $\pi_0$ & $\pi_1$ & $\pi_2$\\
			\hline
			$\calT C_2, \calC C_2 , \Gamma$ ($\chi_{C_2}=-1$) & BDI & \multirow{1}{*}{$R_1$} & \multirow{1}{*}{$\mZ_2$} & \multirow{1}{*}{$\mZ_2$} & \multirow{1}{*}{$0$}\\
			\hline
		\end{tabular}
	\end{center}
\end{table}

\subsection{Inversion symmetric class BDI systems with $\chi_I=+1$}
Next, we discuss even-parity supeconductors in 2D and 3D class BDI. We find that symmetry indicators for these symmetry classes can also indicate nodal superconductors.
\subsubsection{2D}
We start with two-dimensional systems. There are four time-reversal invariant momenta (TRIMs): $\Gamma=(0,0), X=(\pi,0), Y=(0,\pi)$, and $M=(\pi,\pi)$. As with the case of $C_2$ with $\chi_{C_2} = -1$, the BdG Hamiltonian and $U_{\bk}^{\text{BdG}}(I) $ can be block-diagonalized by
\begin{align}
H_{\bk}^{\text{BdG}} &= \text{diag}\left(H_{\bk}^{+},  H_{\bk}^{-}\right) \quad \ (\bk \in \text{TRIMs}),\\
U_{\bk}^{\text{BdG}}(I) &= \text{diag}\left(U_{\bk}^{+}(I) \mathds{1}_{N_+},  U_{\bk}^{-}(I) \mathds{1}_{N_-}\right),
\end{align}
where $U_{\bk}^{\alpha}(I) = \alpha\ (\alpha=\pm 1)$ and $\mathds{1}_{N_\alpha}$ denotes the dimension of $H_{\bk}^{\alpha}$.
Then, EAZ classes for $H_{\bk}^{+}$ and $ H_{\bk}^{-}$ are class BDI, and so band structures are characterized by the Pfaffian invariant $p_{\bk}^{\alpha}$. There exist compatibility relations
\begin{align}
\sum_{\alpha=\pm 1}p_{\bk_1}^{\alpha} = \sum_{\alpha=\pm 1}p_{\bk_2}^{\alpha} \mod 2\ \ (\bk_1, \bk_2 \in \text{TRIMs}).
\end{align}
Again, we perform the same procedure as the above discussions.
\begin{align}
\calC' &= \begin{pmatrix}
1 & 1 & 0 & 0 & -1 & -1 & 0 & 0 \\
1 & 1 & -1 & -1 & 0 & 0 & 0 & 0 \\
0 & 0 & 1 & 1 & 0 & 0 & -1 & -1 \\
0 & 0 & 0 & 0 & 1 & 1 & -1 & -1 \\
\end{pmatrix};\\
\tilde{\calP} &=\text{diag}(2,2,2,2),
\end{align}
Then, we get
\begin{align}
\BS &={\rm span}
\left\{
\begin{array}{cccccccc}
\bb_1=(1 & 0 & 1 & 0 & 1 & 0 & 0 & 1 )^\text{T} \\
\bb_2=(1 & 0 & 1 & 0 & 1 & 0 & 1 & 0 )^\text{T}  \\
\bb_3=(0 & 0 & 0 & 0 & 1 & 1 & 0 & 0 )^\text{T}  \\
\bb_4=(0 & 0 & 1 & 1 & 0 & 0 & 0 & 0)^\text{T}  \\
\bb_5=(1 & 1 & 0 & 0 & 0 & 0 & 0 & 0 )^\text{T}  \\
\end{array}
\right \} \nonumber,\\
&= \left \{ \sum_{i=1}^{5}  r_i \bb_i\ :\ r_i \in \mZ_2  \right\}\simeq (\mZ_2)^5,\\
\AI &={\rm span}
\left\{
\begin{array}{cccccccc}
\ba_1=( 1 & 1 & 1 & 1 & 1 & 1 & 1 & 1)^\text{T} \\
\ba_2=(1 & 1 & 0 & 0 & 1 & 1 & 0 & 0  )^\text{T}  \\
\ba_3=( 1 & 0 & 1 & 0 & 1 & 0 & 1 & 0 )^\text{T}  \\
\ba_4=( 0 & 1 & 1 & 0 & 1 & 0 & 0 & 1)^\text{T}  \\
\end{array}
\right \} \nonumber,\\
&= \left \{ \sum_{i=1}^{4}  r_i \ba_i\ :\ r_i \in \mZ_2  \right\}\simeq (\mZ_2)^4.
\end{align}

From $\BS$ and $\AI$, we find the symmetry indicator group is $\XBdG = \mZ_2$, which can be understood by
\begin{align}
\label{eq:2D-index_BDI}
z_{2}^{2\text{D}} &=\frac{1}{2}\sum_{\bk \in \text{2D TRIMs}}\sum_{\alpha=\pm 1} p_{\bk}^{\alpha} \mod 2.
\end{align}

For this symmetry setting, the 1D winding number should be trivial, and there are no topological phases in the periodic table~\cite{Ryu_2010} for 2D class BDI. Hence, there are no weak and strong topological phases. Since any higher-order topological phases do not exist in the symmetry setting~\cite{PhysRevB.97.205136}, this index should indicate nodal superconductors. As seen in Table~\ref{tab:BDI}, the homotopy group is $\pi_0(\text{O}(n))=\mZ_2$~\cite{AZ-node}. Therefore, nodal lines at generic points can be stable. 

To see this, let us discuss the following example
\begin{align}
\label{eq:2D_nodalBDI}
H_{\bk}^{\text{BdG}} &=\left(4-2\cos k_x -2 \cos k_y -\mu \right)\sigma_z\tau_z+ \sin k_x \sigma_y \tau_z -\delta \sigma_0 \tau_z,
\end{align}
which has $z_{2}^{2\text{D}}=1$. As shown in Fig.~\ref{fig:NodalBDI} (a), this model has nodal lines. 

\begin{table}[H]
	\begin{center}
		\caption{\label{tab:BDI}Classification of node structures at generic points (GPs) for class BDI.}
		\begin{tabular}{c|c|c|c|c|c}
			\hline
			Symmetry & EAZ & classifying space & $\pi_0$ & $\pi_1$ & $\pi_2$\\
			\hline
			$\calT I, \calC I , \Gamma$ ($\chi_I=+1$) & BDI & \multirow{1}{*}{$R_1$} & \multirow{1}{*}{$\mZ_2$} & \multirow{1}{*}{$\mZ_2$} & \multirow{1}{*}{$0$}\\
			\hline
		\end{tabular}
	\end{center}
\end{table}

\subsubsection{3D}
For three-dimensional systems, the symmetry indicator group is $\XBdG = (\mZ_2)^4$. Three $\mZ_2$ factors of $\XBdG$ are originated from the lower dimension. The remaining $\mZ_2$ factor is characterized by
\begin{align}
\label{eq:3D-index_BDI}
z_{2}^{3\text{D}} &=\frac{1}{2}\sum_{\bk \in \text{3D TRIMs}}\sum_{\alpha=\pm 1} p_{\bk}^{\alpha} \mod 2.
\end{align}

As with the 2D case, fully gapped nontrivial phases cannot be realized. For 3D, $\pi_0(\text{O}(n))=\mZ_2$ implies that nodal surfaces, called the Bogoliubov Fermi surfaces~\cite{PhysRevLett.118.127001}, can be stable~\cite{AZ-node}. 

\begin{figure}[t]
	\begin{center}
		\includegraphics[width=0.8\columnwidth]{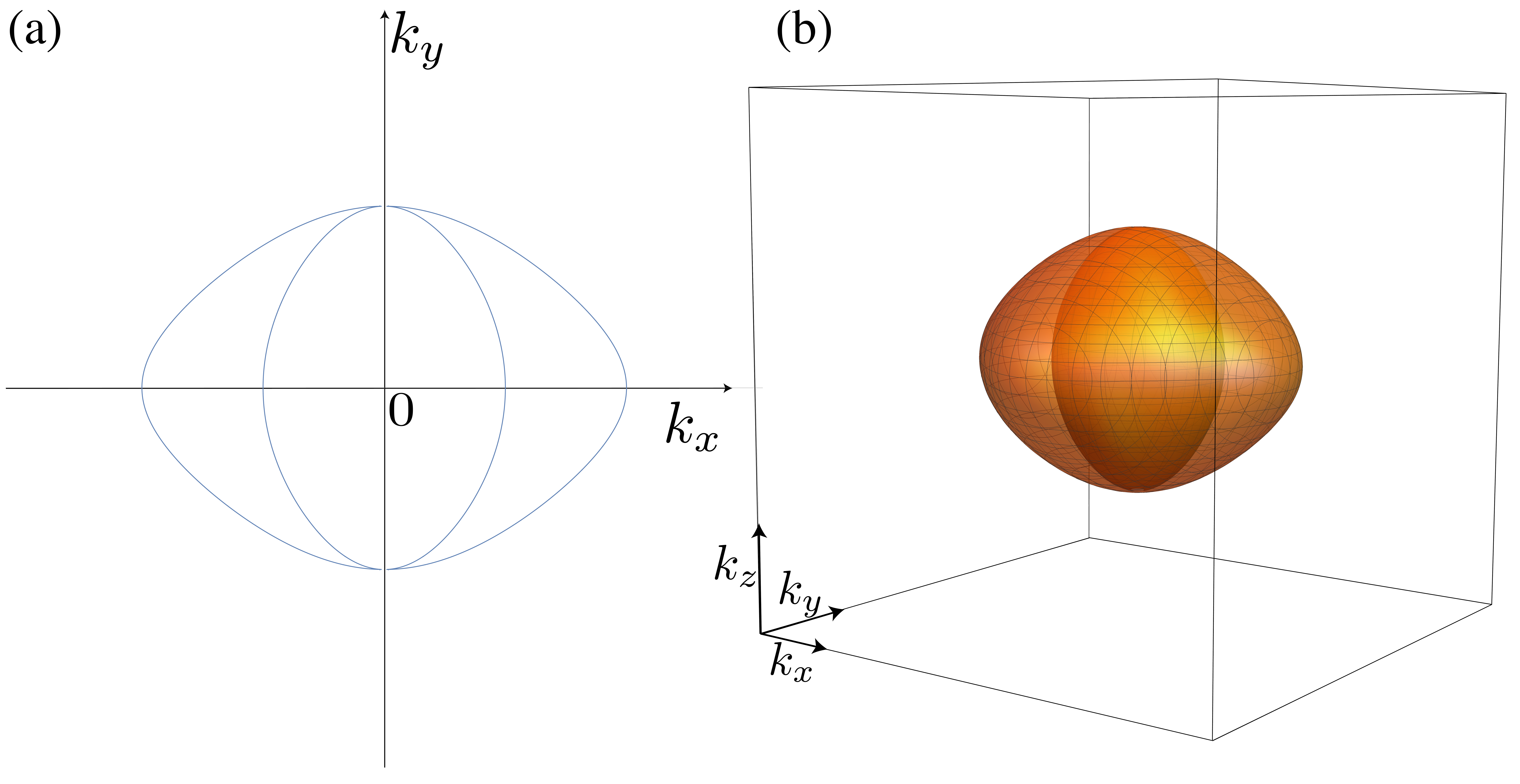}
		\caption{\label{fig:NodalBDI}(a) Nodal lines in $H_{\bk}^{\text{BdG}}$ defined in Eq.~\eqref{eq:2D_nodalBDI} for $\delta=1$ and $\mu=2$. (b) Bogoliubov Fermi surfaces in $H_{\bk}^{\text{BdG}}$ defined in Eq.~\eqref{eq:3D_nodalBDI} for $\delta=1$ and $\mu=2$.}
	\end{center}
\end{figure}

To demonstrate the topology, let us consider the following example
\begin{align}
\label{eq:3D_nodalBDI}
H_{\bk}^{\text{BdG}} &=\left(6-2\cos k_x -2 \cos k_y -2\cos k_z-\mu \right)\sigma_z\tau_z + \sin k_z \sigma_y \tau_z -\delta \sigma_0 \tau_z.
\end{align}
We calculate the index in Eq.~\eqref{eq:3D-index_BDI}, and we get $z_{2}^{3\text{D}}=1$.
As seen in Fig.~\ref{fig:NodalBDI}(b), we find surface nodes.

\subsection{2D $C_4$ symmetric systems in class C with $\chi_{C_4}=-i$}
\label{app:2DC4-C}
As discussed in the main text, we find $\XBdG = \mZ_4$ in $C_4$ symmetric systems with $\chi_{C_4}=-i$. 
In this symmetry setting, there are no Pfaffian invariants at high-symmetry points $\Gamma=(0,0), Y=(0, \pi)$, and $M=(\pi,\pi)$. Then, the band labels are ordered as follows:
\begin{align}
\label{BL-E1-C}
\bm{n} &= (N_{\Gamma}^{-1}, N_{\Gamma}^{-i},N_{\Gamma}^{i}, N_{\Gamma}^{1},N_{X}^{-1},N_{X}^{+1},N_{M}^{-1}, N_{M}^{-i},N_{M}^{i}, N_{M}^{1}), 
\end{align}
where $N_{\bk=\Gamma,M}^{\alpha}$ represent integer-valued band labels for irreducible representations $U_{\bk}^{\alpha}=\alpha\ (\alpha=\pm1,\pm i)$, and $N_{X}^{\alpha}$ denotes those for irreducible representations $U_{X}^{\alpha}=\alpha\ (\alpha=\pm1)$.
As the same as the above case, the Pfaffian invariant is defined globally since $({C_{4}}^{2}\calC)^2 = +1$.
From conservation of the Pfaffian invariant, these indices should satisfy the following compatibility relations
\begin{align}
N_{\Gamma}^{-1}+N_{\Gamma}^{-i}&=N_{M}^{-1}+N_{M}^{-i}\mod 2,\\
&=N_{X}^{-1}\mod 2.
\end{align}
Furthermore, the particle-hole symmetry relates two band labels as
\begin{align}
N_{\bk}^{\alpha}&=-N_{\bk}^{-i \alpha}\mod 2\ \ (\bk \in \Gamma,M),\\
N_{X}^{1}&=-N_{X}^{-1}\mod 2. 
\end{align}

Then, $\calC'$ and $\tilde{\calP}$ are
\begin{align}
\calC' &= \begin{pmatrix}
1 & 1 & 0 & 0 & -1 & 0 & 0 & 0 & 0 & 0  \\
1 & 1 & 0 & 0 & 0 & 0 & -1 & -1 & 0 & 0 \\
1 & 0 & 1 & 0 & 0 & 0 & 0 & 0 & 0 & 0 \\
0 & 1 & 0 & 1 & 0 & 0 & 0 & 0 & 0 & 0 \\
0 & 0 & 0 & 0 & 1 & 1 & 0 & 0 & 0 & 0 \\
0 & 0 & 0 & 0 & 0 & 0 & 1 & 0 & 1 & 0 \\
0 & 0 & 0 & 0 & 0 & 0 & 0 & 1 & 0 & 1 \\
\end{pmatrix};\\
\tilde{\calP} &=\text{diag}(2,2,0,0,0,0,0).
\end{align}
From $\ker \left(\calC' \ \ \tilde{\calP}\right) $, we get $\{\text{BS}\} = \mZ^5$, and 
\begin{align}
\BS &=\mathrm{span}
\left\{
\begin{array}{cccccccccc}
\bb_1=(2 & 0 & -2 & 0 & 0 & 0 & 0 & 0 & 0 & 0)^T \\
\bb_2=(0 & 0 & 0 & 0 & -2 & 2 & 0 & 0 & 0 & 0)^T \\
\bb_3=(-1 & 0 & 1 & 0 & -1 & 1 & 0 & -1 & 0 & 1)^T \\
\bb_4=(-1 & 0 & 1 & 0 & -1 & 1 & -1 & 0 & 1 & 0)^T \\
\bb_5=(1 & -1 & -1 & 1 & 0 & 0 & 0 & 0 & 0 & 0)^T \\
\end{array}
\right \} \nonumber,\\
&= \left \{ \sum_{i=1}^{5}  r_i \bb_i\ :\ r_i \in \mZ  \right\}.
\end{align}
We also find 
\begin{align}
\AI &=\mathrm{span}
\left\{
\begin{array}{cccccccccc}
\ba_1=(8 & 0 & -8 & 0 & 0 & 0 & 0 & 0 & 0 & 0)^T \\
\ba_2=(-10 & 2 & 10 & -2 & -2 & 2 & 2 & 2 & -2 & -2)^T \\
\ba_3=(3 & -1 & -3 & 1 & 2 & -2 & -1 & -1 & 1 & 1)^T \\
\ba_4=(1 & 0 & -1 & 0 & -1 & 1 & 1 & 0 & -1 & 0)^T \\
\ba_5=(-1 & 0 & 1 & 0 & -1 & 1 & 0 & 1 & 0 & -1)^T \\
\end{array}
\right \} \nonumber,\\
&= \left \{ \sum_{i=1}^{5}  s_i \ba_i\ :\ s_i \in \mZ  \right\}.
\end{align}
From $\BS$ and $\AI$, we get $\XBdG=\mZ_4$. 
Let us discuss the following toy model
\begin{align}
H_{\bk}^{\text{BdG}} &=\left(4-2\cos k_x -2 \cos k_y -\mu \right)\sigma_0\tau_z + \sin k_x \sigma_y \tau_x -\sin k_y\sigma_y \tau_y,\\
U^{\text{BdG}} (C_4) &=\text{diag}(\sigma_0, -i\sigma_0),\\
U^{\text{BdG}} (\calC) &= -i\sigma_0\tau_y.
\end{align}
This model for $\mu=1$ has the Chern number $\text{Ch}=2$. When we expand $\bm{n}$ for the model by the basis of $\AI$, we find a coefficient is $-3/4$. Therefore, two stacking of the model is the $2\in\mZ_4$ entry of $\XBdG$ with $\text{Ch}=4$.

%
%
%
\clearpage
\section{Detailed information of DFT calculations}
\label{app:DFT}
In this section, notations of irreducible representations and high-symmetry points are followed by Ref.~\cite{Bilbao}. Fig. 1 of the main text illustrates how our formulation is applied to searches for topological superconductors. Here, we show detailed calculations step by step.  


\subsection{CaPtAs}
\label{app:CaPtAs}
We discuss the topology of CaPtAs, whose space group is $I4_{1}md$ (SG 109). A recent experiment has reported nodal superconductivity and the time-reversal breaking~\cite{PhysRevLett.124.207001}. Breaking TRS implies that the order parameter belongs to two-dimensional representations. Furthermore, the order parameter also breaks some crystalline symmetries, i.e., $I4_{1}md$ (SG 109) is reduced. There are only two compatible possibilities with broken TRS: ${^1}E\ (\chi_{C_4}=-i)$ and $^2E\ (\chi_{C_4}=i)$ representations of the point group $C_4$. Since the case for $^2E$ is the same as that for $^1E$, we consider only the latter case. For these symmetry settings, the material can be a Weyl superconductor. We explain this one by one. 
\subsubsection{$^1E$ representation of $I4_1$}

\begin{enumerate}
	\item[(i)]\textbf{Getting irreducible representations in the normal phase} 
	
	From the DFT calculation with the spin-orbit coupling, we get irreducible representations in $I4_1md$ (SG 109). We can always decompose irreducible representations in $I4_1md$ into those in $I4_1$, and we have 
	\begin{align}
	\label{eq:SG80}
	\bm{n}_{\text{normal}} &= \left(n_{\Gamma}^{5},n_{\Gamma}^{6}, n_{\Gamma}^{7},n_{\Gamma}^{8},n_{\text{M}}^{5},n_{\text{M}}^{6}, n_{\text{M}}^{7}, n_{\text{M}}^{8},n_{\text{N}}^{2}, n_{\text{X}}^{3}, n_{\text{X}}^{4}, n_{\text{P}}^{3}, n_{\text{P}}^{3}, n_{\text{PA}}^{3},n_{\text{PA}}^{4}, n_{\text{T}}^{5},n_{\text{T}}^{6}, n_{\text{T}}^{7},n_{\text{T}}^{8},n_{\text{T}_0}^{5},n_{\text{T}_0}^{6}, n_{\text{T}_0}^{7},n_{\text{T}_0}^{8}\right)\nonumber\\
	&= (64,63,64,63,62, 62, 62, 62, 254, 132, 132,132, 132,132, 132,64,63,64,63,64,63,64,63),
	\end{align}
	where $n_{K}^{\alpha}$ is the number of irreducible representatinos $U_{K}^{\alpha}$ in the normal phase. 	
	
	\item[(ii)] \textbf{Band labels of the superconducting phase}
	
	We tabulate EAZ classes and band indices in Table~\ref{app:tab-C4}. In the weak-pairing limit, $p_{\bk}^{\alpha}=n_{\bk}^{\alpha} \mod 2$~\cite{SI_Luka} and $N_{\bk}^{\alpha}=n_{\bk}^{\alpha}- n_{-\bk}^{\bar{\alpha}}$~\cite{Ono-Po-Watanabe2020}. Then, we construct $\bm{n}$ from $\bm{n}_{\text{normal}}$ as follows:
	\begin{align}
	\bm{n} &= \left(p_{\Gamma}^{7},p_{\Gamma}^{8},p_{\text{M}}^{5},p_{\text{M}}^{6},p_{\text{N}}^{2}, p_{\text{X}}^{3}, p_{\text{X}}^{4}, N_{\Gamma}^{5},N_{\Gamma}^{6}, N_{\text{M}}^{7}, N_{\text{M}}^{8},N_{\text{P}}^{3}, N_{\text{P}}^{3}, N_{\text{PA}}^{3},N_{\text{PA}}^{4}\right) \nonumber\\
	\label{app:qeC4}
	&= (0,1,0,0,0,0,0,1,-1,0,0,0,0,0,0).
	\end{align}
	
	\begin{table}[H]
		\begin{center}
			\caption{\label{app:tab-C4}EAZ classes at high-symmetry points of $I4_1$.}
			\begin{tabular}{c|c|c|c}
				\hline
				Irrep & $(W^{\alpha}_{\bk}(\calT) , W^{\alpha}_{\bk}(\calC), W^{\alpha}_{\bk}(\Gamma))$ & EAZ class & topological invariant \\
				\hline\hline
				$\Gamma_5$ & $(0,0,0)$ & class A & $N_{\Gamma}^{5} = n_{\Gamma}^{5} - n_{\Gamma}^{6}$ \\
				$\Gamma_6$ & $(0,0,0)$ & class A & $N_{\Gamma}^{6}= n_{\Gamma}^{6} - n_{\Gamma}^{5}$ \\
				$\Gamma_7$ & $(0,1,0)$ & class D & $p_{\Gamma}^{7} = n_{\Gamma}^{7}\mod 2$ \\
				$\Gamma_8$ & $(0,1,0)$ & class D & $p_{\Gamma}^{8}= n_{\Gamma}^{8}\mod 2$ \\
				M$_5$ & $(0,1,0)$ & class D & $p_{\text{M}}^{5} = n_{\text{M}}^{5}\mod 2 $ \\
				M$_6$ & $(0,1,0)$ & class D & $p_{\text{M}}^{6}= n_{\text{M}}^{6}\mod 2 $ \\
				M$_7$ & $(0,0,0)$ & class A & $N_{\text{M}}^{7} = n_{\text{M}}^{7} - n_{\text{M}}^{8}$ \\
				M$_8$ & $(0,0,0)$ & class A & $N_{\text{M}}^{8}= n_{\text{M}}^{8} - n_{\text{M}}^{7}$ \\
				N$_2$ & $(0,1,0)$ & class D & $p_{\text{N}}^{2}= n_{\text{N}}^{2}\mod 2 $ \\
				X$_3$ & $(0,1,0)$ & class D & $p_{\text{X}}^{3}= n_{\text{X}}^{3}\mod 2$ \\
				X$_4$ & $(0,1,0)$ & class D & $p_{\text{X}}^{4}= n_{\text{X}}^{4}\mod 2$ \\
				P$_3$ & $(0,0,0)$ & class A & $N_{\text{P}}^{3}=n_{\text{P}}^{3} - n_{\text{P}}^{4}$ \\
				P$_4$ & $(0,0,0)$ & class A & $N_{\text{P}}^{4}=n_{\text{P}}^{4} - n_{\text{P}}^{3}$ \\
				PA$_3$ & $(0,0,0)$ & class A & $N_{\text{PA}}^{3}=n_{\text{PA}}^{3} - n_{\text{PA}}^{4}$ \\
				PA$_4$ & $(0,0,0)$ & class A & $N_{\text{PA}}^{4}=n_{\text{PA}}^{4} - n_{\text{PA}}^{3}$ \\
				\hline
			\end{tabular}
		\end{center}
	\end{table}	 
	\item[(iii)]  \textbf{Diagnosing topological phases based on symmetry indicators}
	
	Due to the compatibility relations at the line between $\Gamma$ and $M=(0,0,1)$, all integer invariants at $\Gamma$ and $M$ should vanish, as seen in Eq.~\eqref{eq:BS_SG80}. However, , as one can find in Eq.~\eqref{app:qeC4}, $N_{\Gamma}^{5}\neq 0$ and $N_{\Gamma}^{6}\neq 0$. This indicates that the superconductor should be a nodal superconductor.
	\begin{align}
	\label{eq:BS_SG80}
	\BS &=\mathrm{span}
	\left\{
	\begin{array}{ccccccccccccccc}
	\bb_1=(0 & 0 & 0 & 0 & 1 & 0 & 0 & 0 & 0 & 0 & 0 & 0 &  0 & 0 & 0  )^T  \\
	\bb_2=(1 & 0 & 1 & 0 & 0 & 1 & 0 & 0 & 0 & 0 & 0 & 0 &  0 & 0 & 0  )^T  \\
	\bb_3=(1 & 0 & 0 & 1 & 0 & 1 & 0 & 0 & 0 & 0 & 0 & 0 &  0 & 0 & 0  )^T  \\
	\bb_4=(0 & 0 & 0 & 0 & 0 & 1 & 1 & 0 & 0 & 0 & 0 & 0 &  0 & 0 & 0  )^T  \\
	\bb_5=(1 & 1 & 0 & 0 & 0 & 0 & 0 & 0 & 0 & 0 & 0 & 0 &  0 & 0 & 0 )^T  \\
	\end{array}
	\right\},\\
	&= \left \{ \sum_{i=1}^{5}  r_i \bb_i\ :\ r_i \in \mZ_2  \right\},\nonumber
	\end{align}
	
	\item[(iv)]  \textbf{Further analysis}
	
	From the above discussions, we cannot judge if the nodes are Weyl points from only the violation compatibility relations. To understand the origin of the nodes, we further analyze its topology.
	To simplify the following discussions, let us forget about body-centered translations. Then, we can map all information in $I4_1$  [Eq.~\eqref{eq:SG80}] to those in $P4_1$.
	\begin{align}
	\bm{n}_{\text{normal}} &= \left(n_{\Gamma}^{5},n_{\Gamma}^{6}, n_{\Gamma}^{7},n_{\Gamma}^{8},n_{\text{Z}}^{5},n_{\text{Z}}^{6}, n_{\text{Z}}^{7},n_{\text{Z}}^{8},n_{X}^{3}, n_{X}^{4},n_{R}^{3}, n_{R}^{4}, n_{\text{M}}^{5},n_{\text{M}}^{6}, n_{\text{M}}^{7}, n_{\text{M}}^{8}, n_{\text{A}}^{5}, n_{\text{A}}^{6}, n_{\text{A}}^{7}, n_{\text{A}}^{8}\right)\nonumber\\
	&= (125,126,126,125,127, 127, 126, 128, 254, 254, 254, 254, 132, 132,132, 132,132, 132,132, 132).
	\end{align}
	
	In the weak-pairing limit with assuming $\chi_{C_4}=+i$, 
	\begin{align}
	\bm{n} &= \left(p_{\Gamma}^{7},p_{\Gamma}^{8},p_{\text{X}}^{3},p_{\text{X}}^{4},p_{\text{M}}^{5},p_{\text{M}}^{6}, N_{\Gamma}^{5},N_{\Gamma}^{6}, N_{\text{Z}}^{5},N_{\text{Z}}^{6}, N_{\text{Z}}^{7},N_{\text{Z}}^{8}, N_{\text{R}}^{3},N_{\text{R}}^{4}, N_{\text{M}}^{7}, N_{\text{M}}^{8},N_{\text{A}}^{5}, N_{\text{A}}^{6}, N_{\text{A}}^{7},N_{\text{A}}^{8}\right) \nonumber\\
	&= (0,1,0,0,0,0,-1,1,1,-1,-1,1,0,0,0,0,0,0,0,0).
	\end{align}
	
	According to Refs.~\cite{PhysRevB.86.115112,Wang:2016aa}, we can compute the Chern numbers at $k_z = 0, \pi$ planes $\left(\text{Ch}_{k_z=0,\pi}\right)$ from eigenvalues of screw symmetries. From these results, we get
	\begin{align}
	\left(\text{Ch}_{k_z=0}, \text{Ch}_{k_z=\pi}\right) = (2,0) \mod 4.
	\end{align}
	Hence, we conclude Weyl nodes should exist between the $k_z=0$ and the $k_z=\pi$ planes. Note that we cannot exclude the possibility of coexistence of point nodes and surface nodes. In systems without the inversion and the time-reversal symmetries, surface nodes can also be stable. 
\end{enumerate}

\subsection{CaSb$_2$}
\label{app:CaSb2}
This material is a nodal-line metal even in the presence of the spin-orbit coupling~\cite{CaSb2_DFT}. Its space group is $P2_{1}/m$ (SG 11), which contains the inversion symmetry $I$, the two-fold screw symmetry $S_{2y}$, and the glide symmetry $G_y$. 
Recently, superconductivity has been reported~\cite{PhysRevMaterials.4.041801}. 
There are four one-dimensional representations: $B_{g}(\chi_I = +1,\chi_{G_{y}}=-1)$, $A_{g}(\chi_I = +1,\chi_{G_{y}}=+1)$, $B_{u}(\chi_I = -1,\chi_{G_{y}}=+1)$, and $A_{u}(\chi_I = -1,\chi_{G_{y}}=-1)$. 
Here, we analyze all possible pairings as follows.

\subsubsection{$B_u$ representations}
For this pairing symmetry, the symmetry indicator group is $\XBdG = \mZ_2 \times (\mZ_4)^2 \times \mZ_8$. After constructing $\bm{n}$ from DFT results, we find that $\bm{n}$ is an element of $\BS$, but not $\AI$. In addition, we compute topological indices that characterize $\XBdG$ in Ref.~\cite{Ono-Po-Watanabe2020}, and we see that this material is the entry $(1,1,0,2) \in\XBdG$.
\begin{enumerate}
	\item[(i)] \textbf{Getting irreducible representations in the normal phase} 
	
	\quad From the DFT calculation with the spin-orbit coupling, we get
	\begin{align}
	\bm{n}_{\text{normal}} &= \left(n_{\Gamma}^{3}, n_{\Gamma}^{4},n_{\Gamma}^{5},n_{\Gamma}^{6}, n_{\text{B}}^{3}, n_{\text{B}}^{4},n_{\text{B}}^{5},n_{\text{B}}^{6},n_{\text{Y}}^{3}, n_{\text{Y}}^{4},n_{\text{Y}}^{5},n_{\text{Y}}^{6}, n_{\text{Z}}^{2}, n_{\text{C}}^{2}, n_{\text{D}}^{2}, n_{\text{A}}^{3}, n_{\text{A}}^{4},n_{\text{A}}^{5},n_{\text{A}}^{6}, n_{\text{E}}^{2}\right) \nonumber\\
	\label{eq:noraml}
	&= (21,21,18,18,20,20,20,20,20,20,19,19,42,40,42,19,19,21,21,40), 
	\end{align}
	where $n_{K}^{\alpha}$ is the number of irreducible representations $U_{K}^{\alpha}$ in the normal phase. 
	
	\item[(ii)] \textbf{Band labels of the superconducting phase} 
	
	\quad In this symmetry setting, all EAZ classes at TRIMs in $k_y=\pi$ plane are class CII, and those in $k_y = 0$ plane are class A. Using the weak-pairing assumption, we get
	\begin{align}
	N_{\bk}^{3} &= -N_{\bk}^{5} =n_{\bk}^{3} - n_{\bk}^{5}\quad (\bk \in \Gamma, \text{B, A, Y}), \\
	N_{\bk}^{4} &= -N_{\bk}^{6} =n_{\bk}^{4} - n_{\bk}^{6}\quad (\bk \in \Gamma, \text{B, A, Y})
	\end{align}
	Therefore, the band structure are characterized by
	\begin{align}
	\bm{n} &=\left(N_{\Gamma}^{3}, N_{\Gamma}^{4},N_{\Gamma}^{5},N_{\Gamma}^{6}, N_{\text{B}}^{3}, N_{\text{B}}^{4},N_{\text{B}}^{5},N_{\text{B}}^{6},N_{\text{Y}}^{3}, N_{\text{Y}}^{4}, N_{\text{Y}}^{5}, N_{\text{Y}}^{6}, N_{\text{A}}^{3}, N_{\text{A}}^{4},N_{\text{A}}^{5},N_{\text{A}}^{6}\right) \nonumber \\
	\label{eq:Bu}
	&= (3,3,-3,-3,0, 0, 0, 0, 1,1,-1,-1, -2,-2,2,2).
	\end{align}
	
	\item[(iii)]  \textbf{Diagnosing topological phases based on symmetry indicators}
	
	\quad After solving compatibility relations by following the scheme in the main text, we have
	\begin{align}
	\label{eq:BS_Bu}
	\BS &=\mathrm{span}
	\left\{
	\begin{array}{cccccccccccccccc}
	\bb_1=(0 & 0 & 0 & 0 & 0 & 0 & 0 & 0 & 1 & 1 & -1 & -1 & 0 & 0 & 0 & 0)^T  \\
	\bb_2=(0 & 0 & 0 & 0 & 0 & 0 & 0 & 0 & 0 & 0 & 0 & 0  & 1 & 1 & -1 & -1)^T  \\
	\bb_3=(0 & 0 & 0 & 0 & 1 & 1 & -1 & -1 & 0 & 0 & 0 & 0 &  0 & 0 & 0 & 0)^T  \\
	\bb_4=(1 & 1 & -1 & -1 & 0 & 0 & 0 & 0 & 0 & 0 & 0 & 0 &  0 & 0 & 0 & 0)^T  \\
	\end{array}
	\right\},\\
	&= \left \{ \sum_{i=1}^{4}  r_i \bb_i\ :\ r_i \in \mZ  \right\}. \nonumber
	\end{align}
	We also have 
	\begin{align}
	\label{eq:AI_Bu}
	\AI &=\mathrm{span}
	\left\{
	\begin{array}{cccccccccccccccc}
	\ba_1=(-8 & -8 & 8 & 8 & 0 & 0 & 0 & 0 & 0 & 0 & 0 & 0 &  0 & 0 & 0 & 0)^T  \\
	\ba_2=(4 & 4 & -4 & -4 & -4 & -4 & 4 & 4 & 0 & 0 & 0 & 0 &  0 & 0 & 0 & 0)^T  \\
	\ba_3=(4 & 4 & -4 & -4 & 0 & 0 & 0 & 0 &  -4 & -4 & 4 & 4& 0 & 0 & 0 & 0 )^T  \\
	\ba_4=(-2 & -2 & 2 & 2 & 2 & 2 & -2 & -2 & 2 & 2 & -2 & -2 & -2 & -2 & 2 & 2)^T \\
	\end{array}
	\right \} ,\\
	&= \left \{ \sum_{i=1}^{4}  s_i \ba_i\ :\ s_i \in \mZ  \right\} \nonumber.
	\end{align}
	From Eqs.~\eqref{eq:BS_Bu} and ~\eqref{eq:AI_Bu}, we get $\bm{n} = -2\bb_1-\bb_2+3\bb_4=--\frac{1}{4}\ba_1 +\frac{1}{2}\ba_2+\frac{1}{4}\ba_3+\ba_4$. This implies that $\bm{n} \in \BS$ but $\bm{n}\notin\AI$, i.e., this system is topologically nontrivial. 	
	
	\item[(iv)]  \textbf{Further analysis}
	
	\quad As shown in Ref.~\cite{Ono-Po-Watanabe2020}, we define topological indices which characterize $\XBdG = \mZ_2 \times (\mZ_4)^2 \times \mZ_8$ by
	\begin{align}
	\label{inv_1D}
	\nu^{1\text{D}} &\equiv \frac{1}{4}\sum_{\bm{k}\in 1\text{D TRIMs}}\sum_{\alpha=\pm 1}\alpha N_{(0,k_y,0)}^{\alpha} \mod 2,\\
	\label{inv_2Dx}
	\nu^{2\text{D}}_{x} &\equiv \frac{1}{4}\sum_{\bm{k}\in 2\text{D TRIMs}}\sum_{\alpha=\pm 1}\alpha N_{(0,k_y,k_z)}^{\alpha}\mod 4,\\
	\label{inv_2Dz}
	\nu^{2\text{D}}_{z} &\equiv \frac{1}{4}\sum_{\bm{k}\in 2\text{D TRIMs}}\sum_{\alpha=\pm 1}\alpha N_{(k_x,k_y,0)}^{\alpha}\mod 4,\\
	\label{kappa1_app}
	\kappa_1 &\equiv \frac{1}{4}\sum_{\bm{k}\in 3\text{D TRIMs}}\sum_{\alpha=\pm 1}\alpha N_{\bk}^{\alpha}\mod 8.
	\end{align}
	From Eqs.~\eqref{eq:Bu} -~\eqref{kappa1_app}, we get $(\nu^{1\text{D}}, \nu^{2\text{D}}_{x}, \nu^{2\text{D}}_{z},\kappa_1 )=(1,1,0,2)$, which implies that this superconductor is a second-order topological superconductor. 
\end{enumerate}

\subsubsection{$A_u$ representations}
\begin{figure}[t]
	\begin{center}
		\includegraphics[width=0.6\columnwidth]{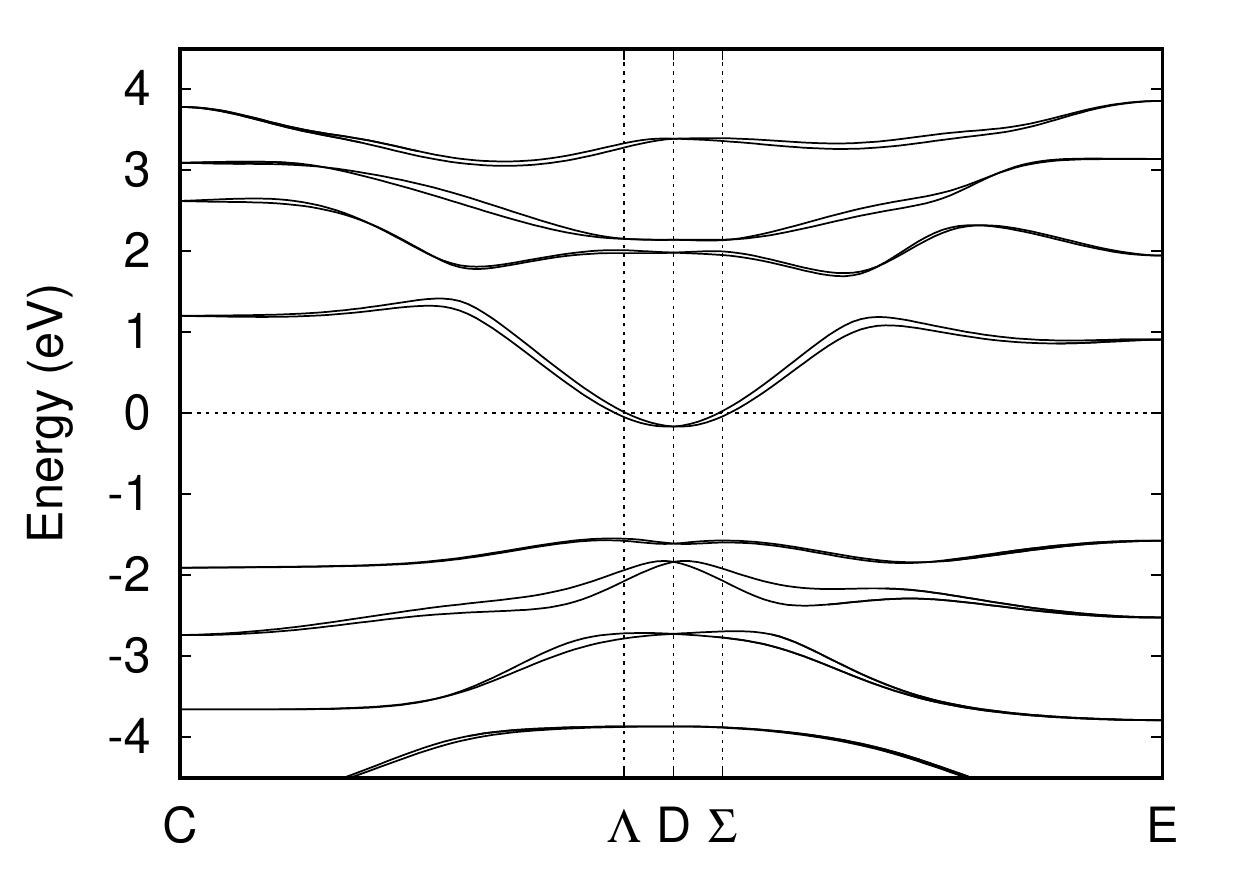}
		\caption{\label{fig:CaSb2}Band structure along $\text{C}-\text{D}-\text{E}$ lines of CaSb$_2$ with the spin-orbit coupling, where $\text{C}=(1/2,1/2,0)$, $\text{D}=(0,1/2,1/2)$, and $\text{E}=(1/2,1/2,1/2)$. $\Lambda=(1/20, 1/2, 9/20)$ and $\Sigma=(1/20, 1/2, 1/2)$ are representative points in the nodal line.
		}
	\end{center}
\end{figure}

For the odd-parity pairing with $\chi_{G_y} = -1$ in nonsymmorphic space groups, nodal lines in glide symmetric plane at $k_y = \pi$ can be stable~\cite{PhysRevB.97.180504}. These nodal lines are understood as violations of compatibility relations in this plane. After computing irreducible representations at lines shown in Fig.~\ref{fig:CaSb2}, we find irreducible representations at $\Lambda$ and $\Sigma$ do not satisfy compatibility relations. Although nodal lines exist in the \textit{strict} weak-pairing limit, these lines can be pair-annihilated by continuous deformations. 
\begin{enumerate}
	\item[(i)] \textbf{Getting irreducible representations in the normal phase}
	
	\quad In addition to TRIMs, we compute irreducible representations at some points in $k_y=\pi$ plane: $\Lambda=(1/20, 1/2, 9/20)$ and $\Sigma=(1/20, 1/2, 1/2)$. Then, we get
	\begin{align}
	\bm{n}_{\text{normal}} &= \left(n_{\Gamma}^{3}, n_{\Gamma}^{4},n_{\Gamma}^{5},n_{\Gamma}^{6}, n_{\text{B}}^{3}, n_{\text{B}}^{4},n_{\text{B}}^{5},n_{\text{B}}^{6},n_{\text{Y}}^{3}, n_{\text{Y}}^{4},n_{\text{Y}}^{5},n_{\text{Y}}^{6}, n_{\text{Z}}^{2}, n_{\text{C}}^{2}, n_{\text{D}}^{2}, n_{\text{A}}^{3}, n_{\text{A}}^{4},n_{\text{A}}^{5},n_{\text{A}}^{6}, n_{\text{E}}^{2}, n_{\Lambda}^{3}, n_{\Lambda}^{4}, n_{\Sigma}^{3}, n_{\Sigma}^{4}\right) \nonumber\\
	\label{eq:noraml_v2}
	&= (21,21,18,18,20,20,20,20,20,20,19,19,42,40,42,19,19,21,21,40, 42,40,42,40).
	\end{align}
	
	\item[(ii)] \textbf{Band labels of the superconducting phase}
	
	\quad We compute the Wigner criteria at $\Lambda$ and $\Sigma$, we find EAZ classes at $\Lambda$ and $\Sigma$ are class AII as tabulated in Table~\ref{tab:CaSb2}. Furtermore, in the weak-pairing limit, 
	\begin{align}
	N_{\bk}^{3} = -N_{\bk}^{4}  = n_{\bk}^{3}\vert_{\text{occ}} -  n_{\bk}^{4}\vert_{\text{occ}}\quad (\bk= \Lambda, \Sigma),
	\end{align}
	and so 
	\begin{align}
	\bm{n} &=\left(N_{\Gamma}^{3}, N_{\Gamma}^{4},N_{\Gamma}^{5},N_{\Gamma}^{6}, N_{\text{B}}^{3}, N_{\text{B}}^{4},N_{\text{B}}^{5},N_{\text{B}}^{6},N_{\text{A}}^{3}, N_{\text{A}}^{4}, N_{\text{A}}^{5}, N_{\text{A}}^{6}, N_{\text{Y}}^{3}, N_{\text{Y}}^{4},N_{\text{Y}}^{5},N_{\text{Y}}^{6},N_{\Lambda}^{3}, N_{\Lambda}^{4}, N_{\Sigma}^{3}, N_{\Sigma}^{4}\right) \nonumber \\
	\label{eq:Au}
	&= (3,3,-3,-3,0, 0, 0, 0, 1,1,-1,-1, -2,-2,2,2, 2,-2,2,-2).
	\end{align}
	
	\begin{table}[H]
		\begin{center}
			\caption{\label{tab:CaSb2}EAZ classes at $\Lambda$ and $\Sigma$.}
			\begin{tabular}{c|c|c|c}
				\hline
				Irrep & $(W^{\alpha}_{\bk}(\calT) , W^{\alpha}_{\bk}(\calC), W^{\alpha}_{\bk}(\Gamma))$ & EAZ class & topological invariant \\
				\hline\hline
				$\Lambda_3$ & $(-1,0,0)$ & class AII & $N_{\Lambda}^{3}$ \\
				$\Lambda_4$ & $(-1,0,0)$ & class AII & $N_{\Lambda}^{4}$ \\
				$\Sigma_3$ & $(-1,0,0)$ & class AII & $N_{\Sigma}^{3}$ \\
				$\Sigma_4$ & $(-1,0,0)$ & class AII & $N_{\Sigma}^{4}$ \\
				\hline
			\end{tabular}
		\end{center}
	\end{table}

	\item[(iii)]  \textbf{Diagnosing topological phases based on symmetry indicators}
	
	\quad If compatibility relations in $k_y$-plane hold, $N_{\bk}^\lambda$ should satisfy
	\begin{align}
	N_{\bk_1}^\lambda = N_{\bk_2}^\lambda = 0\quad (\bk_1, \bk_2 \notin \text{TRIMs at $k_y=\pi$}).
	\end{align}
	Since $N_{\Sigma}^\lambda$ and $N_{\Lambda}^\lambda$ are non-zero, we find that the vector $\bm{n}$ does not belong to $\BS$ and nodal structures should appear in $k_y=\pi$ plane. 
	
	\item[(iv)]  \textbf{Further analysis}
	
	\quad As discussed above, nodal lines in glide symmetric plane at $k_y = \pi$ can be stable~\cite{PhysRevB.97.180504}. The authors of Ref.~\onlinecite{PhysRevB.97.180504} define topological invariants of nodal lines by
	\begin{align}
	\mathcal{N}^{\lambda} \equiv \left(n_{\bm{k}_{\text{out}}}^{\lambda}\right)^{\text{BdG}}- \left(n_{\bm{k}_{\text{in}}}^{\lambda}\right)^{\text{BdG}},
	\end{align}
	where $\left(n_{\bm{k}_{\text{out (in)}}}^{\lambda}\right)^{\text{BdG}}$ denotes the number of occupied bands of the BdG Hamiltonian with the glide eigenvalue $\lambda i = \pm i$ outside (inside) a nodal line. This invariant can be rewritten as
	\begin{align}
	\label{appeq:inv1}
	\mathcal{N}^{-} =N_{\text{D}+\delta}^{3}-N_{\Lambda}^{3} =N_{\text{D}+\delta}^{3}-N_{\Sigma}^{3}  ,\\
	\label{appeq:inv2}
	\mathcal{N}^{+} =N_{\text{D}+\delta}^{4}-N_{\Lambda}^{4}=N_{\text{D}+\delta}^{4}-N_{\Sigma}^{4}  ,
	\end{align}
	where $N_{\text{D}+\delta}^{\lambda=3,4}$ are integer invariants at slightly moved momentum from D.  
	Therefore, $\mathcal{N}^{\lambda} $ can be nontrivial only when the system violates compatibility relations. From Eq.~\eqref{eq:Au}, \eqref{appeq:inv1}, and~\eqref{appeq:inv2}, we get
	\begin{align}
	(\mathcal{N}^{-} ,\mathcal{N}^{+}) = (2,-2). 
	\end{align}
	This implies that nodal lines exist. Next, we consider the stability of these lines. These nodal lines are protected only by 0D topological invariants $\mathcal{N}^\pm$ as discussed in Ref.~\onlinecite{PhysRevB.97.180504}. Furthermore, from the Fermi surfaces in Fig. 2 of the main text, we find that nodal lines do not encircle high-symmetry points.
	Therefore, we conclude that these nodal lines can be gapped out when we get rid of the regions which have nontrivial 0D topological invariants $\mathcal{N}^\pm$. After the pair-annihilation, band labels are the same as the case of $B_u$, and so we get $(\nu^{1\text{D}}, \nu^{2\text{D}}_{x}, \nu^{2\text{D}}_{z},\kappa_1 )=(1,1,0,2)$.
\end{enumerate}

\subsubsection{$B_g$ representations}
For this symmetry class, this material is a symmetry-enforced nodal supercondutor. This is because the vector $\bm{n}$ cannot be expanded by the basis of $\BS$. Since $\mZ_2$-compatibility relations at two-fold rotation symmetric lines $\Gamma$-Z and Y-C are violated, nodal points should appear in these lines. 
\begin{enumerate}
	\item[(i)] \textbf{Getting irreducible representations in the normal phase}
	
	\quad Again, we compute irreducible representations at all TRIMs, $\Lambda$, and $\Sigma$.
	\begin{align}
	\bm{n}_{\text{normal}} &= \left(n_{\Gamma}^{3}, n_{\Gamma}^{4},n_{\Gamma}^{5},n_{\Gamma}^{6}, n_{\text{B}}^{3}, n_{\text{B}}^{4},n_{\text{B}}^{5},n_{\text{B}}^{6},n_{\text{Y}}^{3}, n_{\text{Y}}^{4},n_{\text{Y}}^{5},n_{\text{Y}}^{6}, n_{\text{Z}}^{2}, n_{\text{C}}^{2}, n_{\text{D}}^{2}, n_{\text{A}}^{3}, n_{\text{A}}^{4},n_{\text{A}}^{5},n_{\text{A}}^{6}, n_{\text{E}}^{2}, n_{\Lambda}^{3}, n_{\Lambda}^{4}, n_{\Sigma}^{3}, n_{\Sigma}^{4}\right) \nonumber\\
	\label{eq:noraml_v3}
	&= (21,21,18,18,20,20,20,20,20,20,19,19,42,40,42,19,19,21,21,40, 42,40,42,40).
	\end{align}
	
	\item[(ii)] \textbf{Band labels of the superconducting phase} 
	
	\quad In this symmetry setting, EAZ classes at TRIMs in $k_y=\pi$ plane are class DIII, and those in $k_y = 0$ plane are class D. On the other hand, EAZ classes at generic points in the plane are class AII. Then, the band labels are
	\begin{align}
	\label{eq:Bg}
	\bm{n} &=\left(p_{\Gamma}^{3}, p_{\Gamma}^{4},p_{\Gamma}^{5},p_{\Gamma}^{6}, p_{\text{B}}^{3}, p_{\text{B}}^{4},p_{\text{B}}^{5},p_{\text{B}}^{6},p_{\text{Y}}^{3}, p_{\text{Y}}^{4}, p_{\text{Y}}^{5}, p_{\text{Y}}^{6}, p_{\text{A}}^{3}, p_{\text{A}}^{4},p_{\text{A}}^{5},p_{\text{A}}^{6}, N_{\Lambda}^{3}, N_{\Lambda}^{4}, N_{\Sigma}^{3}, N_{\Sigma}^{4}\right)\nonumber \\
	&= (1,1,0,0,0,0,0,0, 0,0,1,1, 1,1,1,1,2,-2,2,-2),
	\end{align}
	where $p_{\bk}^{\alpha}=n_{\bk}^{\alpha} \mod 2$ in the weak-pairing assumption.
	
	\item[(iii)]  \textbf{Diagnosing topological phases based on symmetry indicators}
	
	\quad After solving compatibility relations, we get
	\begin{align}
	\label{eq:BS_SG11}
	\BS &=\mathrm{span}
	\left\{
	\begin{array}{cccccccccccccccccccc}
	\bb_1=(0 & 0 & 0 & 0 & 0 & 0 & 0 & 0 & 0 & 0 & 0 & 0 & 1 & 1 & 1 & 1 & 0 & 0 & 0 & 0)^T   \\
	\bb_2=(0 & 0 & 0 & 0 & 0 & 0 & 0 & 0 & 1 & 1 & 1 & 1 & 0 & 0 & 0 & 0& 0 & 0 & 0 & 0)^T  \\
	\bb_3=(0 & 0 & 0 & 0 & 1 & 1 & 1 & 1 & 0 & 0 & 0 & 0 & 0 & 0 & 0 & 0 & 0 & 0 & 0 & 0)^T  \\
	\bb_4=(1 & 1 & 1 & 1 & 0 & 0 & 0 & 0 & 0 & 0 & 0 & 0 & 0 & 0 & 0 & 0 & 0 & 0 & 0 & 0)^T  \\
	\end{array}
	\right\},\\
	&= \left \{ \sum_{i=1}^{4}  r_i \bb_i\ :\ r_i \in \mZ_2  \right\}.\nonumber
	\end{align}
	From Eqs.~\eqref{eq:Bg} and \eqref{eq:BS_SG11}, we find that $\bm{n}$ is not included in $\BS$, i.e., various compatibility relations are violated. Therefore, the material with $B_g$ pairing should be gapless. 
	
	\item[(iv)]  \textbf{Further analysis}
	
	\quad As explained in the main text, compatibility relations along the $\Gamma$-Z, $\Gamma$-B, Y-C, and Y-A lines are violated. Furthermore, we show an expected nodal structure assuming the \textit{strict} weak-pairing limit in Fig.~2 of the main text. Here, we discuss the stability of nodal lines in the mirror and glide plane. 
	
	\quad As with the case of $A_u$ representation, we define the following 0D topological invariants~\cite{PhysRevB.97.180504}
	\begin{align}
	\nu^{\lambda} &\equiv p_{\bm{k}_{\text{out}}}^{\lambda} - p_{\bm{k}_{\text{in}}}^{\lambda} \mod 2\quad (\text{in $k_y=0$ plane}),\\
	\mathcal{N}^{\lambda} &\equiv N_{\bm{k}_{\text{out}}}^{\lambda}- N_{\bm{k}_{\text{in}}}^{\lambda}\quad (\text{in $k_y=\pi$ plane}).
	\end{align}
	As discussed in the main text, we find regions where $\nu^{\lambda} = 1$ or $\mathcal{N}^{\lambda} \neq 0$. 
	However, unlike the case of $A_u$ representation, each nodal line in these planes is protected by not only the above 0D invariants but also the 1D winding number $W$~\cite{PhysRevB.97.180504}. According to Ref.~\cite{PhysRevB.97.180504}, $\vert \mathcal{N}^{\lambda}  \vert= \vert W \vert$ always holds. By shrinking the region, a pair of nodal lines change to a single nodal line, and each side of regions separated by the nodal line has $\mathcal{N}^{\lambda} = 0$. From this, $\vert W \vert$ is also trivial, and therefore nodal lines can be pair-annihilated by continuous deformations.

\end{enumerate}

\begin{figure}[t]
	\begin{center}
		\begin{tabular}{c}			
			\begin{minipage}{0.5\hsize}
				\begin{center}		
					\includegraphics[width=0.99\columnwidth]{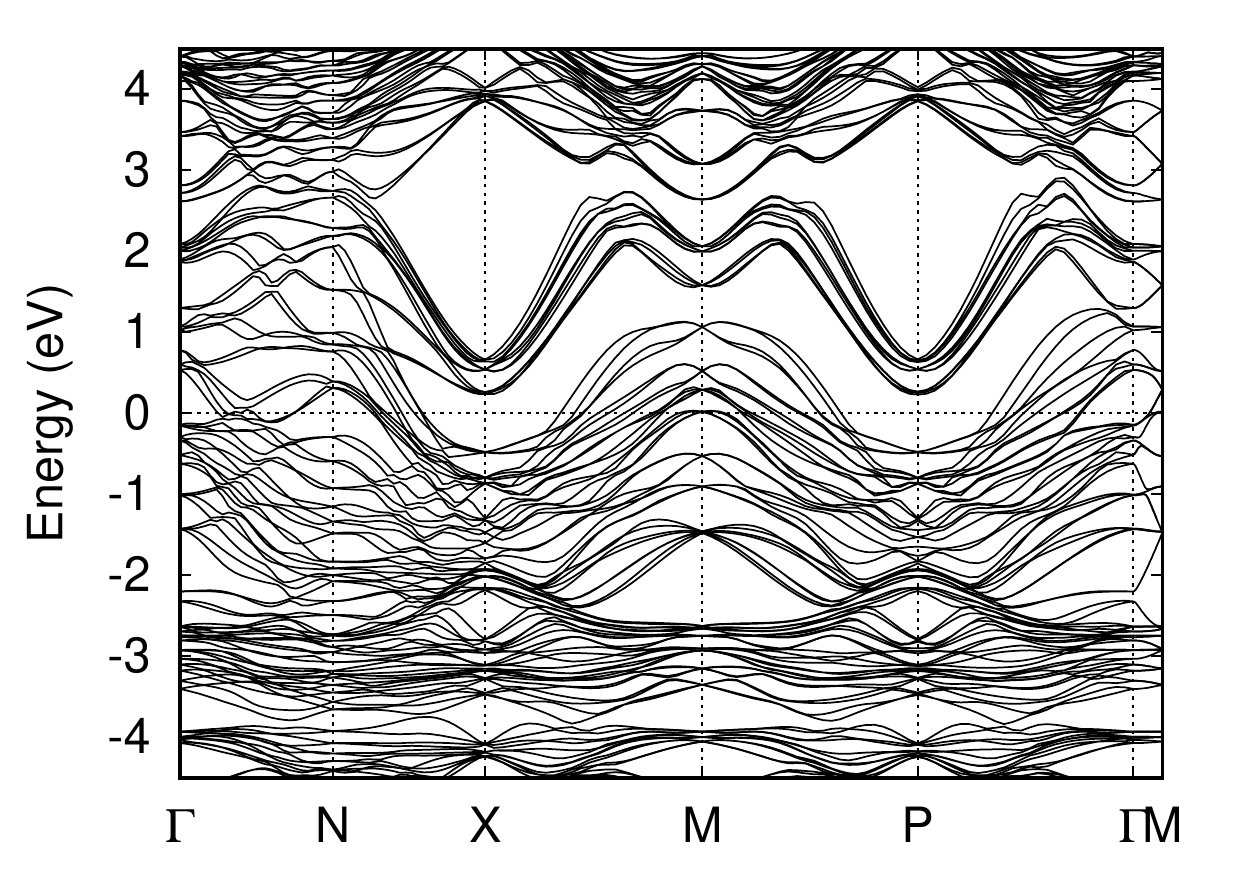}
					\caption{\label{fig:CaPtAs_noSOC}Band structure of CaPtAs with the spin-orbit coupling. }
				\end{center}
			\end{minipage}
			\begin{minipage}{0.5\hsize}
				\begin{center}			
					\includegraphics[width=0.99\columnwidth]{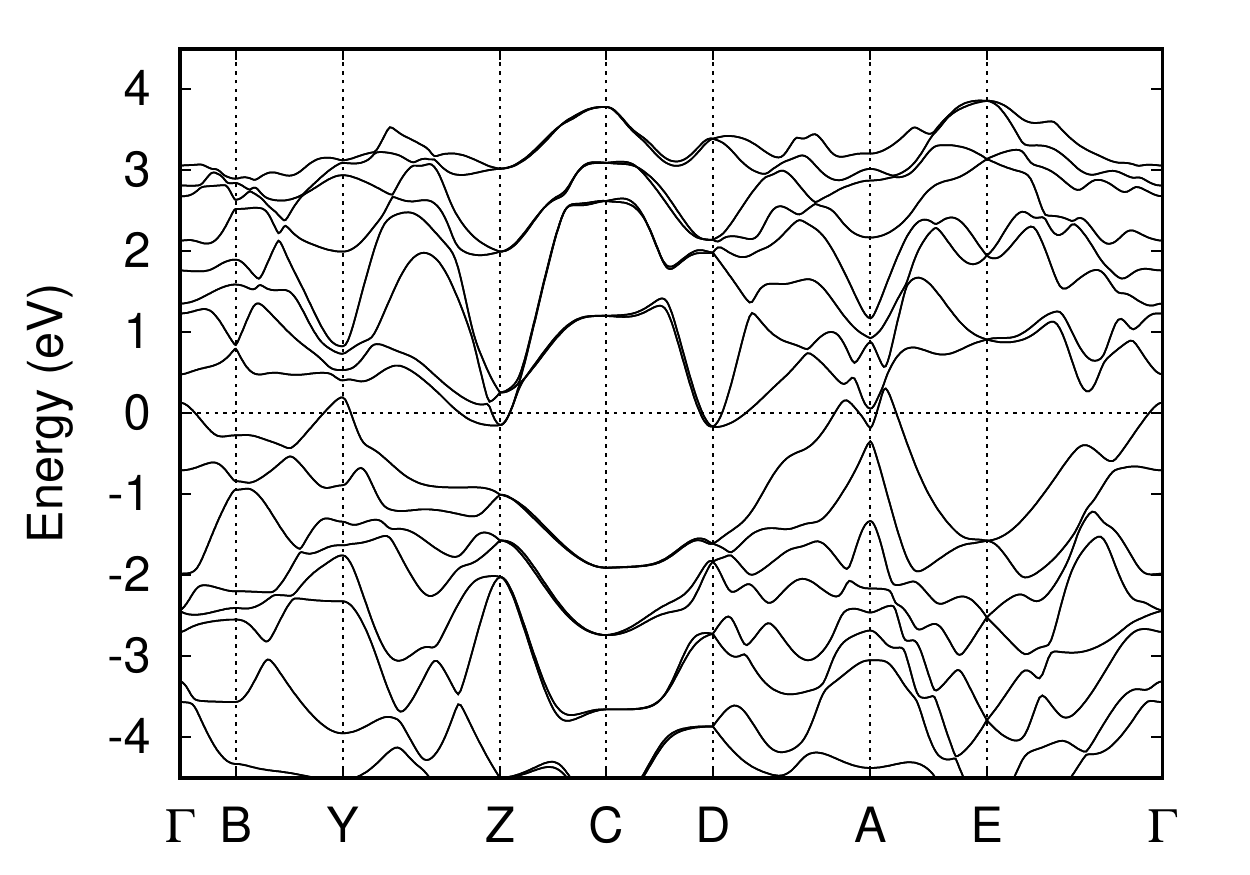}
					\caption{\label{fig:CaSb2_SOC}Band structure of CaSb$_2$ with the spin-orbit coupling.}
				\end{center}
			\end{minipage}
		\end{tabular}
	\end{center}
\end{figure}
\clearpage

\section{Full classification of pairing symmetries and $X_f$}
\label{app:Xlist}
We list all results of $X_f$. For simplicity, we denote $X_f = \mZ_p \times \mZ_q \times\mZ_r\times\ldots$ by $\{p, q,r,\ldots\}$. 
As explained in the main text, we reveal that 16136 symmetry classes (~$58.2\%$ of total number of symmetry classes) are improved.

Here, we classify symmetry classes into the following main three categories:
\begin{enumerate}
	\item[Case1]: $X_f  \neq \XBdG'$,
	\item[Case2]: $X_f  = \XBdG'$ with $(D_P, d_p)=(0,0)$ or $X_f  = \XBdG' = 0$ with $(D_P, d_p)\neq(0,0)$,
	\item[Case3]: $X_f  = \XBdG'\neq 0$ with $(D_P, d_p)\neq(0,0)$,
\end{enumerate}
where $X_f$ and $\XBdG'$ denote the classifications in this work and in the Ref.~\cite{Ono-Po-Watanabe2020}, respectively. Furthermore, the Case 3 is also devided to two small classes:
\begin{enumerate}
	\item[Case 3(a)]: the generators of $\XBdG'$ satisfy the $\mZ_2$-compatibility relations as they are,
	\item[Case 3(b)]: the generators of $\XBdG'$ are not elements of $\BS$ of $\mZ_2$-enriched symmetry indicators. 
\end{enumerate}

In Table~\ref{app:tab-summary-MSG} and \ref{app:tab-summary-SG}, we summarize statistics that show how our work improve the power to diagnose topological phases. We \textit{define} ``improved case'' by Case 1 and Case 3(b).

\begin{table}[H]
	\begin{center}
		\caption{\label{app:tab-summary-MSG} Summary of the statistics of each case. }
		\begin{tabular}{c|cccc|cc}
			\hline
			 & Case 1 & Case 2 & Case 3(a) & Case 3(b) & improved cases & \\
			\hline\hline
			$\calC^2=+1$, spinless & $4280$ & $2065$ & $115$ & $477$ & $4757$ & $\sim 69\%$\\
			$\calC^2=+1$, spinful & $4461$ & $2191$ & $99$ & $186$ & $4647$ & $\sim 67\%$\\
			$\calC^2=-1$, spinless & $2803$ & $3551$ & $166$ & $417$ & $3220$ & $\sim 46\%$\\
			$\calC^2=-1$, spinful & $3203$ & $3326$ & $119$ & $289$ & $3492$ & $\sim 50\%$\\
			\hline
		\end{tabular}
	\end{center}
\end{table}

\begin{table}[H]
	\begin{center}
		\caption{\label{app:tab-summary-SG} Summary of the statistics for each AZ classes in space groups. Note that these symmetry classes are included in the above table. However, for the reader's convenience, we surmmarize results in AZ classes.}
		\begin{tabular}{ccccc}
			\hline
			AZ class & Case 1 & Case 2 & Case 3(a) & Case 3(b) \\
			\hline\hline
			DIII & $566$ & $380$ & $11$ & $23$ \\
			D (spinful)& $722$ & $340$ & $17$ & $39$ \\
			D (spinless)& $620$ & $379$ & $24$ & $95$ \\
			BDI & $540$ & $343$ & $20$ & $77$ \\
			C & $484$ & $560$ & $25$ & $49$ \\
			CI & $414$ & $518$ & $16$ & $32$ \\
			\hline
		\end{tabular}
	\end{center}
\end{table}

In the following tables, we follow Belov-Neronova-Smirnova (BNS) notation. Character tables of one-dimensional irreducible representations in point groups (PGs) and correspondences between MSGs and PGs are listed in Tables~\ref{app:char} and~\ref{app:MSG-PG}.
\clearpage
\begin{table}
	\caption{The list of $X_f$ for spinless systems with $\calC^2=+1$.}
	\label{tab:XBS_D_spinless}
	\begin{align*}
		\begin{array}{cc|cc}
		\mbox{MSG}&&\mbox{Irrep}&\Xf\\
		\hline 
		1.1 & \text{I} & A & \{2,2,2,2,2,2,2\} \\
		1.2 & \text{II} & A & \{2,2,2,2,2,2,2\} \\
		1.3 & \text{IV} & A & \{2,2,2,2\} \\
		2.4 & \text{I} & A_u & \{2,2,2,4,4,4,8\} \\
		2.4 & \text{I} & A_g & \{2,2,2,2\} \\
		2.5 & \text{II} & A_u & \{2,2,2,4,4,4,8\} \\
		2.5 & \text{II} & A_g & \{2,2,2,2\} \\
\end{array}&&
		\begin{array}{cc|cc}
		\mbox{MSG}&&\mbox{Irrep}&\Xf\\
		\hline 
		7.30 & \text{IV} & A'' & \{2,2,2\} \\
		7.30 & \text{IV} & A' & \{2,2,2\} \\
		7.31 & \text{IV} & A'' & \{2,2,2\} \\
		7.31 & \text{IV} & A' & \{2,2,2\} \\
		8.32 & \text{I} & A'' & \{2,2\} \\
		8.32 & \text{I} & A' & \{2,2,2,2,2,2\} \\
		8.33 & \text{II} & A'' & \{2,2\} \\
\end{array}&&
		\begin{array}{cc|cc}
		\mbox{MSG}&&\mbox{Irrep}&\Xf\\
		\hline 
		11.55 & \text{IV} & B_g & \{2\} \\
		11.55 & \text{IV} & A_u & \{2,2,4\} \\
		11.55 & \text{IV} & A_g & \{2,2,2,2\} \\
		11.56 & \text{IV} & B_u & \{2,2,2,2,2,4\} \\
		11.56 & \text{IV} & B_g & \{\} \\
		11.56 & \text{IV} & A_u & \{2,2,2,4\} \\
		11.56 & \text{IV} & A_g & \{2,2,2,2\} \\
\end{array}
	\end{align*}
\end{table}\clearpage

\begin{table}
	\begin{align*}
		\begin{array}{cc|cc}
		\mbox{MSG}&&\mbox{Irrep}&\Xf\\
		\hline 
		13.74 & \text{IV} & B_u & \{2,2,4\} \\
		13.74 & \text{IV} & B_g & \{2,2\} \\
		13.74 & \text{IV} & A_u & \{2,2,2,2,4\} \\
		13.74 & \text{IV} & A_g & \{2\} \\
		14.75 & \text{I} & B_u & \{2,2,2,2,4\} \\
		14.75 & \text{I} & B_g & \{2,2\} \\
		14.75 & \text{I} & A_u & \{2,2,2,2,4\} \\
\end{array}&&
		\begin{array}{cc|cc}
		\mbox{MSG}&&\mbox{Irrep}&\Xf\\
		\hline 
		16.1 & \text{I} & A_ 1 & \{2\} \\
		16.2 & \text{II} & B_ 1 & \{2,2,2,2\} \\
		16.2 & \text{II} & B_ 2 & \{2,2,2,2\} \\
		16.2 & \text{II} & B_ 3 & \{2,2,2,2\} \\
		16.2 & \text{II} & A_ 1 & \{2\} \\
		16.3 & \text{III} & B & \{2,2,4\} \\
		16.3 & \text{III} & A & \{2,2,2,2,2,2\} \\
\end{array}&&
		\begin{array}{cc|cc}
		\mbox{MSG}&&\mbox{Irrep}&\Xf\\
		\hline 
		18.20 & \text{IV} & B_ 1 & \{2,2\} \\
		18.20 & \text{IV} & B_ 2 & \{2,2\} \\
		18.20 & \text{IV} & B_ 3 & \{2,2\} \\
		18.20 & \text{IV} & A_ 1 & \{2,2,2,2\} \\
		18.21 & \text{IV} & B_ 1 & \{2,2\} \\
		18.21 & \text{IV} & B_ 2 & \{4\} \\
		18.21 & \text{IV} & B_ 3 & \{4\} \\
\end{array}
	\end{align*}
\end{table}\clearpage

\begin{table}
	\begin{align*}
		\begin{array}{cc|cc}
		\text{MSG} & \text{} & \text{Irrep} & X_f \\ \hline
		20.37 & \text{IV} & B_ 2 & \{2\} \\
		20.37 & \text{IV} & B_ 3 & \{2\} \\
		20.37 & \text{IV} & A_ 1 & \{2,2\} \\
		21.38 & \text{I} & B_ 1 & \{2,2,2\} \\
		21.38 & \text{I} & B_ 2 & \{2,2\} \\
		21.38 & \text{I} & B_ 3 & \{2,2\} \\
		21.38 & \text{I} & A_ 1 & \{2,2\} \\
\end{array}&&
		\begin{array}{cc|cc}
		\text{MSG} & \text{} & \text{Irrep} & X_f \\ \hline
		24.55 & \text{III} & B & \{2,2,2\} \\
		24.55 & \text{III} & A & \{2,2,2,2\} \\
		24.56 & \text{IV} & B_ 1 & \{2,2\} \\
		24.56 & \text{IV} & B_ 2 & \{2\} \\
		24.56 & \text{IV} & B_ 3 & \{2\} \\
		24.56 & \text{IV} & A_ 1 & \{2,2,2,2\} \\
		25.57 & \text{I} & B_ 1 & \{2,2,2\} \\
\end{array}&&
		\begin{array}{cc|cc}
		\text{MSG} & \text{} & \text{Irrep} & X_f \\ \hline
		26.73 & \text{IV} & A_ 1 & \{2,2,2,2,2,2\} \\
		26.74 & \text{IV} & B_ 1 & \{2\} \\
		26.74 & \text{IV} & B_ 2 & \{2,2,2\} \\
		26.74 & \text{IV} & A_ 2 & \{2\} \\
		26.74 & \text{IV} & A_ 1 & \{2,2,2\} \\
		26.75 & \text{IV} & B_ 1 & \{2\} \\
		26.75 & \text{IV} & B_ 2 & \{2,2,2\} \\
\end{array}
	\end{align*}
\end{table}\clearpage

\begin{table}
	\begin{align*}
		\begin{array}{cc|cc}
		\text{MSG} & \text{} & \text{Irrep} & X_f \\ \hline
		28.92 & \text{IV} & B_ 1 & \{2,2\} \\
		28.92 & \text{IV} & B_ 2 & \{2,2,2,2\} \\
		28.92 & \text{IV} & A_ 2 & \{2\} \\
		28.92 & \text{IV} & A_ 1 & \{2,2,2,2,2,2\} \\
		28.93 & \text{IV} & B_ 1 & \{2,2\} \\
		28.93 & \text{IV} & B_ 2 & \{2,2,2\} \\
		28.93 & \text{IV} & A_ 2 & \{2\} \\
\end{array}&&
		\begin{array}{cc|cc}
		\text{MSG} & \text{} & \text{Irrep} & X_f \\ \hline
		29.109 & \text{IV} & B_ 2 & \{2,2\} \\
		29.109 & \text{IV} & A_ 2 & \{2,2\} \\
		29.109 & \text{IV} & A_ 1 & \{2,2,2\} \\
		29.110 & \text{IV} & B_ 1 & \{2\} \\
		29.110 & \text{IV} & B_ 2 & \{2,2\} \\
		29.110 & \text{IV} & A_ 2 & \{2,2\} \\
		29.110 & \text{IV} & A_ 1 & \{2,2,2\} \\
\end{array}&&
		\begin{array}{cc|cc}
		\text{MSG} & \text{} & \text{Irrep} & X_f \\ \hline
		31.128 & \text{IV} & B_ 1 & \{2\} \\
		31.128 & \text{IV} & B_ 2 & \{2,2,2\} \\
		31.128 & \text{IV} & A_ 2 & \{2,2\} \\
		31.128 & \text{IV} & A_ 1 & \{2,2,2,2\} \\
		31.129 & \text{IV} & B_ 1 & \{\} \\
		31.129 & \text{IV} & B_ 2 & \{2,2,2\} \\
		31.129 & \text{IV} & A_ 2 & \{\} \\
\end{array}
	\end{align*}
\end{table}\clearpage

\begin{table}
	\begin{align*}
	\begin{array}{cc|cc}
	\text{MSG} & \text{} & \text{Irrep} & X_f \\ \hline
	33.144 & \text{I} & A_ 1 & \{2,2,2\} \\
	33.145 & \text{II} & B_ 1 & \{2,2\} \\
	33.145 & \text{II} & B_ 2 & \{2,2\} \\
	33.145 & \text{II} & A_ 2 & \{2\} \\
	33.145 & \text{II} & A_ 1 & \{2,2,2\} \\
	33.146 & \text{III} & A'' & \{2,2,2\} \\
	33.146 & \text{III} & A' & \{2,2,2\} \\
\end{array}&&
	\begin{array}{cc|cc}
	\text{MSG} & \text{} & \text{Irrep} & X_f \\ \hline
	34.163 & \text{IV} & B_ 1 & \{2,2,4\} \\
	34.163 & \text{IV} & B_ 2 & \{2,2,4\} \\
	34.163 & \text{IV} & A_ 2 & \{2,2,2\} \\
	34.163 & \text{IV} & A_ 1 & \{2,2,2\} \\
	34.164 & \text{IV} & B_ 1 & \{2,2,2\} \\
	34.164 & \text{IV} & B_ 2 & \{2,2,2\} \\
	34.164 & \text{IV} & A_ 2 & \{2,2\} \\
\end{array}&&
	\begin{array}{cc|cc}
	\text{MSG} & \text{} & \text{Irrep} & X_f \\ \hline
	37.181 & \text{II} & B_ 2 & \{2,4\} \\
	37.181 & \text{II} & A_ 2 & \{2,2\} \\
	37.181 & \text{II} & A_ 1 & \{2,2,2,2\} \\
	37.182 & \text{III} & A'' & \{2,2,2\} \\
	37.182 & \text{III} & A' & \{2,2,2,2\} \\
	37.183 & \text{III} & B & \{2,4\} \\
	37.183 & \text{III} & A & \{2,2,2,2\} \\
\end{array}
	\end{align*}
\end{table}\clearpage

\begin{table}
	\begin{align*}
		\begin{array}{cc|cc}
		\text{MSG} & \text{} & \text{Irrep} & X_f \\ \hline
		39.201 & \text{IV} & B_ 1 & \{2,2\} \\
		39.201 & \text{IV} & B_ 2 & \{2,2\} \\
		39.201 & \text{IV} & A_ 2 & \{2\} \\
		39.201 & \text{IV} & A_ 1 & \{2,2,2,2,2\} \\
		39.202 & \text{IV} & B_ 1 & \{2,2\} \\
		39.202 & \text{IV} & B_ 2 & \{2\} \\
		39.202 & \text{IV} & A_ 2 & \{2,2\} \\
\end{array}&&
		\begin{array}{cc|cc}
		\text{MSG} & \text{} & \text{Irrep} & X_f \\ \hline
		42.219 & \text{I} & A_ 1 & \{2,2,2,2,2\} \\
		42.220 & \text{II} & B_ 1 & \{2,2\} \\
		42.220 & \text{II} & B_ 2 & \{2,2\} \\
		42.220 & \text{II} & A_ 2 & \{2\} \\
		42.220 & \text{II} & A_ 1 & \{2,2,2,2,2\} \\
		42.221 & \text{III} & A'' & \{2\} \\
		42.221 & \text{III} & A' & \{2,2,2,2,2\} \\
\end{array}&&
		\begin{array}{cc|cc}
		\text{MSG} & \text{} & \text{Irrep} & X_f \\ \hline
		45.239 & \text{IV} & A_ 2 & \{2,2,2\} \\
		45.239 & \text{IV} & A_ 1 & \{2,2,2,2\} \\
		45.240 & \text{IV} & B_ 1 & \{2,2\} \\
		45.240 & \text{IV} & B_ 2 & \{2,2\} \\
		45.240 & \text{IV} & A_ 2 & \{2,2,2,2\} \\
		45.240 & \text{IV} & A_ 1 & \{2,2,2,2\} \\
		46.241 & \text{I} & B_ 1 & \{2\} \\
\end{array}
	\end{align*}
\end{table}\clearpage

\begin{table}
	\begin{align*}
		\begin{array}{cc|cc}
		\text{MSG} & \text{} & \text{Irrep} & X_f \\ \hline
		47.254 & \text{IV} & B_{2g} & \{2,2,2\} \\
		47.254 & \text{IV} & B_{3u} & \{2,2,2,2,2,2,2,2,2\} \\
		47.254 & \text{IV} & B_{3g} & \{\} \\
		47.254 & \text{IV} & A_u & \{2\} \\
		47.254 & \text{IV} & A_g & \{\} \\
		47.255 & \text{IV} & B_{1u} & \{2,2,2,2,2\} \\
		47.255 & \text{IV} & B_{1g} & \{2,2,2\} \\
\end{array}&&
		\begin{array}{cc|cc}
		\text{MSG} & \text{} & \text{Irrep} & X_f \\ \hline
		48.263 & \text{IV} & A_u & \{2,2,2,2,2\} \\
		48.263 & \text{IV} & A_g & \{2,2\} \\
		48.264 & \text{IV} & B_{1u} & \{2,4\} \\
		48.264 & \text{IV} & B_{1g} & \{2,2,2\} \\
		48.264 & \text{IV} & B_{2u} & \{2,4\} \\
		48.264 & \text{IV} & B_{2g} & \{2,2,2\} \\
		48.264 & \text{IV} & B_{3u} & \{2,4\} \\
\end{array}&&
		\begin{array}{cc|cc}
		\text{MSG} & \text{} & \text{Irrep} & X_f \\ \hline
		49.274 & \text{IV} & B_{1g} & \{2,2,2\} \\
		49.274 & \text{IV} & B_{2u} & \{2,2,2,4\} \\
		49.274 & \text{IV} & B_{2g} & \{2,2,2\} \\
		49.274 & \text{IV} & B_{3u} & \{2,2,2,2,4\} \\
		49.274 & \text{IV} & B_{3g} & \{2,2\} \\
		49.274 & \text{IV} & A_u & \{2,2,2\} \\
		49.274 & \text{IV} & A_g & \{2,2\} \\
\end{array}
	\end{align*}
\end{table}\clearpage

\begin{table}
	\begin{align*}
		\begin{array}{cc|cc}
		\text{MSG} & \text{} & \text{Irrep} & X_f \\ \hline
		50.284 & \text{IV} & B_{3u} & \{2,2,2,4\} \\
		50.284 & \text{IV} & B_{3g} & \{2,2,2\} \\
		50.284 & \text{IV} & A_u & \{2,2,2,2\} \\
		50.284 & \text{IV} & A_g & \{2,2\} \\
		50.285 & \text{IV} & B_{1u} & \{2,2,4\} \\
		50.285 & \text{IV} & B_{1g} & \{2,2,2,2\} \\
		50.285 & \text{IV} & B_{2u} & \{4,4\} \\
\end{array}&&
		\begin{array}{cc|cc}
		\text{MSG} & \text{} & \text{Irrep} & X_f \\ \hline
		51.293 & \text{III} & A_ 1 & \{2,2,2,2,2,2\} \\
		51.294 & \text{III} & B_u & \{2,2,2,4,4\} \\
		51.294 & \text{III} & B_g & \{2,2\} \\
		51.294 & \text{III} & A_u & \{2,2,2,2,2,4\} \\
		51.294 & \text{III} & A_g & \{2,2,2\} \\
		51.295 & \text{III} & B_u & \{2,2,2,2,2,4\} \\
		51.295 & \text{III} & B_g & \{\} \\
\end{array}&&
		\begin{array}{cc|cc}
		\text{MSG} & \text{} & \text{Irrep} & X_f \\ \hline
		51.303 & \text{IV} & A_u & \{2,2\} \\
		51.303 & \text{IV} & A_g & \{2,2\} \\
		51.304 & \text{IV} & B_{1u} & \{2,2,2,2\} \\
		51.304 & \text{IV} & B_{1g} & \{\} \\
		51.304 & \text{IV} & B_{2u} & \{2,2,2\} \\
		51.304 & \text{IV} & B_{2g} & \{2\} \\
		51.304 & \text{IV} & B_{3u} & \{2,2,2\} \\
\end{array}
	\end{align*}
\end{table}\clearpage

\begin{table}
	\begin{align*}
		\begin{array}{cc|cc}
		\text{MSG} & \text{} & \text{Irrep} & X_f \\ \hline
		52.315 & \text{IV} & B_{1g} & \{2,2\} \\
		52.315 & \text{IV} & B_{2u} & \{2,4\} \\
		52.315 & \text{IV} & B_{2g} & \{2,2\} \\
		52.315 & \text{IV} & B_{3u} & \{4\} \\
		52.315 & \text{IV} & B_{3g} & \{2,2\} \\
		52.315 & \text{IV} & A_u & \{2,2,2\} \\
		52.315 & \text{IV} & A_g & \{2,2,2\} \\
\end{array}&&
		\begin{array}{cc|cc}
		\text{MSG} & \text{} & \text{Irrep} & X_f \\ \hline
		53.323 & \text{III} & B_ 1 & \{2,2,2\} \\
		53.323 & \text{III} & B_ 2 & \{2,2,2\} \\
		53.323 & \text{III} & A_ 2 & \{2,2,2\} \\
		53.323 & \text{III} & A_ 1 & \{2,2,2\} \\
		53.324 & \text{III} & B_ 2 & \{2\} \\
		53.324 & \text{III} & B_ 1 & \{2,2,2,2\} \\
		53.324 & \text{III} & A_ 2 & \{2,2\} \\
\end{array}&&
		\begin{array}{cc|cc}
		\text{MSG} & \text{} & \text{Irrep} & X_f \\ \hline
		53.334 & \text{IV} & B_{2g} & \{2\} \\
		53.334 & \text{IV} & B_{3u} & \{2,2,2\} \\
		53.334 & \text{IV} & B_{3g} & \{2\} \\
		53.334 & \text{IV} & A_u & \{2,2,2\} \\
		53.334 & \text{IV} & A_g & \{2,2\} \\
		53.335 & \text{IV} & B_{1u} & \{2,2,4\} \\
		53.335 & \text{IV} & B_{1g} & \{2,2\} \\
\end{array}
	\end{align*}
\end{table}\clearpage

\begin{table}
	\begin{align*}
		\begin{array}{cc|cc}
		\text{MSG} & \text{} & \text{Irrep} & X_f \\ \hline
		54.345 & \text{III} & B_ 3 & \{2,2,4\} \\
		54.345 & \text{III} & A_ 1 & \{2,2,2\} \\
		54.346 & \text{IV} & B_{1u} & \{2,2,2,2\} \\
		54.346 & \text{IV} & B_{1g} & \{2,2\} \\
		54.346 & \text{IV} & B_{2u} & \{2,2,4\} \\
		54.346 & \text{IV} & B_{2g} & \{2,2\} \\
		54.346 & \text{IV} & B_{3u} & \{2,2,2,4\} \\
\end{array}&&
		\begin{array}{cc|cc}
		\text{MSG} & \text{} & \text{Irrep} & X_f \\ \hline
		55.353 & \text{I} & B_{3g} & \{\} \\
		55.353 & \text{I} & A_u & \{2,2,2\} \\
		55.353 & \text{I} & A_g & \{2,2,2\} \\
		55.354 & \text{II} & B_{1u} & \{2,2,2\} \\
		55.354 & \text{II} & B_{1g} & \{\} \\
		55.354 & \text{II} & B_{2u} & \{2,2,2,2,2,2\} \\
		55.354 & \text{II} & B_{2g} & \{\} \\
\end{array}&&
		\begin{array}{cc|cc}
		\text{MSG} & \text{} & \text{Irrep} & X_f \\ \hline
		55.364 & \text{IV} & B_{1u} & \{2,2,2\} \\
		55.364 & \text{IV} & B_{1g} & \{2,2\} \\
		55.364 & \text{IV} & B_{2u} & \{2,2,2,2\} \\
		55.364 & \text{IV} & B_{2g} & \{2\} \\
		55.364 & \text{IV} & B_{3u} & \{2,2,2,2\} \\
		55.364 & \text{IV} & B_{3g} & \{2\} \\
		55.364 & \text{IV} & A_u & \{2,2,2\} \\
\end{array}
	\end{align*}
\end{table}\clearpage

\begin{table}
	\begin{align*}
		\begin{array}{cc|cc}
		\text{MSG} & \text{} & \text{Irrep} & X_f \\ \hline
		56.374 & \text{IV} & B_{2g} & \{2\} \\
		56.374 & \text{IV} & B_{3u} & \{2,4\} \\
		56.374 & \text{IV} & B_{3g} & \{2\} \\
		56.374 & \text{IV} & A_u & \{2,2,2,2\} \\
		56.374 & \text{IV} & A_g & \{2,2,2\} \\
		56.375 & \text{IV} & B_{1u} & \{2,2,2,2\} \\
		56.375 & \text{IV} & B_{1g} & \{2,2,2\} \\
\end{array}&&
		\begin{array}{cc|cc}
		\text{MSG} & \text{} & \text{Irrep} & X_f \\ \hline
		57.385 & \text{III} & B_ 3 & \{2,2\} \\
		57.385 & \text{III} & A_ 1 & \{2,2,2,2\} \\
		57.386 & \text{IV} & B_{1u} & \{2,2,2\} \\
		57.386 & \text{IV} & B_{1g} & \{2,2\} \\
		57.386 & \text{IV} & B_{2u} & \{2,2,2,2\} \\
		57.386 & \text{IV} & B_{2g} & \{2,2\} \\
		57.386 & \text{IV} & B_{3u} & \{2,2,2\} \\
\end{array}&&
		\begin{array}{cc|cc}
		\text{MSG} & \text{} & \text{Irrep} & X_f \\ \hline
		58.393 & \text{I} & B_{3g} & \{\} \\
		58.393 & \text{I} & A_u & \{2,2,2\} \\
		58.393 & \text{I} & A_g & \{2,2\} \\
		58.394 & \text{II} & B_{1u} & \{2,2,2\} \\
		58.394 & \text{II} & B_{1g} & \{2\} \\
		58.394 & \text{II} & B_{2u} & \{2,2,2,2,4\} \\
		58.394 & \text{II} & B_{2g} & \{\} \\
\end{array}
	\end{align*}
\end{table}\clearpage

\begin{table}
	\begin{align*}
		\begin{array}{cc|cc}
		\text{MSG} & \text{} & \text{Irrep} & X_f \\ \hline
		58.404 & \text{IV} & B_{1u} & \{2,2,2\} \\
		58.404 & \text{IV} & B_{1g} & \{2,2\} \\
		58.404 & \text{IV} & B_{2u} & \{2,2,2,2,4\} \\
		58.404 & \text{IV} & B_{2g} & \{\} \\
		58.404 & \text{IV} & B_{3u} & \{2,2,2,2,4\} \\
		58.404 & \text{IV} & B_{3g} & \{\} \\
		58.404 & \text{IV} & A_u & \{2,2,2\} \\
\end{array}&&
		\begin{array}{cc|cc}
		\text{MSG} & \text{} & \text{Irrep} & X_f \\ \hline
		59.414 & \text{IV} & B_{2g} & \{2\} \\
		59.414 & \text{IV} & B_{3u} & \{2\} \\
		59.414 & \text{IV} & B_{3g} & \{2,2\} \\
		59.414 & \text{IV} & A_u & \{2\} \\
		59.414 & \text{IV} & A_g & \{2,2,2\} \\
		59.415 & \text{IV} & B_{1u} & \{2,2,2,2,2\} \\
		59.415 & \text{IV} & B_{1g} & \{\} \\
\end{array}&&
		\begin{array}{cc|cc}
		\text{MSG} & \text{} & \text{Irrep} & X_f \\ \hline
		60.425 & \text{III} & B_ 3 & \{2\} \\
		60.425 & \text{III} & A_ 1 & \{2,2\} \\
		60.426 & \text{IV} & B_{1u} & \{2,2,4\} \\
		60.426 & \text{IV} & B_{1g} & \{2,2,2\} \\
		60.426 & \text{IV} & B_{2u} & \{2,2,2,2\} \\
		60.426 & \text{IV} & B_{2g} & \{2,2,2,2\} \\
		60.426 & \text{IV} & B_{3u} & \{2,4\} \\
\end{array}
	\end{align*}
\end{table}\clearpage

\begin{table}
	\begin{align*}
		\begin{array}{cc|cc}
		\text{MSG} & \text{} & \text{Irrep} & X_f \\ \hline
		61.433 & \text{I} & B_{3g} & \{2,2\} \\
		61.433 & \text{I} & A_u & \{4\} \\
		61.433 & \text{I} & A_g & \{2,2\} \\
		61.434 & \text{II} & B_{1u} & \{2,2,2\} \\
		61.434 & \text{II} & B_{1g} & \{2\} \\
		61.434 & \text{II} & B_{2u} & \{2,2,2\} \\
		61.434 & \text{II} & B_{2g} & \{2\} \\
\end{array}&&
		\begin{array}{cc|cc}
		\text{MSG} & \text{} & \text{Irrep} & X_f \\ \hline
		62.443 & \text{III} & B_ 1 & \{2\} \\
		62.443 & \text{III} & B_ 2 & \{2,2,2\} \\
		62.443 & \text{III} & A_ 2 & \{2\} \\
		62.443 & \text{III} & A_ 1 & \{2,2,2\} \\
		62.444 & \text{III} & B_ 2 & \{2,2,2\} \\
		62.444 & \text{III} & B_ 1 & \{2,2\} \\
		62.444 & \text{III} & A_ 2 & \{2,2\} \\
\end{array}&&
		\begin{array}{cc|cc}
		\text{MSG} & \text{} & \text{Irrep} & X_f \\ \hline
		62.454 & \text{IV} & B_{2g} & \{2,2\} \\
		62.454 & \text{IV} & B_{3u} & \{2,2,2\} \\
		62.454 & \text{IV} & B_{3g} & \{\} \\
		62.454 & \text{IV} & A_u & \{2,2\} \\
		62.454 & \text{IV} & A_g & \{2,2\} \\
		62.455 & \text{IV} & B_{1u} & \{2,2,2\} \\
		62.455 & \text{IV} & B_{1g} & \{2\} \\
\end{array}
	\end{align*}
\end{table}\clearpage

\begin{table}
	\begin{align*}
		\begin{array}{cc|cc}
		\text{MSG} & \text{} & \text{Irrep} & X_f \\ \hline
		63.465 & \text{III} & B_ 3 & \{2\} \\
		63.465 & \text{III} & A_ 1 & \{2,2\} \\
		63.466 & \text{IV} & B_{1u} & \{2,2,2,2\} \\
		63.466 & \text{IV} & B_{1g} & \{2\} \\
		63.466 & \text{IV} & B_{2u} & \{2,2,2,2\} \\
		63.466 & \text{IV} & B_{2g} & \{\} \\
		63.466 & \text{IV} & B_{3u} & \{2,2,2\} \\
\end{array}&&
		\begin{array}{cc|cc}
		\text{MSG} & \text{} & \text{Irrep} & X_f \\ \hline
		64.476 & \text{III} & B_g & \{2,2\} \\
		64.476 & \text{III} & A_u & \{2,2,2,4\} \\
		64.476 & \text{III} & A_g & \{2,2,2\} \\
		64.477 & \text{III} & B_ 1 & \{2,2\} \\
		64.477 & \text{III} & B_ 2 & \{2,2\} \\
		64.477 & \text{III} & B_ 3 & \{2,2\} \\
		64.477 & \text{III} & A_ 1 & \{2,2,2\} \\
\end{array}&&
		\begin{array}{cc|cc}
		\text{MSG} & \text{} & \text{Irrep} & X_f \\ \hline
		65.487 & \text{III} & B_ 1 & \{2,2,2\} \\
		65.487 & \text{III} & B_ 2 & \{2,2\} \\
		65.487 & \text{III} & B_ 3 & \{2,2\} \\
		65.487 & \text{III} & A_ 1 & \{2,2\} \\
		65.488 & \text{IV} & B_{1u} & \{2,2,2,2,2,2\} \\
		65.488 & \text{IV} & B_{1g} & \{\} \\
		65.488 & \text{IV} & B_{2u} & \{2,2,2,2\} \\
\end{array}
	\end{align*}
\end{table}\clearpage

\begin{table}
	\begin{align*}
		\begin{array}{cc|cc}
		\text{MSG} & \text{} & \text{Irrep} & X_f \\ \hline
		66.497 & \text{III} & A_ 1 & \{2,2,2\} \\
		66.498 & \text{IV} & B_{1u} & \{2,2,2,2\} \\
		66.498 & \text{IV} & B_{1g} & \{2\} \\
		66.498 & \text{IV} & B_{2u} & \{2,2,2,4\} \\
		66.498 & \text{IV} & B_{2g} & \{2,2\} \\
		66.498 & \text{IV} & B_{3u} & \{2,2,2,4\} \\
		66.498 & \text{IV} & B_{3g} & \{2,2\} \\
\end{array}&&
		\begin{array}{cc|cc}
		\text{MSG} & \text{} & \text{Irrep} & X_f \\ \hline
		67.508 & \text{IV} & B_{2u} & \{2,2,2,2\} \\
		67.508 & \text{IV} & B_{2g} & \{2,2\} \\
		67.508 & \text{IV} & B_{3u} & \{2,2,2,2\} \\
		67.508 & \text{IV} & B_{3g} & \{2,2\} \\
		67.508 & \text{IV} & A_u & \{2,2,2\} \\
		67.508 & \text{IV} & A_g & \{2\} \\
		67.509 & \text{IV} & B_{1u} & \{2,2,2,2,2\} \\
\end{array}&&
		\begin{array}{cc|cc}
		\text{MSG} & \text{} & \text{Irrep} & X_f \\ \hline
		68.518 & \text{IV} & B_{3g} & \{2,2\} \\
		68.518 & \text{IV} & A_u & \{2,2,2,2\} \\
		68.518 & \text{IV} & A_g & \{2,2\} \\
		68.519 & \text{IV} & B_{1u} & \{2,2,2,2,2\} \\
		68.519 & \text{IV} & B_{1g} & \{2,2,2\} \\
		68.519 & \text{IV} & B_{2u} & \{2,2,2,4\} \\
		68.519 & \text{IV} & B_{2g} & \{2,2,2\} \\
\end{array}
	\end{align*}
\end{table}\clearpage

\begin{table}
	\begin{align*}
		\begin{array}{cc|cc}
		\text{MSG} & \text{} & \text{Irrep} & X_f \\ \hline
		70.528 & \text{II} & B_{1u} & \{2,4\} \\
		70.528 & \text{II} & B_{1g} & \{2\} \\
		70.528 & \text{II} & B_{2u} & \{2,4\} \\
		70.528 & \text{II} & B_{2g} & \{2\} \\
		70.528 & \text{II} & B_{3u} & \{2,4\} \\
		70.528 & \text{II} & B_{3g} & \{2\} \\
		70.528 & \text{II} & A_u & \{2,4\} \\
\end{array}&&
		\begin{array}{cc|cc}
		\text{MSG} & \text{} & \text{Irrep} & X_f \\ \hline
		71.538 & \text{IV} & A_g & \{\} \\
		72.539 & \text{I} & B_{1u} & \{2,2,2,2\} \\
		72.539 & \text{I} & B_{1g} & \{2\} \\
		72.539 & \text{I} & B_{2u} & \{2,2,2,2\} \\
		72.539 & \text{I} & B_{2g} & \{2\} \\
		72.539 & \text{I} & B_{3u} & \{2,2,2,2\} \\
		72.539 & \text{I} & B_{3g} & \{2\} \\
\end{array}&&
		\begin{array}{cc|cc}
		\text{MSG} & \text{} & \text{Irrep} & X_f \\ \hline
		73.549 & \text{II} & B_{2u} & \{2,2,2,2\} \\
		73.549 & \text{II} & B_{2g} & \{2\} \\
		73.549 & \text{II} & B_{3u} & \{2,2,2,2\} \\
		73.549 & \text{II} & B_{3g} & \{2\} \\
		73.549 & \text{II} & A_u & \{2,2,2,2\} \\
		73.549 & \text{II} & A_g & \{2,2,2\} \\
		73.550 & \text{III} & B_ 1 & \{2,2\} \\
\end{array}
	\end{align*}
\end{table}\clearpage

\begin{table}
	\begin{align*}
		\begin{array}{cc|cc}
		\text{MSG} & \text{} & \text{Irrep} & X_f \\ \hline
		74.561 & \text{IV} & B_{1g} & \{2,2,2\} \\
		74.561 & \text{IV} & B_{2u} & \{2,2,2,2\} \\
		74.561 & \text{IV} & B_{2g} & \{2\} \\
		74.561 & \text{IV} & B_{3u} & \{2,2,2,2\} \\
		74.561 & \text{IV} & B_{3g} & \{2\} \\
		74.561 & \text{IV} & A_u & \{2,2,2,2\} \\
		74.561 & \text{IV} & A_g & \{2\} \\
\end{array}&&
		\begin{array}{cc|cc}
		\text{MSG} & \text{} & \text{Irrep} & X_f \\ \hline
		78.23 & \text{IV} & B & \{2,2,2\} \\
		78.23 & \text{IV} & A & \{2,2,2\} \\
		78.24 & \text{IV} & B & \{2,2\} \\
		78.24 & \text{IV} & A & \{2,2\} \\
		79.25 & \text{I} & B & \{2,2,2,2\} \\
		79.25 & \text{I} & \text{}^1E & \{2,4\} \\
		79.25 & \text{I} & \text{}^2E & \{2,4\} \\
\end{array}&&
		\begin{array}{cc|cc}
		\text{MSG} & \text{} & \text{Irrep} & X_f \\ \hline
		83.48 & \text{IV} & B_g & \{2,4\} \\
		83.48 & \text{IV} & A_u & \{2,2,2,2,2\} \\
		83.48 & \text{IV} & A_g & \{2,4\} \\
		83.49 & \text{IV} & B_u & \{2,2,2,2,2\} \\
		83.49 & \text{IV} & B_g & \{2,2,2\} \\
		83.49 & \text{IV} & A_u & \{2,2,2,2,2\} \\
		83.49 & \text{IV} & A_g & \{2,2,2\} \\
\end{array}
	\end{align*}
\end{table}\clearpage

\begin{table}
	\begin{align*}
		\begin{array}{cc|cc}
		\text{MSG} & \text{} & \text{Irrep} & X_f \\ \hline
		85.65 & \text{IV} & B_u & \{2,2,2,2\} \\
		85.65 & \text{IV} & B_g & \{2,2\} \\
		85.65 & \text{IV} & A_u & \{2,2,2,2,2\} \\
		85.65 & \text{IV} & A_g & \{2,2\} \\
		85.66 & \text{IV} & B_u & \{2,2,2,2,2\} \\
		85.66 & \text{IV} & B_g & \{2,2\} \\
		85.66 & \text{IV} & A_u & \{2,2,2,2,2\} \\
\end{array}&&
		\begin{array}{cc|cc}
		\text{MSG} & \text{} & \text{Irrep} & X_f \\ \hline
		88.81 & \text{I} & \text{}^1E_g & \{4\} \\
		88.81 & \text{I} & \text{}^2E_u & \{2,2,4\} \\
		88.81 & \text{I} & \text{}^2E_g & \{4\} \\
		88.81 & \text{I} & A_u & \{2,2,4\} \\
		88.81 & \text{I} & A_g & \{2,2\} \\
		88.82 & \text{II} & B_u & \{2,4\} \\
		88.82 & \text{II} & B_g & \{2,2\} \\
\end{array}&&
		\begin{array}{cc|cc}
		\text{MSG} & \text{} & \text{Irrep} & X_f \\ \hline
		90.100 & \text{IV} & B_ 2 & \{2,2\} \\
		90.100 & \text{IV} & B_ 1 & \{2,2,2\} \\
		90.100 & \text{IV} & A_ 2 & \{2,2\} \\
		90.100 & \text{IV} & A_ 1 & \{2,2\} \\
		90.101 & \text{IV} & B_ 2 & \{2\} \\
		90.101 & \text{IV} & B_ 1 & \{2,2\} \\
		90.101 & \text{IV} & A_ 2 & \{2,2\} \\
\end{array}
	\end{align*}
\end{table}\clearpage

\begin{table}
	\begin{align*}
		\begin{array}{cc|cc}
		\text{MSG} & \text{} & \text{Irrep} & X_f \\ \hline
		92.117 & \text{IV} & A_ 1 & \{2,2\} \\
		92.118 & \text{IV} & B_ 2 & \{2\} \\
		92.118 & \text{IV} & B_ 1 & \{2\} \\
		92.118 & \text{IV} & A_ 2 & \{2\} \\
		92.118 & \text{IV} & A_ 1 & \{2\} \\
		93.119 & \text{I} & B_ 2 & \{2\} \\
		93.119 & \text{I} & B_ 1 & \{\} \\
\end{array}&&
		\begin{array}{cc|cc}
		\text{MSG} & \text{} & \text{Irrep} & X_f \\ \hline
		95.135 & \text{I} & A_ 2 & \{2,2\} \\
		95.135 & \text{I} & A_ 1 & \{2,2\} \\
		95.136 & \text{II} & B_ 2 & \{2\} \\
		95.136 & \text{II} & B_ 1 & \{2\} \\
		95.136 & \text{II} & A_ 2 & \{2,2\} \\
		95.136 & \text{II} & A_ 1 & \{2,2\} \\
		95.137 & \text{III} & B_ 1 & \{2,2\} \\
\end{array}&&
		\begin{array}{cc|cc}
		\text{MSG} & \text{} & \text{Irrep} & X_f \\ \hline
		97.153 & \text{III} & A_ 1 & \{2,2\} \\
		97.154 & \text{III} & B & \{2,2,2,2\} \\
		97.154 & \text{III} & \text{}^1E & \{2,4\} \\
		97.154 & \text{III} & \text{}^2E & \{2,4\} \\
		97.154 & \text{III} & A & \{2,2,2,2\} \\
		97.155 & \text{III} & B_ 1 & \{2,2\} \\
		97.155 & \text{III} & A_ 1 & \{2,2\} \\
\end{array}
	\end{align*}
\end{table}\clearpage

\begin{table}
	\begin{align*}
		\begin{array}{cc|cc}
		\text{MSG} & \text{} & \text{Irrep} & X_f \\ \hline
		100.172 & \text{II} & B_ 2 & \{2,2,2\} \\
		100.172 & \text{II} & B_ 1 & \{2\} \\
		100.172 & \text{II} & A_ 2 & \{2\} \\
		100.172 & \text{II} & A_ 1 & \{2,2,2,2\} \\
		100.173 & \text{III} & A_ 2 & \{2\} \\
		100.173 & \text{III} & A_ 1 & \{2,2,2\} \\
		100.174 & \text{III} & A_ 2 & \{2,2\} \\
\end{array}&&
		\begin{array}{cc|cc}
		\text{MSG} & \text{} & \text{Irrep} & X_f \\ \hline
		102.190 & \text{III} & A_ 1 & \{2,2\} \\
		102.191 & \text{III} & B & \{2,2,2,2\} \\
		102.191 & \text{III} & \text{}^1E & \{4\} \\
		102.191 & \text{III} & \text{}^2E & \{4\} \\
		102.191 & \text{III} & A & \{2,2,2,2\} \\
		102.192 & \text{IV} & B_ 2 & \{2,2,2\} \\
		102.192 & \text{IV} & B_ 1 & \{2\} \\
\end{array}&&
		\begin{array}{cc|cc}
		\text{MSG} & \text{} & \text{Irrep} & X_f \\ \hline
		104.208 & \text{IV} & A_ 2 & \{2,2\} \\
		104.208 & \text{IV} & A_ 1 & \{2,2,2\} \\
		104.209 & \text{IV} & B_ 2 & \{2,2,2\} \\
		104.209 & \text{IV} & B_ 1 & \{2\} \\
		104.209 & \text{IV} & A_ 2 & \{2\} \\
		104.209 & \text{IV} & A_ 1 & \{2,2,2\} \\
		104.210 & \text{IV} & B_ 2 & \{2,2,2\} \\
\end{array}
	\end{align*}
\end{table}\clearpage

\begin{table}
	\begin{align*}
		\begin{array}{cc|cc}
		\text{MSG} & \text{} & \text{Irrep} & X_f \\ \hline
		106.226 & \text{IV} & B_ 1 & \{2\} \\
		106.226 & \text{IV} & A_ 2 & \{2\} \\
		106.226 & \text{IV} & A_ 1 & \{2,2,2\} \\
		107.227 & \text{I} & B_ 2 & \{2\} \\
		107.227 & \text{I} & B_ 1 & \{2,2\} \\
		107.227 & \text{I} & A_ 2 & \{\} \\
		107.227 & \text{I} & A_ 1 & \{2,2,2,2,2,2\} \\
\end{array}&&
		\begin{array}{cc|cc}
		\text{MSG} & \text{} & \text{Irrep} & X_f \\ \hline
		110.245 & \text{I} & B_ 2 & \{2,2,2\} \\
		110.245 & \text{I} & B_ 1 & \{2,2,2\} \\
		110.245 & \text{I} & A_ 2 & \{2,2,2\} \\
		110.245 & \text{I} & A_ 1 & \{2,2,2\} \\
		110.246 & \text{II} & B_ 2 & \{2,2\} \\
		110.246 & \text{II} & B_ 1 & \{2,2\} \\
		110.246 & \text{II} & A_ 2 & \{2,2\} \\
\end{array}&&
		\begin{array}{cc|cc}
		\text{MSG} & \text{} & \text{Irrep} & X_f \\ \hline
		112.263 & \text{III} & A & \{2,2,2,2,2\} \\
		112.264 & \text{IV} & B_ 2 & \{2,2,2\} \\
		112.264 & \text{IV} & A_ 2 & \{2,2\} \\
		112.264 & \text{IV} & B_ 1 & \{2\} \\
		112.264 & \text{IV} & A_ 1 & \{2,2\} \\
		112.265 & \text{IV} & B_ 2 & \{2,2,2,2\} \\
		112.265 & \text{IV} & A_ 2 & \{2,2,2\} \\
\end{array}
	\end{align*}
\end{table}\clearpage

\begin{table}
	\begin{align*}
		\begin{array}{cc|cc}
		\text{MSG} & \text{} & \text{Irrep} & X_f \\ \hline
		114.281 & \text{IV} & B_ 1 & \{2,2,2\} \\
		114.281 & \text{IV} & A_ 1 & \{2,2,2,2\} \\
		114.282 & \text{IV} & B_ 2 & \{2,2,2\} \\
		114.282 & \text{IV} & A_ 2 & \{2,2\} \\
		114.282 & \text{IV} & B_ 1 & \{2,2\} \\
		114.282 & \text{IV} & A_ 1 & \{2,2,2\} \\
		115.283 & \text{I} & B_ 2 & \{2,2,2,2,2,2\} \\
\end{array}&&
		\begin{array}{cc|cc}
		\text{MSG} & \text{} & \text{Irrep} & X_f \\ \hline
		117.299 & \text{I} & A_ 2 & \{2,2,2\} \\
		117.299 & \text{I} & B_ 1 & \{2,2\} \\
		117.299 & \text{I} & A_ 1 & \{2,2\} \\
		117.300 & \text{II} & B_ 2 & \{2,2,2\} \\
		117.300 & \text{II} & A_ 2 & \{2,2\} \\
		117.300 & \text{II} & B_ 1 & \{2\} \\
		117.300 & \text{II} & A_ 1 & \{2\} \\
\end{array}&&
		\begin{array}{cc|cc}
		\text{MSG} & \text{} & \text{Irrep} & X_f \\ \hline
		119.317 & \text{III} & B_ 1 & \{2\} \\
		119.317 & \text{III} & A_ 1 & \{2,2\} \\
		119.318 & \text{III} & A_ 2 & \{\} \\
		119.318 & \text{III} & A_ 1 & \{2,2,2,2\} \\
		119.319 & \text{III} & \text{}^1E & \{2,2,4\} \\
		119.319 & \text{III} & \text{}^2E & \{2,2,4\} \\
		119.319 & \text{III} & B & \{2,2,2,2\} \\
\end{array}
	\end{align*}
\end{table}\clearpage

\begin{table}
	\begin{align*}
		\begin{array}{cc|cc}
		\text{MSG} & \text{} & \text{Irrep} & X_f \\ \hline
		122.336 & \text{III} & A_ 1 & \{2,2\} \\
		122.337 & \text{III} & \text{}^1E & \{2,2,2\} \\
		122.337 & \text{III} & \text{}^2E & \{2,2,2\} \\
		122.337 & \text{III} & B & \{2,2,2\} \\
		122.337 & \text{III} & A & \{2,2,2\} \\
		122.338 & \text{IV} & B_ 2 & \{2,2\} \\
		122.338 & \text{IV} & A_ 2 & \{2,2\} \\
\end{array}&&
		\begin{array}{cc|cc}
		\text{MSG} & \text{} & \text{Irrep} & X_f \\ \hline
		123.348 & \text{IV} & A_{1u} & \{2\} \\
		123.348 & \text{IV} & A_{1g} & \{\} \\
		123.349 & \text{IV} & B_{2u} & \{2,2\} \\
		123.349 & \text{IV} & B_{2g} & \{\} \\
		123.349 & \text{IV} & B_{1u} & \{2,2\} \\
		123.349 & \text{IV} & B_{1g} & \{\} \\
		123.349 & \text{IV} & A_{2u} & \{2,2,2,2,2,2\} \\
\end{array}&&
		\begin{array}{cc|cc}
		\text{MSG} & \text{} & \text{Irrep} & X_f \\ \hline
		124.359 & \text{III} & B_ 1 & \{2\} \\
		124.359 & \text{III} & A_ 2 & \{2,2,2\} \\
		124.359 & \text{III} & A_ 1 & \{2,2\} \\
		124.360 & \text{IV} & B_{2u} & \{2,2,2\} \\
		124.360 & \text{IV} & B_{2g} & \{4\} \\
		124.360 & \text{IV} & B_{1u} & \{2,2\} \\
		124.360 & \text{IV} & B_{1g} & \{4\} \\
\end{array}
	\end{align*}
\end{table}\clearpage

\begin{table}
	\begin{align*}
		\begin{array}{cc|cc}
		\text{MSG} & \text{} & \text{Irrep} & X_f \\ \hline
		125.369 & \text{III} & \text{}^2E_u & \{2,4,8\} \\
		125.369 & \text{III} & \text{}^2E_g & \{2,4\} \\
		125.369 & \text{III} & A_u & \{2,2,2,2,4\} \\
		125.369 & \text{III} & A_g & \{2,2,4\} \\
		125.370 & \text{III} & B_ 2 & \{2,2,2\} \\
		125.370 & \text{III} & A_ 2 & \{2,2,2\} \\
		125.370 & \text{III} & B_ 1 & \{2,2\} \\
\end{array}&&
		\begin{array}{cc|cc}
		\text{MSG} & \text{} & \text{Irrep} & X_f \\ \hline
		126.379 & \text{III} & A_g & \{\} \\
		126.380 & \text{III} & B_ 2 & \{2,2,2\} \\
		126.380 & \text{III} & A_ 2 & \{2,2,2\} \\
		126.380 & \text{III} & B_ 1 & \{2\} \\
		126.380 & \text{III} & A_ 1 & \{2\} \\
		126.381 & \text{III} & B_u & \{2,2,2,2,2\} \\
		126.381 & \text{III} & B_g & \{2,2,2\} \\
\end{array}&&
		\begin{array}{cc|cc}
		\text{MSG} & \text{} & \text{Irrep} & X_f \\ \hline
		127.389 & \text{III} & A_ 2 & \{2\} \\
		127.389 & \text{III} & A_ 1 & \{2,2,2,2\} \\
		127.390 & \text{III} & B_{1u} & \{2,2,2,2\} \\
		127.390 & \text{III} & B_{1g} & \{\} \\
		127.390 & \text{III} & A_u & \{2,2\} \\
		127.390 & \text{III} & A_g & \{\} \\
		127.391 & \text{III} & B_{1u} & \{2,2,2\} \\
\end{array}
	\end{align*}
\end{table}\clearpage

\begin{table}
	\begin{align*}
		\begin{array}{cc|cc}
		\text{MSG} & \text{} & \text{Irrep} & X_f \\ \hline
		128.400 & \text{II} & B_{2g} & \{2\} \\
		128.400 & \text{II} & B_{1u} & \{2,2\} \\
		128.400 & \text{II} & B_{1g} & \{2\} \\
		128.400 & \text{II} & A_{2u} & \{2,2,2,2\} \\
		128.400 & \text{II} & A_{2g} & \{2\} \\
		128.400 & \text{II} & A_{1u} & \{2,2\} \\
		128.400 & \text{II} & A_{1g} & \{2\} \\
\end{array}&&
		\begin{array}{cc|cc}
		\text{MSG} & \text{} & \text{Irrep} & X_f \\ \hline
		129.411 & \text{I} & B_{2u} & \{2\} \\
		129.411 & \text{I} & B_{2g} & \{\} \\
		129.411 & \text{I} & B_{1u} & \{2\} \\
		129.411 & \text{I} & B_{1g} & \{2,2\} \\
		129.411 & \text{I} & A_{2u} & \{2,2,2,2,2,2\} \\
		129.411 & \text{I} & A_{2g} & \{\} \\
		129.411 & \text{I} & A_{1u} & \{\} \\
\end{array}&&
		\begin{array}{cc|cc}
		\text{MSG} & \text{} & \text{Irrep} & X_f \\ \hline
		129.421 & \text{IV} & A_{1g} & \{2,2,2\} \\
		129.422 & \text{IV} & B_{2u} & \{2\} \\
		129.422 & \text{IV} & B_{2g} & \{\} \\
		129.422 & \text{IV} & B_{1u} & \{2\} \\
		129.422 & \text{IV} & B_{1g} & \{2,2\} \\
		129.422 & \text{IV} & A_{2u} & \{2,2,2,2,2,2\} \\
		129.422 & \text{IV} & A_{2g} & \{\} \\
\end{array}
	\end{align*}
\end{table}\clearpage
\begin{table}
	\begin{align*}
\begin{array}{cc|cc}
	\text{MSG} & \text{} & \text{Irrep} & X_f \\ \hline
	130.432 & \text{IV} & A_{1u} & \{2\} \\
	130.432 & \text{IV} & A_{1g} & \{2,2\} \\
	130.433 & \text{IV} & B_{2u} & \{2\} \\
	130.433 & \text{IV} & B_{2g} & \{\} \\
	130.433 & \text{IV} & B_{1u} & \{2,2\} \\
	130.433 & \text{IV} & B_{1g} & \{2\} \\
	130.433 & \text{IV} & A_{2u} & \{2,2,2\} \\
\end{array}&&
\begin{array}{cc|cc}
	\text{MSG} & \text{} & \text{Irrep} & X_f \\ \hline
	131.443 & \text{III} & B_1 & \{\} \\
	131.443 & \text{III} & A_2 & \{2,2\} \\
	131.443 & \text{III} & A_1 & \{2\} \\
	131.444 & \text{IV} & B_{2u} & \{2,2,2\} \\
	131.444 & \text{IV} & B_{2g} & \{\} \\
	131.444 & \text{IV} & B_{1u} & \{2\} \\
	131.444 & \text{IV} & B_{1g} & \{\} \\
\end{array}&&
\begin{array}{cc|cc}
	\text{MSG} & \text{} & \text{Irrep} & X_f \\ \hline
	132.453 & \text{III} & \text{}^2E_u & \{2,2,4,8\} \\
	132.453 & \text{III} & \text{}^2E_g & \{2,4\} \\
	132.453 & \text{III} & A_u & \{2,2,2,2,2\} \\
	132.453 & \text{III} & A_g & \{2,4\} \\
	132.454 & \text{III} & B_2 & \{2,2,2\} \\
	132.454 & \text{III} & A_2 & \{2,2,2\} \\
	132.454 & \text{III} & B_1 & \{2,2\} \\
\end{array}
	\end{align*}
\end{table}\clearpage

\begin{table}
\begin{align*}
\begin{array}{cc|cc}
	\text{MSG} & \text{} & \text{Irrep} & X_f \\ \hline
	133.463 & \text{III} & A_g & \{2\} \\
	133.464 & \text{III} & B_2 & \{2,2,2,2\} \\
	133.464 & \text{III} & A_2 & \{2,2,2,2\} \\
	133.464 & \text{III} & B_1 & \{2,2\} \\
	133.464 & \text{III} & A_1 & \{2,2\} \\
	133.465 & \text{III} & B_u & \{2,2,2,2,2,2\} \\
	133.465 & \text{III} & B_g & \{2,2,2,2\} \\
\end{array}&&
\begin{array}{cc|cc}
	\text{MSG} & \text{} & \text{Irrep} & X_f \\ \hline
	134.473 & \text{III} & A_2 & \{2\} \\
	134.473 & \text{III} & A_1 & \{2,2,2,2\} \\
	134.474 & \text{III} & B_{1u} & \{2,2,2,2\} \\
	134.474 & \text{III} & B_{1g} & \{2\} \\
	134.474 & \text{III} & A_u & \{2,2,2\} \\
	134.474 & \text{III} & A_g & \{2\} \\
	134.475 & \text{III} & B_{1u} & \{2,2\} \\
\end{array}&&
\begin{array}{cc|cc}
	\text{MSG} & \text{} & \text{Irrep} & X_f \\ \hline
	135.484 & \text{II} & B_{2g} & \{\} \\
	135.484 & \text{II} & B_{1u} & \{2,2\} \\
	135.484 & \text{II} & B_{1g} & \{2\} \\
	135.484 & \text{II} & A_{2u} & \{2,2,2\} \\
	135.484 & \text{II} & A_{2g} & \{\} \\
	135.484 & \text{II} & A_{1u} & \{2\} \\
	135.484 & \text{II} & A_{1g} & \{2\} \\
\end{array}
	\end{align*}
\end{table}\clearpage

\begin{table}
\begin{align*}
\begin{array}{cc|cc}
	\text{MSG} & \text{} & \text{Irrep} & X_f \\ \hline
	136.495 & \text{I} & B_{2u} & \{2\} \\
	136.495 & \text{I} & B_{2g} & \{\} \\
	136.495 & \text{I} & B_{1u} & \{2,2,2\} \\
	136.495 & \text{I} & B_{1g} & \{\} \\
	136.495 & \text{I} & A_{2u} & \{2,2,2\} \\
	136.495 & \text{I} & A_{2g} & \{\} \\
	136.495 & \text{I} & A_{1u} & \{2\} \\
\end{array}&&
\begin{array}{cc|cc}
	\text{MSG} & \text{} & \text{Irrep} & X_f \\ \hline
	136.505 & \text{IV} & A_{1g} & \{\} \\
	136.506 & \text{IV} & B_{2u} & \{2\} \\
	136.506 & \text{IV} & B_{2g} & \{\} \\
	136.506 & \text{IV} & B_{1u} & \{2,2,2\} \\
	136.506 & \text{IV} & B_{1g} & \{\} \\
	136.506 & \text{IV} & A_{2u} & \{2,2,2\} \\
	136.506 & \text{IV} & A_{2g} & \{\} \\
\end{array}&&
\begin{array}{cc|cc}
	\text{MSG} & \text{} & \text{Irrep} & X_f \\ \hline
	137.516 & \text{IV} & A_{1u} & \{\} \\
	137.516 & \text{IV} & A_{1g} & \{2,2\} \\
	137.517 & \text{IV} & B_{2u} & \{2,2\} \\
	137.517 & \text{IV} & B_{2g} & \{\} \\
	137.517 & \text{IV} & B_{1u} & \{2,2\} \\
	137.517 & \text{IV} & B_{1g} & \{2,2,2\} \\
	137.517 & \text{IV} & A_{2u} & \{2,2,2,2\} \\
\end{array}
	\end{align*}
\end{table}\clearpage

\begin{table}
\begin{align*}
\begin{array}{cc|cc}
	\text{MSG} & \text{} & \text{Irrep} & X_f \\ \hline
	138.527 & \text{III} & B_1 & \{2,2\} \\
	138.527 & \text{III} & A_2 & \{2,2\} \\
	138.527 & \text{III} & A_1 & \{2\} \\
	138.528 & \text{IV} & B_{2u} & \{2\} \\
	138.528 & \text{IV} & B_{2g} & \{\} \\
	138.528 & \text{IV} & B_{1u} & \{2\} \\
	138.528 & \text{IV} & B_{1g} & \{2\} \\
\end{array}&&
\begin{array}{cc|cc}
	\text{MSG} & \text{} & \text{Irrep} & X_f \\ \hline
	139.537 & \text{III} & \text{}^2E_u & \{2,2,8,8\} \\
	139.537 & \text{III} & \text{}^2E_g & \{4\} \\
	139.537 & \text{III} & A_u & \{2,2,2,2,2\} \\
	139.537 & \text{III} & A_g & \{4,4\} \\
	139.538 & \text{III} & B_2 & \{2,2,2,2\} \\
	139.538 & \text{III} & A_2 & \{2\} \\
	139.538 & \text{III} & B_1 & \{\} \\
\end{array}&&
\begin{array}{cc|cc}
	\text{MSG} & \text{} & \text{Irrep} & X_f \\ \hline
	140.548 & \text{III} & A_1 & \{2\} \\
	140.549 & \text{III} & B_2 & \{\} \\
	140.549 & \text{III} & B_1 & \{2\} \\
	140.549 & \text{III} & A_2 & \{2,2\} \\
	140.549 & \text{III} & A_1 & \{2,2\} \\
	140.550 & \text{IV} & B_{2u} & \{2\} \\
	140.550 & \text{IV} & B_{2g} & \{\} \\
\end{array}
	\end{align*}
\end{table}\clearpage

\begin{table}
\begin{align*}
\begin{array}{cc|cc}
	\text{MSG} & \text{} & \text{Irrep} & X_f \\ \hline
	141.560 & \text{IV} & B_{1u} & \{2,2\} \\
	141.560 & \text{IV} & B_{1g} & \{2\} \\
	141.560 & \text{IV} & A_{2u} & \{2,2,2\} \\
	141.560 & \text{IV} & A_{2g} & \{2,2\} \\
	141.560 & \text{IV} & A_{1u} & \{2,2\} \\
	141.560 & \text{IV} & A_{1g} & \{\} \\
	142.561 & \text{I} & B_{2u} & \{2,2\} \\
\end{array}&&
\begin{array}{cc|cc}
	\text{MSG} & \text{} & \text{Irrep} & X_f \\ \hline
	143.1 & \text{I} & \text{}^2E & \{2,2,6\} \\
	143.1 & \text{I} & A & \{2,2,6\} \\
	143.2 & \text{II} & A & \{2,2,2\} \\
	143.3 & \text{IV} & A & \{2,2\} \\
	144.4 & \text{I} & \text{}^1E & \{2,2,2\} \\
	144.4 & \text{I} & \text{}^2E & \{2,2,2\} \\
	144.4 & \text{I} & A & \{2,2,2\} \\
\end{array}&&
\begin{array}{cc|cc}
	\text{MSG} & \text{} & \text{Irrep} & X_f \\ \hline
	151.31 & \text{III} & \text{}^1E & \{2,2,2\} \\
	151.31 & \text{III} & \text{}^2E & \{2,2,2\} \\
	151.31 & \text{III} & A & \{2,2,2\} \\
	151.32 & \text{IV} & A_2 & \{2\} \\
	151.32 & \text{IV} & A_1 & \{2,2\} \\
	152.33 & \text{I} & A_2 & \{2\} \\
	152.33 & \text{I} & A_1 & \{2,2\} \\
\end{array}
	\end{align*}
\end{table}\clearpage

\begin{table}
\begin{align*}
\begin{array}{cc|cc}
	\text{MSG} & \text{} & \text{Irrep} & X_f \\ \hline
	158.59 & \text{III} & \text{}^1E & \{2,6\} \\
	158.59 & \text{III} & \text{}^2E & \{2,6\} \\
	158.59 & \text{III} & A & \{2,6\} \\
	158.60 & \text{IV} & A_2 & \{2\} \\
	158.60 & \text{IV} & A_1 & \{2,2\} \\
	159.61 & \text{I} & A_2 & \{2,2\} \\
	159.61 & \text{I} & A_1 & \{2,2\} \\
\end{array}&&
\begin{array}{cc|cc}
	\text{MSG} & \text{} & \text{Irrep} & X_f \\ \hline
	163.81 & \text{III} & A_1 & \{2,2\} \\
	163.82 & \text{III} & A_2 & \{2\} \\
	163.82 & \text{III} & A_1 & \{2,2\} \\
	163.83 & \text{III} & \text{}^1E_u & \{2,24\} \\
	163.83 & \text{III} & \text{}^1E_g & \{2,6\} \\
	163.83 & \text{III} & \text{}^2E_u & \{2,24\} \\
	163.83 & \text{III} & \text{}^2E_g & \{2,6\} \\
\end{array}&&
\begin{array}{cc|cc}
	\text{MSG} & \text{} & \text{Irrep} & X_f \\ \hline
	166.98 & \text{II} & A_{1u} & \{2,4\} \\
	166.98 & \text{II} & A_{1g} & \{2\} \\
	166.99 & \text{III} & A_2 & \{\} \\
	166.99 & \text{III} & A_1 & \{2,2,2,2,2\} \\
	166.100 & \text{III} & A_2 & \{2\} \\
	166.100 & \text{III} & A_1 & \{2,2\} \\
	166.101 & \text{III} & \text{}^1E_u & \{2,4,8\} \\
\end{array}
	\end{align*}
\end{table}\clearpage

\begin{table}
\begin{align*}
\begin{array}{cc|cc}
	\text{MSG} & \text{} & \text{Irrep} & X_f \\ \hline
	170.117 & \text{I} & \text{}^1E_1 & \{2,2\} \\
	170.117 & \text{I} & \text{}^2E_1 & \{2,2\} \\
	170.117 & \text{I} & A & \{2,2\} \\
	170.118 & \text{II} & B & \{2\} \\
	170.118 & \text{II} & A & \{2,2\} \\
	170.119 & \text{III} & A & \{2,2\} \\
	170.120 & \text{IV} & B & \{2\} \\
\end{array}&&
\begin{array}{cc|cc}
	\text{MSG} & \text{} & \text{Irrep} & X_f \\ \hline
	175.137 & \text{I} & A_g & \{6,6,6\} \\
	175.138 & \text{II} & B_u & \{4,4,4\} \\
	175.138 & \text{II} & B_g & \{2,2\} \\
	175.138 & \text{II} & A_u & \{2,2,2,2,2,2,4\} \\
	175.138 & \text{II} & A_g & \{2,2,2\} \\
	175.139 & \text{III} & A'' & \{2,2\} \\
	175.139 & \text{III} & A' & \{2,2,2\} \\
\end{array}&&
\begin{array}{cc|cc}
	\text{MSG} & \text{} & \text{Irrep} & X_f \\ \hline
	178.155 & \text{I} & B_1 & \{\} \\
	178.155 & \text{I} & B_2 & \{\} \\
	178.155 & \text{I} & A_2 & \{2\} \\
	178.155 & \text{I} & A_1 & \{2\} \\
	178.156 & \text{II} & B_1 & \{\} \\
	178.156 & \text{II} & B_2 & \{\} \\
	178.156 & \text{II} & A_2 & \{2\} \\
\end{array}
	\end{align*}
\end{table}\clearpage

\begin{table}
\begin{align*}
\begin{array}{cc|cc}
	\text{MSG} & \text{} & \text{Irrep} & X_f \\ \hline
	180.172 & \text{IV} & B_2 & \{4\} \\
	180.172 & \text{IV} & A_2 & \{2,2\} \\
	180.172 & \text{IV} & A_1 & \{2,2\} \\
	181.173 & \text{I} & B_1 & \{2\} \\
	181.173 & \text{I} & B_2 & \{2\} \\
	181.173 & \text{I} & A_2 & \{2,2\} \\
	181.173 & \text{I} & A_1 & \{2\} \\
\end{array}&&
\begin{array}{cc|cc}
	\text{MSG} & \text{} & \text{Irrep} & X_f \\ \hline
	183.189 & \text{III} & \text{}^2E_1 & \{12\} \\
	183.189 & \text{III} & A & \{2,2,2,6\} \\
	183.190 & \text{IV} & B_2 & \{2\} \\
	183.190 & \text{IV} & B_1 & \{2,2\} \\
	183.190 & \text{IV} & A_2 & \{\} \\
	183.190 & \text{IV} & A_1 & \{2,2,2,2,2,2,2,2\} \\
	184.191 & \text{I} & B_2 & \{4\} \\
\end{array}&&
\begin{array}{cc|cc}
	\text{MSG} & \text{} & \text{Irrep} & X_f \\ \hline
	186.207 & \text{III} & \text{}^2E_2 & \{2,6\} \\
	186.207 & \text{III} & \text{}^1E_2 & \{2,6\} \\
	186.207 & \text{III} & \text{}^1E_1 & \{6\} \\
	186.207 & \text{III} & \text{}^2E_1 & \{6\} \\
	186.207 & \text{III} & A & \{2,6\} \\
	186.208 & \text{IV} & B_2 & \{\} \\
	186.208 & \text{IV} & B_1 & \{2,2\} \\
\end{array}
	\end{align*}
\end{table}\clearpage

\begin{table}
\begin{align*}
\begin{array}{cc|cc}
	\text{MSG} & \text{} & \text{Irrep} & X_f \\ \hline
	189.224 & \text{III} & A_2 & \{2\} \\
	189.224 & \text{III} & A_1 & \{2,2\} \\
	189.225 & \text{III} & \text{}^1\text{E'} & \{6,6,6\} \\
	189.225 & \text{III} & \text{}^1\text{E''} & \{6\} \\
	189.225 & \text{III} & \text{}^2\text{E'} & \{6,6,6\} \\
	189.225 & \text{III} & \text{}^2\text{E''} & \{6\} \\
	189.225 & \text{III} & A'' & \{6\} \\
\end{array}&&
\begin{array}{cc|cc}
	\text{MSG} & \text{} & \text{Irrep} & X_f \\ \hline
	191.238 & \text{III} & A_{2g} & \{\} \\
	191.238 & \text{III} & A_{1u} & \{2,4\} \\
	191.238 & \text{III} & A_{1g} & \{2\} \\
	191.239 & \text{III} & A_{2u} & \{2,2,2,2,2,4\} \\
	191.239 & \text{III} & A_{2g} & \{\} \\
	191.239 & \text{III} & A_{1u} & \{2,4\} \\
	191.239 & \text{III} & A_{1g} & \{2,2\} \\
\end{array}&&
\begin{array}{cc|cc}
	\text{MSG} & \text{} & \text{Irrep} & X_f \\ \hline
	192.249 & \text{III} & A_{2u} & \{2,2,4\} \\
	192.249 & \text{III} & A_{2g} & \{\} \\
	192.249 & \text{III} & A_{1u} & \{2,4\} \\
	192.249 & \text{III} & A_{1g} & \{2\} \\
	192.250 & \text{III} & B_u & \{12,12\} \\
	192.250 & \text{III} & B_g & \{12\} \\
	192.250 & \text{III} & \text{}^2E_{2u} & \{2,2,12\} \\
\end{array}
	\end{align*}
\end{table}\clearpage

\begin{table}
\begin{align*}
\begin{array}{cc|cc}
	\text{MSG} & \text{} & \text{Irrep} & X_f \\ \hline
	193.259 & \text{III} & A_{1g} & \{2\} \\
	193.260 & \text{III} & B_u & \{6,12\} \\
	193.260 & \text{III} & B_g & \{3\} \\
	193.260 & \text{III} & \text{}^2E_{2u} & \{2,12\} \\
	193.260 & \text{III} & \text{}^2E_{2g} & \{6,6\} \\
	193.260 & \text{III} & \text{}^1E_{2u} & \{2,12\} \\
	193.260 & \text{III} & \text{}^1E_{2g} & \{6,6\} \\
\end{array}&&
\begin{array}{cc|cc}
	\text{MSG} & \text{} & \text{Irrep} & X_f \\ \hline
	194.270 & \text{III} & \text{}^2E_{2u} & \{2,12\} \\
	194.270 & \text{III} & \text{}^2E_{2g} & \{6,6\} \\
	194.270 & \text{III} & \text{}^1E_{2u} & \{2,12\} \\
	194.270 & \text{III} & \text{}^1E_{2g} & \{6,6\} \\
	194.270 & \text{III} & \text{}^1E_{1u} & \{6,12\} \\
	194.270 & \text{III} & \text{}^1E_{1g} & \{3\} \\
	194.270 & \text{III} & \text{}^2E_{1u} & \{6,12\} \\
\end{array}&&
\begin{array}{cc|cc}
	\text{MSG} & \text{} & \text{Irrep} & X_f \\ \hline
	201.19 & \text{II} & A_g & \{2\} \\
	201.20 & \text{III} & A & \{2,2,2\} \\
	201.21 & \text{IV} & A_u & \{2,2,2\} \\
	201.21 & \text{IV} & A_g & \{2\} \\
	202.22 & \text{I} & \text{}^2E_u & \{2,2\} \\
	202.22 & \text{I} & \text{}^2E_g & \{\} \\
	202.22 & \text{I} & \text{}^1E_u & \{2,2\} \\
\end{array}
	\end{align*}
\end{table}\clearpage

\begin{table}
\begin{align*}
\begin{array}{cc|cc}
	\text{MSG} & \text{} & \text{Irrep} & X_f \\ \hline
	207.43 & \text{IV} & A_1 & \{2,2\} \\
	208.44 & \text{I} & A_2 & \{\} \\
	208.44 & \text{I} & A_1 & \{2\} \\
	208.45 & \text{II} & A_2 & \{\} \\
	208.45 & \text{II} & A_1 & \{2\} \\
	208.46 & \text{III} & \text{}^2E & \{2,2\} \\
	208.46 & \text{III} & \text{}^1E & \{2,2\} \\
\end{array}&&
\begin{array}{cc|cc}
	\text{MSG} & \text{} & \text{Irrep} & X_f \\ \hline
	215.71 & \text{II} & A_1 & \{2,2\} \\
	215.72 & \text{III} & \text{}^2E & \{2,2\} \\
	215.72 & \text{III} & \text{}^1E & \{2,2\} \\
	215.72 & \text{III} & A & \{2,2\} \\
	215.73 & \text{IV} & A_2 & \{\} \\
	215.73 & \text{IV} & A_1 & \{2,2\} \\
	216.74 & \text{I} & A_2 & \{\} \\
\end{array}&&
\begin{array}{cc|cc}
	\text{MSG} & \text{} & \text{Irrep} & X_f \\ \hline
	221.96 & \text{III} & A_2 & \{2\} \\
	221.96 & \text{III} & A_1 & \{2\} \\
	221.97 & \text{IV} & A_{2u} & \{2,2\} \\
	221.97 & \text{IV} & A_{2g} & \{\} \\
	221.97 & \text{IV} & A_{1u} & \{2\} \\
	221.97 & \text{IV} & A_{1g} & \{\} \\
	222.98 & \text{I} & A_{2u} & \{2,2\} \\
\end{array}
	\end{align*}
\end{table}\clearpage

\begin{table}
\begin{align*}
\begin{array}{cc|cc}
	\text{MSG} & \text{} & \text{Irrep} & X_f \\ \hline
	224.113 & \text{III} & \text{}^1E_g & \{2\} \\
	224.113 & \text{III} & A_u & \{2,2,2\} \\
	224.113 & \text{III} & A_g & \{2\} \\
	224.114 & \text{III} & A_2 & \{2\} \\
	224.114 & \text{III} & A_1 & \{2\} \\
	224.115 & \text{IV} & A_{2u} & \{2,2\} \\
	224.115 & \text{IV} & A_{2g} & \{\} \\
\end{array}&&
\begin{array}{cc|cc}
	\text{MSG} & \text{} & \text{Irrep} & X_f \\ \hline
	227.131 & \text{III} & \text{}^2E_u & \{2,4\} \\
	227.131 & \text{III} & \text{}^2E_g & \{2\} \\
	227.131 & \text{III} & \text{}^1E_u & \{2,4\} \\
	227.131 & \text{III} & \text{}^1E_g & \{2\} \\
	227.131 & \text{III} & A_u & \{2,4\} \\
	227.131 & \text{III} & A_g & \{2\} \\
	227.132 & \text{III} & A_2 & \{\} \\
\end{array}&&
\begin{array}{cccc}
	\text{MSG} & \text{} & \text{Irrep} & X_f \\
	\hline
	230.148 & \text{III} & \text{}^2E_g & \{2\} \\
	230.148 & \text{III} & \text{}^1E_u & \{2,2,2\} \\
	230.148 & \text{III} & \text{}^1E_g & \{2\} \\
	230.148 & \text{III} & A_u & \{2,2,2\} \\
	230.148 & \text{III} & A_g & \{2\} \\
	230.149 & \text{III} & A_2 & \{2\} \\
	230.149 & \text{III} & A_1 & \{2\} \\
	&&&\\
	&&&\\
	&&&\\
	&&&\\
	&&&\\
	&&&\\
	&&&\\
	&&&\\
	&&&\\
	&&&\\
	&&&\\
	&&&\\
	&&&\\
	&&&\\
	&&&\\
	&&&\\
	&&&\\
	&&&\\
	&&&\\
	&&&\\
	&&&\\
	&&&\\
	&&&\\
	&&&\\
	&&&\\
	&&&\\
	&&&\\
	&&&\\
	&&&\\
	&&&\\
	&&&\\
	&&&\\
	&&&\\
	&&&\\
	&&&\\
	&&&\\
	&&&\\
	&&&\\
	&&&\\
	&&&\\
	&&&\\
	&&&\\
	&&&\\
	&&&\\
	&&&\\
	&&&\\
	&&&\\
	&&&\\
	&&&\\
	&&&\\
	&&&\\
	&&&\\
	&&&\\
	&&&\\
	&&&\\
	&&&\\
\end{array}
\end{align*}
\end{table}
\input{spinfulClassD}
\input{spinlessClassC}
\input{spinfulClassC}
\clearpage

Here, $n_{hkl}$ denotes the $n\ (n=2,3,4,6)$-fold rotation symmety along the $(hkl)$-direction, and $m_{hkl}$ represents the mirror symmetry with the mirror plane orthgonal to  the $(hkl)$-direction. In addition, $\bar{1}$ and $\bar{n}_{hkl}$ are the inversion and the $n$-fold rotoinversion (the product of inversion and rotation symmetries) symmetries, respectively. 
\begin{table}[H]
\caption{\label{app:char}Chracter tables of one-dimensional irreducible representations of PGs. }
\begin{align*}
	&
\begin{array}{c|c}
	\hline
		C_1& e \\
		\hline\hline
		A & 1 \\
		\hline
	\end{array}\\
	&\begin{array}{c|cc}
	\hline
		C_i & e & \bar{1} \\
		\hline\hline
		A_g & 1 & 1 \\
		A_u & 1 & -1 \\
		\hline
	\end{array}\\
	&\begin{array}{c|cc}
	\hline
	C'_2 & e & 2_{100} \\
	\hline\hline
	A & 1 & 1 \\
	B & 1 & -1 \\
	\hline
	\end{array}\\
	&\begin{array}{c|cc}
	\hline
	C_2 & e & 2_{010} \\
	\hline\hline
	A & 1 & 1 \\
	B & 1 & -1 \\
	\hline
	\end{array}\\	
	&\begin{array}{c|cc}
	\hline
	C''_2 & e & 2_{001} \\
	\hline\hline
	A & 1 & 1 \\
	B & 1 & -1 \\
	\hline
	\end{array}\\
	&\begin{array}{c|cc}
	\hline
	C'_s & e & m_{100} \\
	\hline\hline
	A' & 1 & 1 \\
	A'' & 1 & -1 \\
	\hline
	\end{array}\\
	&\begin{array}{c|cc}
	\hline
	C_s & e & m_{010} \\
	\hline\hline
	A' & 1 & 1 \\
	A'' & 1 & -1 \\
	\hline
	\end{array}\\
	&\begin{array}{c|cccc}
	\hline
	C'_{2h} & e & 2_{100} &  \bar{1} & m_{100} \\
	\hline\hline
	A_g & 1 & 1 & 1 & 1 \\
	A_u & 1 & 1 & -1 & -1 \\
	B_g & 1 & -1 & 1 & -1 \\
	B_u & 1 & -1 & -1 & 1 \\
	\hline
	\end{array}\\
	&\begin{array}{c|cccc}
	\hline
	C_{2h}& e & 2_{010} & \bar{1} & m_{010} \\
	\hline\hline
	A_g & 1 & 1 & 1 & 1 \\
	A_u & 1 & 1 & -1 & -1 \\
	B_g & 1 & -1 & 1 & -1 \\
	B_u & 1 & -1 & -1 & 1 \\
	\hline
	\end{array}\\
	&\begin{array}{c|cccc}
	\hline
	C''_{2h} & e & 2_{001} &  \bar{1} & m_{001} \\
	\hline\hline
	A_g & 1 & 1 & 1 & 1 \\
	A_u & 1 & 1 & -1 & -1 \\
	B_g & 1 & -1 & 1 & -1 \\
	B_u & 1 & -1 & -1 & 1 \\
	\hline
	\end{array}\\
	\end{align*}
	\end{table}
	\begin{align*}
	&\begin{array}{c|cccc}
	\hline
	D_2 & e & 2_{100} & 2_{010} & 2_{001} \\
	\hline\hline
	A_1 & 1 & 1 & 1 & 1 \\
	B_3 & 1 & 1 & -1 & -1 \\
	B_2 & 1 & -1 & 1 & -1 \\
	B_1 & 1 & -1 & -1 & 1 \\
	\hline
	\end{array}\\
	&\begin{array}{c|cccc}
	\hline
	D'_2 & e & 2_{001} & 2_{1\bar{1}0} & 2_{110} \\
	\hline\hline
	A_1 & 1 & 1 & 1 & 1 \\
	B_1 & 1 & 1 & -1 & -1 \\
	\hline
	\end{array}\\
	&\begin{array}{c|cccc}
	\hline
	C'_{2v} & e & 2_{100} & m_{010} & m_{001} \\
	\hline\hline
	A_1 & 1 & 1 & 1 & 1 \\
	A_2 & 1 & 1 & -1 & -1 \\
	B_2 & 1 & -1 & 1 & -1 \\
	B_1 & 1 & -1 & -1 & 1 \\
	\hline
	\end{array}\\
	&\begin{array}{c|cccc}
	\hline
	C''_{2v} & e& 2_{010} & m_{100} & m_{001} \\
	\hline\hline
	A_1 & 1 & 1 & 1 & 1 \\
	A_2 & 1 & 1 & -1 & -1 \\
	B_1 & 1 & -1 & 1 & -1 \\
	B_2 & 1 & -1 & -1 & 1 \\
	\hline
	\end{array}\\
	&\begin{array}{c|cccc}
	\hline
	C_{2v} & e & 2_{001} & m_{100} & m_{010} \\
	\hline\hline
	A_1 & 1 & 1 & 1 & 1 \\
	A_2 & 1 & 1 & -1 & -1 \\
	B_2 & 1 & -1 & 1 & -1 \\
	B_1 & 1 & -1 & -1 & 1 \\
	\hline
	\end{array}\\
	&\begin{array}{c|cccc}
	\hline
	C'''_{2v} & e & 2_{001} & m_{1-10} & m_{110} \\
	\hline\hline
	A_1 & 1 & 1 & 1 & 1 \\
	A_2 & 1 & 1 & -1 & -1 \\
	\hline
	\end{array}\\
	&\begin{array}{c|cccccccc}
	\hline
	D_{2h} & e & 2_{100} & 2_{010} & 2_{001} & \bar{1} & m_{100} & m_{010} & m_{001} \\
	\hline\hline
	A_g & 1 & 1 & 1 & 1 & 1 & 1 & 1 & 1 \\
	A_u & 1 & 1 & 1 & 1 & -1 & -1 & -1 & -1 \\
	B_{3g} & 1 & 1 & -1 & -1 & 1 & 1 & -1 & -1 \\
	B_{3u} & 1 & 1 & -1 & -1 & -1 & -1 & 1 & 1 \\
	B_{2g} & 1 & -1 & 1 & -1 & 1 & -1 & 1 & -1 \\
	B_{2u} & 1 & -1 & 1 & -1 & -1 & 1 & -1 & 1 \\
	B_{1g} & 1 & -1 & -1 & 1 & 1 & -1 & -1 & 1 \\
	B_{1u} & 1 & -1 & -1 & 1 & -1 & 1 & 1 & -1 \\
	\hline
	\end{array}\\
	&\begin{array}{c|cccccccc}
	\hline
	D'_{2h} & e & 2_{001} & 2_{1-10} & 2_{110} & \bar{1}& m_{001} & m_{1-10} & m_{110} \\
	\hline\hline
	A_g & 1 & 1 & 1 & 1 & 1 & 1 & 1 & 1 \\
	A_u & 1 & 1 & 1 & 1 & -1 & -1 & -1 & -1 \\
	B_{1g} & 1 & 1 & -1 & -1 & 1 & 1 & -1 & -1 \\
	B_{1u} & 1 & 1 & -1 & -1 & -1 & -1 & 1 & 1 \\
	\hline
	\end{array}\\
	&\begin{array}{c|cccc}
	\hline
	C_4 & e & 2_{001} & 4^{+}_{001} & 4^{-}_{001} \\
	\hline\hline
	A & 1 & 1 & 1 & 1 \\
	B & 1 & 1 & -1 & -1 \\
	^2E & 1 & -1 &  i & -i \\
	^1E & 1 & -1 & -i &  i \\
	\hline
	\end{array}\\
	&\begin{array}{c|cccc}
	\hline
	S_4 & e & 2_{001} & \bar{4}^{+}_{001} & \bar{4}^{-}_{001} \\
	\hline\hline
	A & 1 & 1 & 1 & 1 \\
	B & 1 & 1 & -1 & -1 \\
	^2E & 1 & -1 & i & -i \\
	^1E & 1 & -1 & -i & +i \\
	\hline
	\end{array}
\end{align*}
\begin{align*}
	&\begin{array}{c|cccccccc}
	\hline
	C_{4h} & e & 2_{001} & 4^{+}_{001} & 4^{-}_{001} & \bar{1} & m_{001} & \bar{4}^{+}_{001} & \bar{4}^{-}_{001} \\
	\hline\hline
	A_g & 1 & 1 & 1 & 1 & 1 & 1 & 1 & 1 \\
	A_u & 1 & 1 & 1 & 1 & -1 & -1 & -1 & -1 \\
	B_g & 1 & 1 & -1 & -1 & 1 & 1 & -1 & -1 \\
	B_u & 1 & 1 & -1 & -1 & -1 & -1 & 1 & 1 \\
	^2E_g & 1 & -1 & i & -i & 1 & -1 & i & -i \\
	^2E_u & 1 & -1 & i & -i & -1 & 1 & -i & i \\
	^1E_g & 1 & -1 & -i & i & 1 & -1 & -i & i \\
	^1E_u & 1 & -1 & -i & i & -1 & 1 & i & -i \\
	\hline
	\end{array}\\
	&\begin{array}{c|cccccccc}
	\hline
	D_2 & e & 2_{100} & 2_{010} & 2_{001} & 2_{1\bar{1}0} & 4^{+}_{001} & 4^{-}_{001} & 2_{110} \\
	\hline\hline
	A_1 & 1 & 1 & 1 & 1 & 1 & 1 & 1 & 1 \\
	B_1 & 1 & 1 & 1 & 1 & -1 & -1 & -1 & -1 \\
	B_2 & 1 & -1 & -1 & 1 & 1 & -1 & -1 & 1 \\
	A_2 & 1 & -1 & -1 & 1 & -1 & 1 & 1 & -1 \\
	\hline
	\end{array}\\
	&\begin{array}{c|cccccccc}
	\hline
	C_{4v} & e & 2_{001} & 4^{+}_{001} & 4^{-}_{001} & m_{100} & m_{010} & m_{1\bar{1}0} & m_{110} \\
	\hline\hline
	A_1 & 1 & 1 & 1 & 1 & 1 & 1 & 1 & 1 \\
	A_2 & 1 & 1 & 1 & 1 & -1 & -1 & -1 & -1 \\
	B_1 & 1 & 1 & -1 & -1 & 1 & 1 & -1 & -1 \\
	B_2 & 1 & 1 & -1 & -1 & -1 & -1 & 1 & 1 \\
	\hline
	\end{array}\\
	&\begin{array}{c|cccccccc}
	\hline
	D_{2d}^{1} & e & 2_{100} & 2_{010} & 2_{001} & m_{1\bar{1}0} &\bar{4}^{+}_{001}  & \bar{4}^{-}_{001}  & m_{110} \\
	\hline\hline
	A_1 & 1 & 1 & 1 & 1 & 1 & 1 & 1 & 1 \\
	B_1 & 1 & 1 & 1 & 1 & -1 & -1 & -1 & -1 \\
	B_2 & 1 & -1 & -1 & 1 & 1 & -1 & -1 & 1 \\
	A_2 & 1 & -1 & -1 & 1 & -1 & 1 & 1 & -1 \\
	\hline
	\end{array}\\
	&\begin{array}{c|cccccccc}
	\hline
	D_{2d}^{5}& e & 2_{001} & 2_{1\bar{1}0} & 2_{110} & m_{100} & m_{010} & \bar{4}^{+}_{001} & \bar{4}^{-}_{001}\\
	\hline\hline
	A_1 & 1 & 1 & 1 & 1 & 1 & 1 & 1 & 1 \\
	B_1 & 1 & 1 & 1 & 1 & -1 & -1 & -1 & -1 \\
	B_2 & 1 & 1 & -1 & -1 & 1 & 1 & -1 & -1 \\
	A_2 & 1 & 1 & -1 & -1 & -1 & -1 & 1 & 1 \\
	\hline
	\end{array}\\
	&\begin{array}{c|cccccccccccccccc}
	\hline
	D_{4h}& e & 2_{100} & 2_{010} & 2_{001} & 2_{1\bar{1}0} & 4^{+}_{001} & 4^{-}_{001}& 2_{110} & \bar{1} &
	m_{100} & m_{010} & m_{001} & m_{1\bar{1}0} & \bar{4}^{+}_{001}& \bar{4}^{-}_{001} & m_{110} \\
	\hline\hline
	A_{1g} & 1 & 1 & 1 & 1 & 1 & 1 & 1 & 1 & 1 & 1 & 1 & 1 & 1 & 1 & 1 & 1 \\
	A_{1u} & 1 & 1 & 1 & 1 & 1 & 1 & 1 & 1 & -1 & -1 & -1 & -1 & -1 & -1 & -1 & -1
	\\
	B_{1g} & 1 & 1 & 1 & 1 & -1 & -1 & -1 & -1 & 1 & 1 & 1 & 1 & -1 & -1 & -1 & -1
	\\
	B_{1u} & 1 & 1 & 1 & 1 & -1 & -1 & -1 & -1 & -1 & -1 & -1 & -1 & 1 & 1 & 1 & 1
	\\
	B_{2g} & 1 & -1 & -1 & 1 & 1 & -1 & -1 & 1 & 1 & -1 & -1 & 1 & 1 & -1 & -1 & 1
	\\
	B_{2u} & 1 & -1 & -1 & 1 & 1 & -1 & -1 & 1 & -1 & 1 & 1 & -1 & -1 & 1 & 1 & -1
	\\
	A_{2g} & 1 & -1 & -1 & 1 & -1 & 1 & 1 & -1 & 1 & -1 & -1 & 1 & -1 & 1 & 1 & -1
	\\
	A_{2u} & 1 & -1 & -1 & 1 & -1 & 1 & 1 & -1 & -1 & 1 & 1 & -1 & 1 & -1 & -1 & 1
	\\
	\hline
	\end{array}\\
	&\begin{array}{c|ccc}
	\hline
	C_3 & e & 3^{+}_{001} & 3^{-}_{001} \\
	\hline\hline
	 A & 1 & 1 & 1 \\
	^2E & 1 & \omega & \omega^{*}\\
	^1E & 1 & \omega^{*} & \omega\\
	\hline
	\end{array}\\
	&\left(\omega=e^{i\frac{2\pi}{3}}\right)\\
	&\begin{array}{c|cccccc}
	\hline
	C_{3i}& e & 3^{+}_{001} & 3^{-}_{001} & \bar{1} & \bar{3}^{+}_{001} & \bar{3}^{-}_{001} \\
	\hline\hline
	A_g & 1 & 1 & 1 & 1 & 1 & 1 \\
	A_u & 1 & 1 & 1 & -1 & -1 & -1 \\
	^2E_g & 1 & \omega & \omega^{*} & 1 & \omega & \omega^{*} \\
	^2E_u & 1 & \omega & \omega^{*} & -1 & -\omega & -\omega^{*}  \\
	^1E_g & 1 & \omega^{*} & \omega & 1 & \omega^{*} & \omega \\
	^1E_u & 1 & \omega^{*} & \omega & -1 & -\omega^{*} & -\omega \\
	\hline
	\end{array}\\
	&\begin{array}{c|cccccc}
	\hline
	D_{3}^{1}& e & 3^{+}_{001} & 3^{-}_{001} & 2_{210} & 2_{120} & 2_{1\bar{1}0} \\
	\hline\hline
	A_1 & 1 & 1 & 1 & 1 & 1 & 1 \\
	A_2 & 1 & 1 & 1 & -1 & -1 & -1 \\
	\hline
	\end{array}\\
	&\begin{array}{c|cccccc}
	\hline
	D_{3}^{2} & e & 3^{+}_{001} & 3^{-}_{001} & 2_{100} & 2_{110} & 2_{010} \\
	\hline\hline
	A_1 & 1 & 1 & 1 & 1 & 1 & 1 \\
	A_2 & 1 & 1 & 1 & -1 & -1 & -1 \\
	\hline
\end{array}
\end{align*}
\begin{align*}
	&\begin{array}{c|cccccc}
	\hline
	C_{3v}^{1} & e & 3^{+}_{001} & 3^{-}_{001} & m_{100} & m_{110} & m_{010} \\
	\hline\hline
	A_1 & 1 & 1 & 1 & 1 & 1 & 1 \\
	A_2 & 1 & 1 & 1 & -1 & -1 & -1 \\
	\hline
	\end{array}\\
	&\begin{array}{c|cccccc}
	\hline
	C_{3v}^{2}& e & 3^{+}_{001} & 3^{-}_{001} & m_{210} & m_{120} & m_{1\bar{1}0} \\
	\hline\hline
	A_1 & 1 & 1 & 1 & 1 & 1 & 1 \\
	A_2 & 1 & 1 & 1 & -1 & -1 & -1 \\
	\hline
	\end{array}\\
	&\begin{array}{c|cccccccccccc}
	\hline
	 D_{3d}^{1}& e & 3^{+}_{001} & 3^{-}_{001} & 2_{100} & 2_{110} & 2_{010} & \bar{1} & \bar{3}^{+}_{001} & \bar{3}^{-}_{001} & m_{100} & m_{110} & m_{010} \\
	\hline\hline
	A_{1g} & 1 & 1 & 1 & 1 & 1 & 1 & 1 & 1 & 1 & 1 & 1 & 1 \\
	A_{1u} & 1 & 1 & 1 & 1 & 1 & 1 & -1 & -1 & -1 & -1 & -1 & -1 \\
	A_{2g} & 1 & 1 & 1 & -1 & -1 & -1 & 1 & 1 & 1 & -1 & -1 & -1 \\
	A_{2u} & 1 & 1 & 1 & -1 & -1 & -1 & -1 & -1 & -1 & 1 & 1 & 1 \\
	\hline
	\end{array}\\
	&\begin{array}{c|cccccccccccc}
	\hline
	D_{3}^{3} & e & 3^{+}_{001} & 3^{-}_{001} & 2_{210} & 2_{120} & 2_{1\bar{1}0} &\bar{1} & \bar{3}^{+}_{001} & \bar{3}^{-}_{001} & m_{210} & m_{120} & m_{1\bar{1}0} \\
	\hline\hline
	A_{1g} & 1 & 1 & 1 & 1 & 1 & 1 & 1 & 1 & 1 & 1 & 1 & 1 \\
	A_{1u} & 1 & 1 & 1 & 1 & 1 & 1 & -1 & -1 & -1 & -1 & -1 & -1 \\
	A_{2g} & 1 & 1 & 1 & -1 & -1 & -1 & 1 & 1 & 1 & -1 & -1 & -1 \\
	A_{2u} & 1 & 1 & 1 & -1 & -1 & -1 & -1 & -1 & -1 & 1 & 1 & 1 \\
	\hline
	\end{array}\\
	&\begin{array}{c|cccccc}
	\hline
	C_6 & e & 6^{+}_{001} & 3^{+}_{001} & 2_{001} & 3^{-}_{001} & 6^{-}_{001}\\
	\hline\hline 
	A & 1 & 1 & 1 & 1 & 1 & 1 \\
	^2E_1 & 1 & \eta& \eta^{2} & -1 & -\eta & -\eta^{2}\\
	^1E_1 & 1 & -\eta^{2}& -\eta & -1 & \eta^{2} & \eta\\
	^1E_2 & 1 & \eta^{2} & -\eta & 1 & \eta^{2} & -\eta \\
	^2E_2 & 1 & -\eta & \eta^{2} & 1 & -\eta & \eta^{2} \\
	B & 1 & -1 & 1 & -1 & 1 & -1 \\
	\hline
	\end{array}\\
	&\left(\eta=e^{i\frac{2\pi}{6}}\right)\\
	&\begin{array}{c|cccccc}
	\hline
	C_{3h}& e & 3^{+}_{001} & 3^{-}_{001} & \bar{6}^{+}_{001} & m_{001} & \bar{6}^{-}_{001} \\
	\hline\hline
	A' & 1 & 1 & 1 & 1 & 1 & 1 \\
	A'' & 1 & 1 & 1 & -1 & -1 & -1 \\
	^2E'' & 1 & \eta^{2} & -\eta & \eta & -1 & -\eta^{2}\\
	^2E' & 1 & \eta^{2} & -\eta & -\eta & 1 & \eta^{2} \\
	^1E'' & 1 & -\eta & \eta^{2} & -\eta^{2} & -1 & \eta
	i \\
	^2E' & 1 & -\eta & \eta^{2} & \eta^{2} & 1 & -\eta \\
	\hline
	\end{array}\\
	&\begin{array}{c|cccccccccccc}
	\hline
	C_{6h}& e & 6^{+}_{001} & 3^{+}_{001} & 2_{001} & 3^{-}_{001} & 6^{-}_{001} & \bar{1} & \bar{6}^{+}_{001} &
	\bar{3}^{+}_{001} & m_{001} & \bar{3}^{-}_{001} & \bar{6}^{-}_{001} \\
	\hline\hline
	A_g & 1 & 1 & 1 & 1 & 1 & 1 & 1 & 1 & 1 & 1 & 1 & 1 \\
	A_u & 1 & 1 & 1 & 1 & 1 & 1 & -1 & -1 & -1 & -1 & -1 & -1 \\
	^2E_{1g} & 1 & \eta& \eta^{2} & -1 & -\eta & -\eta^{2}&
	1 & \eta& \eta^{2} & -1 & -\eta & -\eta^{2}\\
	^2E_{1u} & 1 & \eta& \eta^{2} & -1 & -\eta & -\eta^{2}&
	-1 & -\eta & -\eta^{2}& 1 & \eta& \eta^{2} \\
	^1E_{1g} & 1 & -\eta^{2}& -\eta & -1 & \eta^{2} & \eta&
	1 & -\eta^{2}& -\eta & -1 & \eta^{2} & \eta\\
	^1E_{1u} & 1 & -\eta^{2}& -\eta & -1 & \eta^{2} & \eta&
	-1 & \eta^{2} & \eta& 1 & -\eta^{2}& -\eta \\
	^1E_{2g} & 1 & \eta^{2} & -\eta & 1 & \eta^{2} & -\eta & 1 &
	\eta^{2} & -\eta & 1 & \eta^{2} & -\eta \\
	^1E_{2u} & 1 & \eta^{2} & -\eta & 1 & \eta^{2} & -\eta & -1 &
	-\eta^{2}& \eta& -1 & -\eta^{2}& \eta\\
	^2E_{2g} & 1 & -\eta & \eta^{2} & 1 & -\eta & \eta^{2} & 1 &
	-\eta & \eta^{2} & 1 & -\eta & \eta^{2} \\
	^2E_{2u} & 1 & -\eta & \eta^{2} & 1 & -\eta & \eta^{2} & -1 &
	\eta& -\eta^{2}& -1 & \eta& -\eta^{2}\\
	B_g & 1 & -1 & 1 & -1 & 1 & -1 & 1 & -1 & 1 & -1 & 1 & -1 \\
	B_u & 1 & -1 & 1 & -1 & 1 & -1 & -1 & 1 & -1 & 1 & -1 & 1 \\
	\hline
	\end{array}\\
	&\begin{array}{c|cccccccccccc}
	\hline
	D_6& e & 6^{+}_{001} & 3^{+}_{001} & 2_{001} & 3^{-}_{001} & 6^{-}_{001} & 2_{100} & 2_{210} &
	2_{110} & 2_{120} & 2_{010} & 2_{1\bar{1}0} \\
	\hline\hline
	A_1 & 1 & 1 & 1 & 1 & 1 & 1 & 1 & 1 & 1 & 1 & 1 & 1 \\
	A_2 & 1 & 1 & 1 & 1 & 1 & 1 & -1 & -1 & -1 & -1 & -1 & -1 \\
	B_2 & 1 & -1 & 1 & -1 & 1 & -1 & 1 & -1 & 1 & -1 & 1 & -1 \\
	B_1 & 1 & -1 & 1 & -1 & 1 & -1 & -1 & 1 & -1 & 1 & -1 & 1 \\
	\hline
	\end{array}\\
	&\begin{array}{c|cccccccccccc}
	\hline
	C_{6v} & e & 6^{+}_{001} & 3^{+}_{001} & 2_{001} & 3^{-}_{001} & 6^{-}_{001} & m_{100} & m_{210} &
	m_{110} & m_{120} & m_{010} & m_{1\bar{1}0} \\
	\hline\hline
	A_1 & 1 & 1 & 1 & 1 & 1 & 1 & 1 & 1 & 1 & 1 & 1 & 1 \\
	A_2 & 1 & 1 & 1 & 1 & 1 & 1 & -1 & -1 & -1 & -1 & -1 & -1 \\
	B_1 & 1 & -1 & 1 & -1 & 1 & -1 & 1 & -1 & 1 & -1 & 1 & -1 \\
	B_2 & 1 & -1 & 1 & -1 & 1 & -1 & -1 & 1 & -1 & 1 & -1 & 1 \\
	\hline
	\end{array}
\end{align*}
\fontsize{7pt}{7pt}\selectfont
\begin{align*}
	&\begin{array}{c|cccccccccccc}
	\hline
	D_{3h}^{1} & e & 3^{+}_{001} & 3^{-}_{001} & 2_{210} & 2_{120} & 2_{1\bar{1}0} & \bar{6}^{+}_{001} & m_{001}
	& \bar{6}^{-}_{001} & m_{100} & m_{110} & m_{010} \\
	\hline\hline
	A'_1 & 1 & 1 & 1 & 1 & 1 & 1 & 1 & 1 & 1 & 1 & 1 & 1 \\
	A''_1 & 1 & 1 & 1 & 1 & 1 & 1 & -1 & -1 & -1 & -1 & -1 & -1 \\
	A'_2 & 1 & 1 & 1 & -1 & -1 & -1 & 1 & 1 & 1 & -1 & -1 & -1 \\
	A''_2 & 1 & 1 & 1 & -1 & -1 & -1 & -1 & -1 & -1 & 1 & 1 & 1 \\
	\hline
	\end{array}\\
	&\begin{array}{c|cccccccccccc}
	\hline
	D_{3h}^{3} & e & 3^{+}_{001} & 3^{-}_{001}& 2_{100} & 2_{110} & 2_{010} & \bar{6}^{+}_{001}  & m_{001}
	& \bar{6}^{-}_{001}  & m_{210} & m_{120} & m_{1\bar{1}0} \\
	\hline\hline
	A'_1 & 1 & 1 & 1 & 1 & 1 & 1 & 1 & 1 & 1 & 1 & 1 & 1 \\
	A''_1 & 1 & 1 & 1 & 1 & 1 & 1 & -1 & -1 & -1 & -1 & -1 & -1 \\
	A'_2 & 1 & 1 & 1 & -1 & -1 & -1 & 1 & 1 & 1 & -1 & -1 & -1 \\
	A''_2 & 1 & 1 & 1 & -1 & -1 & -1 & -1 & -1 & -1 & 1 & 1 & 1 \\
	\hline
	\end{array}\\
	&\begin{array}{c|cccccccccccccccccccccccc}
	\hline
	D_{6h} & e & 6^{+}_{001} & 3^{+}_{001} & 2_{001} & 3^{-}_{001} & 6^{-}_{001} & 2_{100} &
	2_{210} & 2_{110} & 2_{120} & 2_{010} & 2_{1\bar{1}0} & \bar{1} & \bar{6}^{+}_{001} & \bar{3}^{+}_{001} &
	m_{001} & \bar{3}^{-}_{001} & \bar{6}^{-}_{001} & m_{100} & m_{210} & m_{110} & m_{120} & m_{010}
	& m_{1-10} \\
	\hline\hline
	A_{1g} & 1 & 1 & 1 & 1 & 1 & 1 & 1 & 1 & 1 & 1 & 1 & 1 & 1 & 1 & 1 & 1 &
	1 & 1 & 1 & 1 & 1 & 1 & 1 & 1 \\
	A_{1u} & 1 & 1 & 1 & 1 & 1 & 1 & 1 & 1 & 1 & 1 & 1 & 1 & -1 & -1 & -1 &
	-1 & -1 & -1 & -1 & -1 & -1 & -1 & -1 & -1 \\
	A_{2g} & 1 & 1 & 1 & 1 & 1 & 1 & -1 & -1 & -1 & -1 & -1 & -1 & 1 & 1 & 1 &
	1 & 1 & 1 & -1 & -1 & -1 & -1 & -1 & -1 \\
	A_{2u} & 1 & 1 & 1 & 1 & 1 & 1 & -1 & -1 & -1 & -1 & -1 & -1 & -1 & -1 &
	-1 & -1 & -1 & -1 & 1 & 1 & 1 & 1 & 1 & 1 \\
	B_{2g} & 1 & -1 & 1 & -1 & 1 & -1 & 1 & -1 & 1 & -1 & 1 & -1 & 1 & -1 & 1
	& -1 & 1 & -1 & 1 & -1 & 1 & -1 & 1 & -1 \\
	B_{2u} & 1 & -1 & 1 & -1 & 1 & -1 & 1 & -1 & 1 & -1 & 1 & -1 & -1 & 1 & -1
	& 1 & -1 & 1 & -1 & 1 & -1 & 1 & -1 & 1 \\
	B_{1g} & 1 & -1 & 1 & -1 & 1 & -1 & -1 & 1 & -1 & 1 & -1 & 1 & 1 & -1 & 1
	& -1 & 1 & -1 & -1 & 1 & -1 & 1 & -1 & 1 \\
	B_{1u} & 1 & -1 & 1 & -1 & 1 & -1 & -1 & 1 & -1 & 1 & -1 & 1 & -1 & 1 & -1
	& 1 & -1 & 1 & 1 & -1 & 1 & -1 & 1 & -1 \\
	\hline
	\end{array}\\
	&\begin{array}{c|cccccccccccc}
	\hline
	T & e & 2_{100} & 2_{010} & 2_{001} & 3^{-}_{111} & 3^{-}_{\bar{1} \bar{1} 1} & 3^{-}_{1\bar{1} \bar{1} } & 3^{-}_{\bar{1} 1\bar{1} } & 3^{+}_{111} & 3^{+}_{\bar{1} 1\bar{1} } & 3^{+}_{\bar{1}\bar{1} 1} & 3^{+}_{1\bar{1} \bar{1} } \\
	\hline\hline
	A & 1 & 1 & 1 & 1 & 1 & 1 & 1 & 1 & 1 & 1 & 1 & 1 \\
	^2E & 1 & 1 & 1 & 1 &\omega & \omega& \omega& \omega& \omega^{*}& \omega^{*}& \omega^{*}&
	\omega^{*}\\
	^1E & 1 & 1 & 1 & 1 & \omega^{*}& \omega^{*}& \omega^{*}& \omega^{*}& \omega& \omega& \omega&
	\omega\\
	\hline
	\end{array}\\
	&\begin{array}{c|cccccccccccccccccccccccc}
	\hline
	T_h& e & 2_{100} & 2_{010} & 2_{001} & 3^{-}_{111} & 3^{-}_{\bar{1} \bar{1} 1} & 3^{-}_{1\bar{1}\bar{1} } & 3^{-}_{\bar{1} 1\bar{1} } & 3^{+}_{111} & 3^{+}_{\bar{1} 1\bar{1} } & 3^{+}_{\bar{1} \bar{1} 1} & 3^{+}_{1\bar{1}\bar{1} } & \bar{1}  &
	m_{100} & m_{010} & m_{001} & \bar{3}^{-}_{111} & \bar{3}^{-}_{\bar{1} \bar{1} 1} & \bar{3}^{-}_{1\bar{1} \bar{1} } & \bar{3}^{-}_{\bar{1} 1\bar{1} } &\bar{3}^{+}_{111} & \bar{3}^{+}_{\bar{1} 1\bar{1} } & \bar{3}^{+}_{\bar{1} \bar{1} 1} & \bar{3}^{+}_{1\bar{1} \bar{1} } \\
	\hline\hline
	 A_g & 1 & 1 & 1 & 1 & 1 & 1 & 1 & 1 & 1 & 1 & 1 & 1 & 1 & 1 & 1 & 1 & 1 & 1 & 1 & 1 & 1 & 1 & 1 & 1 \\
	A_u & 1 & 1 & 1 & 1 & 1 & 1 & 1 & 1 & 1 & 1 & 1 & 1 & -1 & -1 & -1 & -1 & -1 & -1 & -1 & -1 & -1 & -1 & -1 & -1 \\
	^2E_g & 1 & 1 & 1 & 1 & \omega & \omega & \omega & \omega & \omega^{*} & \omega^{*} & \omega^{*} & \omega^{*} & 1 & 1 & 1 & 1 & \omega & \omega & \omega & \omega & \omega^{*} & \omega^{*} & \omega^{*} & \omega^{*} \\
	^2E_u & 1 & 1 & 1 & 1 & \omega & \omega & \omega & \omega & \omega^{*} & \omega^{*} & \omega^{*} & \omega^{*} & -1 & -1 & -1 & -1 & -\omega & -\omega & -\omega & -\omega & -\omega^{*} & -\omega^{*} & -\omega^{*} & -\omega^{*} \\
	^1E_g & 1 & 1 & 1 & 1 & \omega^{*} & \omega^{*} & \omega^{*} & \omega^{*} & \omega & \omega & \omega & \omega & 1 & 1 & 1 & 1 & \omega^{*} & \omega^{*} & \omega^{*} & \omega^{*} & \omega & \omega & \omega & \omega \\
	^1E_u & 1 & 1 & 1 & 1 & \omega^{*} & \omega^{*} & \omega^{*} & \omega^{*} & \omega & \omega & \omega & \omega & -1 & -1 & -1 & -1 & -\omega^{*} & -\omega^{*} & -\omega^{*} & -\omega^{*} & -\omega & -\omega & -\omega & -\omega \\
	\hline
	\end{array}\\
	&\begin{array}{c|cccccccccccccccccccccccc}
	\hline
	O & e & 2_{100} & 2_{010} & 2_{001} & 3^{-}_{111} & 3^{-}_{\bar{1}\bar{1}1} & 3^{-}_{1\bar{1}\bar{1}} & 3^{-}_{\bar{1}1\bar{1}} & 3^{+}_{111} & 3^{+}_{\bar{1}1\bar{1}} & 3^{+}_{\bar{1}\bar{1}1} & 3^{+}_{1\bar{1}\bar{1}} &
	2_{1\bar{1}0} & 4^{+}_{001} & 4^{-}_{001} & 2_{110} & 2_{01\bar{1}} & 2_{011} & 4^{+}_{100} & 4^{-}_{100} & 2_{-101} & 4^{-}_{010} & 2_{101} & 4^{+}_{010} \\
	\hline\hline
	A_1 & 1 & 1 & 1 & 1 & 1 & 1 & 1 & 1 & 1 & 1 & 1 & 1 & 1 & 1 & 1 & 1 & 1 & 1 & 1 & 1 & 1 & 1 & 1 & 1 \\
	A_2 & 1 & 1 & 1 & 1 & 1 & 1 & 1 & 1 & 1 & 1 & 1 & 1 & -1 & -1 & -1 & -1 & -1 & -1 & -1 & -1 & -1 & -1 & -1 & -1 \\
	\hline
	\end{array}\\
	&\begin{array}{c|cccccccccccccccccccccccc}
	\hline
	T_d& e & 2_{100} & 2_{010} & 2_{001} &  3^{-}_{111} &  3^{-}_{\bar{1}\bar{1}1} &  3^{-}_{1\bar{1}\bar{1}} &  3^{-}_{\bar{1}1\bar{1}} &  3^{+}_{111} & 3^{+}_{\bar{1}1\bar{1}} & 3^{+}_{\bar{1}\bar{1}1} & 3^{+}_{1\bar{1}\bar{1}} &
	m_{1\bar{1}0} & \bar{4}^{+}_{001} & \bar{4}^{-}_{001} & m_{110} & m_{01\bar{1}} & m_{011} & \bar{4}^{+}_{100} & \bar{4}^{-}_{100} & m_{\bar{1}01} & \bar{4}^{-}_{010} & m_{101} & \bar{4}^{+}_{010} \\
	\hline\hline
	A_1 & 1 & 1 & 1 & 1 & 1 & 1 & 1 & 1 & 1 & 1 & 1 & 1 & 1 & 1 & 1 & 1 & 1 & 1 & 1 & 1 & 1 & 1 & 1 & 1 \\
	A_2 & 1 & 1 & 1 & 1 & 1 & 1 & 1 & 1 & 1 & 1 & 1 & 1 & -1 & -1 & -1 & -1 & -1 & -1 & -1 & -1 & -1 & -1 & -1 & -1 \\
	\hline
	\end{array}\\
	&\begin{array}{c|cccccccccccccccccccccccc}
	\hline
	O_h &  e & 2_{100} & 2_{010} & 2_{001} & 3^{-}_{111} & 3^{-}_{\bar{1}\bar{1}1} & 3^{-}_{1\bar{1}\bar{1}} & 3^{-}_{\bar{1}1\bar{1}} & 3^{+}_{111} & 3^{+}_{\bar{1}1\bar{1}} & 3^{+}_{\bar{1}\bar{1}1} & 3^{+}_{1\bar{1}\bar{1}} &
	2_{1\bar{1}0} & 4^{+}_{001} & 4^{-}_{001} & 2_{110} & 2_{01\bar{1}} & 2_{011} & 4^{+}_{100} & 4^{-}_{100} & 2_{-101} & 4^{-}_{010} & 2_{101} & 4^{+}_{010} \\
	&\bar{1} & m_{100} & m_{010} & m_{001} &  \bar{3}^{-}_{111} & \bar{3}^{-}_{\bar{1}\bar{1}1} & \bar{3}^{-}_{1\bar{1}\bar{1}} & \bar{3}^{-}_{\bar{1}1\bar{1}} &  \bar{3}^{+}_{111} &  \bar{3}^{+}_{\bar{1}1\bar{1}} & \bar{3}^{+}_{\bar{1}\bar{1}1} & \bar{3}^{+}_{1\bar{1}\bar{1}} &
	m_{1\bar{1}0} & \bar{4}^{+}_{001} & \bar{4}^{-}_{001} & m_{110} & m_{01\bar{1}} & m_{011} & \bar{4}^{+}_{100} & \bar{4}^{-}_{100} & m_{\bar{1}01} & \bar{4}^{-}_{010} & m_{101} & \bar{4}^{+}_{010} \\
	\hline\hline
	A_{1g} & 1 & 1 & 1 & 1 & 1 & 1 & 1 & 1 & 1 & 1 & 1 & 1 & 1 & 1 & 1 & 1 & 1 & 1 & 1 & 1 & 1 & 1 & 1 & 1 \\
	&1 & 1 & 1 & 1 & 1 & 1 & 1 & 1 & 1 & 1 & 1 & 1 & 1 & 1 & 1 & 1 & 1 & 1 & 1 & 1 & 1 & 1 & 1 & 1 \\
	A_{1u} & 1 & 1 & 1 & 1 & 1 & 1 & 1 & 1 & 1 & 1 & 1 & 1 & 1 & 1 & 1 & 1 & 1 & 1 & 1 & 1 & 1 & 1 & 1 & 1 \\
	&-1 & -1 & -1 & -1 & -1 & -1 & -1 & -1 & -1 & -1 & -1 & -1 & -1 & -1 & -1 & -1 & -1 & -1 & -1 & -1 & -1 & -1 & -1 & -1 \\
	A_{2g} & 1 & 1 & 1 & 1 & 1 & 1 & 1 & 1 & 1 & 1 & 1 & 1 & -1 & -1 & -1 & -1 & -1 & -1 & -1 & -1 & -1 & -1 & -1 & -1 \\
	&1 & 1 & 1 & 1 & 1 & 1 & 1 & 1 & 1 & 1 & 1 & 1 & -1 & -1 & -1 & -1 & -1 & -1 & -1 & -1 & -1 & -1 & -1 & -1 \\
\hline
	\end{array}
\end{align*}
\clearpage
\begin{table}[H]
	\caption{\label{app:MSG-PG}Correspondences between MSGs and PGs. Group elements of PGs are listed in Table~\ref{app:char}.}
	\begin{align*}
		\begin{array}{cc||cc||cc||cc||cc||cc||cc||cc}
		\text{MSG} & \text{PG} & \text{MSG} & \text{PG} & \text{MSG} & \text{PG} & \text{MSG} & \text{PG} & \text{MSG} & \text{PG} & \text{MSG} & \text{PG} & \text{MSG} & \text{PG} & \text{MSG} & \text{PG}\\
		\hline
		1.1 & C_1 & 11.57 & C_{2h} & 19.29 & D_2 & 28.92 & C_{2v} & 33.155 & C_{2v} & 41.218 & C_{2v} & 50.281 & C''_{2h} & 54.344 & C_{2h} \\
		1.2 & C_1 & 12.58 & C_{2h} & 19.30 & D_2 & 28.93 & C_{2v} & 34.156 & C_{2v} & 42.219 & C_{2v} & 50.282 & C_{2h} & 54.345 & D_2 \\
		1.3 & C_1 & 12.59 & C_{2h} & 20.31 & D_2 & 28.94 & C_{2v} & 34.157 & C_{2v} & 42.220 & C_{2v} & 50.283 & D_2 & 54.346 & D_{2h} \\
		2.4 & C_i & 12.60 & C_s & 20.32 & D_2 & 28.95 & C_{2v} & 34.158 & C_s & 42.221 & C_s & 50.284 & D_{2h} & 54.347 & D_{2h} \\
		2.5 & C_i & 12.61 & C_2 & 20.33 & C''_2 & 28.96 & C_{2v} & 34.159 & C''_2 & 42.222 & C''_2 & 50.285 & D_{2h} & 54.348 & D_{2h} \\
		2.6 & C_1 & 12.62 & C_i & 20.34 & C'_2 & 28.97 & C_{2v} & 34.160 & C_{2v} & 42.223 & C_{2v} & 50.286 & D_{2h} & 54.349 & D_{2h} \\
		2.7 & C_i & 12.63 & C_{2h} & 20.35 & D_2 & 28.98 & C_{2v} & 34.161 & C_{2v} & 43.224 & C_{2v} & 50.287 & D_{2h} & 54.350 & D_{2h} \\
\end{array}
	\end{align*}
\end{table}
\begin{table}[H]
	\begin{align*}
		\begin{array}{cc||cc||cc||cc||cc||cc||cc||cc}
		\text{MSG} & \text{PG} & \text{MSG} & \text{PG} & \text{MSG} & \text{PG} & \text{MSG} & \text{PG} & \text{MSG} & \text{PG} & \text{MSG} & \text{PG} & \text{MSG} & \text{PG} & \text{MSG} & \text{PG}\\
		\hline
		59.407 & C'_{2v} & 64.470 & D_{2h} & 71.533 & D_{2h} & 81.34 & S_4 & 90.97 & D_2 & 98.160 & C_4 & 106.223 & C_4 & 115.286 & C_{2v} \\
		59.408 & C_{2v} & 64.471 & C'_{2v} & 71.534 & D_{2h} & 81.35 & C''_2 & 90.98 & C_4 & 98.161 & D'_2 & 106.224 & C_{4v} & 115.287 & S_4 \\
		59.409 & C''_{2h} & 64.472 & C''_{2v} & 71.535 & C'_{2v} & 81.36 & S_4 & 90.99 & D'_2 & 98.162 & D_4 & 106.225 & C_{4v} & 115.288 & D_{2d}^{5} \\
		59.410 & C'_{2h} & 64.473 & C_{2v} & 71.536 & C''_{2h} & 81.37 & S_4 & 90.100 & D_4 & 99.163 & C_{4v} & 106.226 & C_{4v} & 115.289 & D_{2d}^{5} \\
		59.411 & D_2 & 64.474 & C''_{2h} & 71.537 & D_2 & 81.38 & S_4 & 90.101 & D_4 & 99.164 & C_{4v} & 107.227 & C_{4v} & 115.290 & D_{2d}^{5} \\
		59.412 & D_{2h} & 64.475 & C'_{2h} & 71.538 & D_{2h} & 82.39 & S_4 & 90.102 & D_4 & 99.165 & C'''_{2v} & 107.228 & C_{4v} & 116.291 & D_{2d}^{5} \\
		59.413 & D_{2h} & 64.476 & C_{2h} & 72.539 & D_{2h} & 82.40 & S_4 & 91.103 & D_4 & 99.166 & C_{2v} & 107.229 & C'''_{2v} & 116.292 & D_{2d}^{5} \\
\end{array}
	\end{align*}
\end{table}

\begin{table}[H]
	\begin{align*}
		\begin{array}{cc||cc||cc||cc||cc||cc||cc||cc}
		\text{MSG} & \text{PG} & \text{MSG} & \text{PG} & \text{MSG} & \text{PG} & \text{MSG} & \text{PG} & \text{MSG} & \text{PG} & \text{MSG} & \text{PG} & \text{MSG} & \text{PG} & \text{MSG} & \text{PG}\\
		\hline
		123.349 & D_{4h} & 129.412 & D_{4h} & 134.475 & D_{2h} & 139.538 & D_{2d}^{5} & 151.31 & C_3 & 165.94 & D_{3}^{2} & 178.157 & D_{3}^{1}& 188.220 & D_{3h}^{1} \\
		123.350 & D_{4h} & 129.413 & C_{4v} & 134.476 & D_{2d}^{1} & 139.539 & D_4 & 151.32 & D_{3}^{1}& 165.95 & C_{3i} & 178.158 & D_{3}^{2} & 189.221 & D_{3h}^{3} \\
		124.351 & D_{4h} & 129.414 & D'_{2h} & 134.477 & C_{4h} & 139.540 & D_{4h} & 152.33 & D_{3}^{2} & 165.96 & D_{3d}^{3} & 178.159 & C_6 & 189.222 & D_{3h}^{3} \\
		124.352 & D_{4h} & 129.415 & D_{2h} & 134.478 & D_{2d}^{5} & 140.541 & D_{4h} & 152.34 & D_{3}^{2} & 166.97 & D_{3d}^{3} & 178.160 & D_6 & 189.223 & C_{3v}^{2} \\
		124.353 & C_{4v} & 129.416 & D_{2d}^{1} & 134.479 & D_4 & 140.542 & D_{4h} & 152.35 & C_3 & 166.98 & D_{3d}^{3} & 179.161 & D_6 & 189.224 & D_{3}^{2} \\
		124.354 & D'_{2h} & 129.417 & C_{4h} & 134.480 & D_{4h} & 140.543 & C_{4v} & 152.36 & D_{3}^{2} & 166.99 & C_{3v}^{1} & 179.162 & D_6 & 189.225 & C_{3h} \\
		124.355 & D_{2h} & 129.418 & D_{2d}^{5} & 134.481 & D_{4h} & 140.544 & D'_{2h} & 153.37 & D_{3}^{1}& 166.100 & D_{3}^{2} & 179.163 & D_{3}^{1}& 189.226 & D_{3h}^{3} \\
\end{array}
	\end{align*}
\end{table}

\begin{table}[H]
	\begin{align*}
	\begin{array}{cc||cc||cc}
	\text{MSG} & \text{PG} & \text{MSG} & \text{PG} & \text{MSG} & \text{PG}\\
	\hline
	198.11 & T & 216.74 & T_d & 228.137 & T_h \\
	199.12 & T & 216.75 & T_d & 228.138 & O \\
	199.13 & T & 216.76 & T & 228.139 & O_h \\
	200.14 & T_h & 216.77 & T_d & 229.140 & O_h \\
	200.15 & T_h & 217.78 & T_d & 229.141 & O_h \\
	200.16 & T & 217.79 & T_d & 229.142 & T_d \\
	200.17 & T_h & 217.80 & T & 229.143 & T_h \\
\end{array}
	\end{align*}
\end{table}


\end{document}